\documentclass[10pt]{article}

\usepackage{authblk}
\usepackage{natbib}
\usepackage{mathtools}
\usepackage{amsmath}
\usepackage{amsthm}
\usepackage{amssymb}
\usepackage{amsbsy}
\usepackage{amsfonts}
\usepackage{amscd}
\usepackage{mathrsfs}
\usepackage{bm}
\usepackage{breqn}
\usepackage{color, soul}
\usepackage{enumerate}
\usepackage{empheq}
\usepackage{rotating}
\usepackage{ftnxtra}
\usepackage{fnpos}
\usepackage[titletoc,toc,title]{appendix}
\usepackage{euscript}
\usepackage{graphicx}
\usepackage{epsfig}
\usepackage{epstopdf}
\DeclareGraphicsExtensions{.pdf,.png,.jpg,.eps}
\usepackage{pstool}
\usepackage{upgreek}
\usepackage{mathrsfs}

\usepackage{psfrag}

\numberwithin{equation}{section}

\theoremstyle{plain}	
\newtheorem{thm}{Theorem}[section]

\newtheorem{lem}[thm]{Lemma}
\newtheorem{prop}[thm]{Proposition}
\newtheorem*{prop*}{Proposition} 
\theoremstyle{definition}	
\newtheorem{defi}[thm]{Definition}
\newtheorem{remark}[thm]{Remark}

\usepackage{graphicx}
\usepackage{lipsum}

\usepackage{caption}
\usepackage{subcaption}

\usepackage{hyperref}
\hypersetup{colorlinks=true, linkcolor=blue}
\hypersetup{colorlinks=true,citecolor=blue}

\DeclareMathAlphabet{\mathpzc}{OT1}{pzc}{m}{it}

\usepackage{amsmath, amsthm, amssymb}

\usepackage{cleveref}

\DeclarePairedDelimiter\abs{\lvert}{\rvert}

\makeatletter
\newsavebox{\@brx}
\newcommand{\llangle}[1][]{\savebox{\@brx}{\(\m@th{#1\langle}\)}%
  \mathopen{\copy\@brx\mkern2mu\kern-0.9\wd\@brx\usebox{\@brx}}}
\newcommand{\rrangle}[1][]{\savebox{\@brx}{\(\m@th{#1\rangle}\)}%
  \mathclose{\copy\@brx\mkern2mu\kern-0.9\wd\@brx\usebox{\@brx}}}%
\let\oldabs\abs
\def\abs{\@ifstar{\oldabs}{\oldabs*}}
\makeatother

\usepackage{accents}

    {\end{bmatrix}}%

\usepackage{enumitem}

\usepackage[utf8]{inputenc}
\usepackage{dutchcal}
\usepackage{multirow}

\renewcommand{\arraystretch}{2.0}

\usepackage{xcolor}

\usepackage{setspace}

\usepackage{hyperref} 

\newcommand{\n}{\accentset{1}{\boldsymbol{\mathsf{n}}}}
\newcommand{\nc}{\accentset{1}{\mathsf{n}}}
\newcommand{\nn}{\accentset{2}{\boldsymbol{\mathsf{n}}}}

\usepackage{adjustbox}
\usepackage{graphicx}

\begin{document}


\title{\textbf{On Universal Deformations of Compressible Cauchy Elastic Solids Reinforced by Inextensible Fibers}}

\author[1,2]{Arash Yavari\thanks{Corresponding author, e-mail: arash.yavari@ce.gatech.edu}}
\affil[1]{\small \textit{School of Civil and Environmental Engineering, Georgia Institute of Technology, Atlanta, GA 30332, USA}}
\affil[2]{\small \textit{The George W. Woodruff School of Mechanical Engineering, Georgia Institute of Technology, Atlanta, GA 30332, USA}}

\maketitle

\begin{abstract}
Universal deformations are those that can be maintained in the absence of body forces and with boundary tractions alone, for all materials within a given constitutive class.
We study the universal deformations of compressible isotropic Cauchy elastic solids reinforced by a single family of inextensible fibers. We consider straight fibers parallel to the Cartesian $Z$-axis in the reference configuration and derive the associated universality constraints, which depend explicitly on the geometry of the deformed fibers.
We study universal deformations in two cases: (i) deformed fibers are straight lines, and (ii) deformed fibers have non-vanishing curvature. For case (i), we provide a complete classification. Assuming that at least one principal invariant of the right Cauchy-Green tensor is not constant, we show that the deformed fiber direction must be an eigenvector of the Finger tensor, and the invariants depend only on the fiber arclength parameter. The universality constraints reduce to geometric restrictions on the orthogonal surfaces, which must be planes, circular cylinders, or spheres. 
This gives one inhomogeneous universal deformation family: the non-isochoric \emph{Family $Z_1$} of combined bending and stretching deformations. In addition, \emph{Family $0Z$} consists of homogeneous deformations that respect the inextensibility constraint.
We further show that if all principal invariants are constant and deformed fibers remain straight, then only homogeneous universal deformations are possible.
For case (ii), when deformed fibers have non-vanishing curvature, the universality constraints become significantly more complex. 
We show that the three principal invariants are functionally dependent and that the binormal to the deformed fibers is an eigenvector of the Finger tensor.
The existence of universal deformations in this case remains an open problem. In particular, we demonstrate that Family $5$ universal deformations of incompressible elasticity, when restricted to satisfy the inextensibility constraint, are no longer universal in fiber-reinforced solids.
Finally, we prove that the universal deformations of Cauchy and hyperelastic solids with the same fiber reinforcement coincide. Our results provide the first systematic classification of universal deformations for compressible isotropic fiber-reinforced solids and include a new inhomogeneous family. These solutions may serve as benchmark problems for numerical methods. 
\end{abstract}

\begin{description}
\item[Keywords:] Universal deformations, fiber-reinforced solids, inextensible fibers, inextensibility constraint, Cauchy elasticity, Hyper-elasticity.
\end{description}

\tableofcontents

\section{Introduction}

A \emph{universal deformation} is one that can be maintained in the absence of body forces for all materials within a given class. In other words, a universal deformation of a body can be maintained by applying only boundary tractions, regardless of the particular material chosen from the specified class---for example, homogeneous compressible isotropic solids or homogeneous incompressible isotropic solids. In the context of nonlinear elasticity, universal deformations have played an important role both experimentally \citep{Rivlin1951} and theoretically \citep{Tadmor2012,Goriely2017}.

The notion of universal deformations was introduced by Jerry Ericksen in two seminal papers \citep{Ericksen1954,Ericksen1955}. In \citep{Ericksen1955}, he showed that for homogeneous compressible isotropic solids, all universal deformations must necessarily be homogeneous. Ericksen's investigation of universal deformations in homogeneous incompressible isotropic solids \citep{Ericksen1954} was motivated by earlier work of Ronald Rivlin \citep{Rivlin1948,Rivlin1949a,Rivlin1949b}. Characterizing universal deformations in the presence of internal constraints is a more difficult problem \citep{Saccomandi2001}. Other than homogeneous isochoric deformations, \citet{Ericksen1954} identified four families of universal deformations in incompressible isotropic elastic solids. Subsequently, a fifth family was discovered independently by \citet{SinghPipkin1965} and \citet{KlingbeilShield1966}. Ericksen had conjectured that deformations with constant principal invariants must be homogeneous; this was later shown to be false by \citet{Fosdick1966}. In fact, the deformations in the fifth family are inhomogeneous but have constant principal invariants.\footnote{Other examples of inhomogeneous deformations with constant principal invariants exist \citep{Yin1983}, but they are not universal.}
It is still unknown whether additional inhomogeneous universal deformations with constant principal invariants exist.

Ericksen’s study of universal deformations has since been extended to various settings, including inhomogeneous isotropic elasticity (both compressible and incompressible) \citep{Yavari2021}, anisotropic elasticity \citep{YavariGoriely2021,Yavari2022Universal}, and anelasticity \citep{YavariGoriely2016,Goodbrake2020}. In linear elasticity, the analogue of universal deformations is the concept of \emph{universal displacements} \citep{Truesdell1966,Gurtin1972,Yavari2020,YavariGoriely2022}. For compressible anisotropic linear elastic solids, universal displacements were classified for all eight symmetry classes in \citep{Yavari2020}. In particular, it was shown that the higher the symmetry group, the larger the space of universal displacements. Thus, isotropic solids admit the largest class of universal displacements, while triclinic solids admit the smallest. This analysis has also been extended to inhomogeneous solids \citep{YavariGoriely2022} and to linear anelasticity \citep{Yavari2022Anelastic-Universality}.

Recently, the study of universal deformations has been extended to Cauchy elasticity, which includes hyperelastic (Green elastic) solids as a special case but does not necessarily assume the existence of an energy function \citep{Yavari2024Cauchy}. For both compressible and incompressible inhomogeneous isotropic Cauchy elastic solids, it was shown that the sets of universal deformations and universal inhomogeneities are identical to those of hyperelasticity, despite the more general constitutive structure.
The universal displacements of anisotropic linear Cauchy elastic solids have also been systematically characterized \citep{YavariSfyris2025}. In contrast to linear hyperelasticity, Cauchy elasticity does not require the existence of an energy function and allows for more general constitutive laws. Despite this greater generality, it was shown that for all eight symmetry classes of linear elasticity, the set of universal displacements in Cauchy elasticity coincides exactly with that of linear hyperelasticity. 

Universal deformations have also been studied in compressible isotropic implicit elasticity, a broader class of elastic solids whose constitutive equations take the implicit form $\boldsymbol{\mathcal{F}}(\boldsymbol{\sigma}, \mathbf{b}) = \mathbf{0}$, where $\boldsymbol{\sigma}$ is the Cauchy stress and $\mathbf{b}$ is the Finger tensor \citep{Morgan1966,Rajagopal2003,Rajagopal2007}. It has been shown that all universal deformations in this setting are homogeneous \citep{Yavari2024ImplicitElasticity}. However, unlike Cauchy or Green elasticity, not every homogeneous deformation is constitutively admissible in implicit elasticity. As a result, the set of universal deformations is material-dependent but always contained within the set of homogeneous deformations. This highlights an important distinction between implicit and classical forms of elasticity.

A class of solids with internal constraints that frequently arises in engineering applications is that of materials reinforced with inextensible fibers \citep{Rivlin1955Fibers,Adkins1956,Pipkin1971,Pipkin1974,Pipkin1979,Pipkin1980,Erdemir2007}.\footnote{These are sometimes called \textit{ideal fiber-reinforced composites} \citep{Rogers1984}.} 
A compressible solid reinforced with inextensible fibers provides a simple yet useful idealization of many natural and engineered materials composed of a soft matrix reinforced by a family of stiff fibers. 
The literature on universal deformations in fiber-reinforced solids remains limited. \citet{Beskos1972} considered homogeneous compressible isotropic solids reinforced with inextensible fibers and examined whether the universal deformations of incompressible isotropic solids remain universal in the fiber-reinforced setting. In particular, Families~$1$ through~$4$ were analyzed. It was shown that certain subsets of these families are indeed universal for specific fiber distributions. All such deformations are homogeneous, except for the shearing of a circular tube with circumferential fibers.
A similar study was carried out for incompressible isotropic hyperelastic solids reinforced by a family of inextensible fibers in \citep{Beskos1973}.
Universal relations for these two classes of materials have been studied in \citep{Saccomandi2002}.
\citet{Beatty1978,Beatty1989} studied homogeneous compressible isotropic solids reinforced with a single family of inextensible fibers and studied the problem of identifying all fiber distributions for which homogeneous deformations are universal. 
He showed that only three such types of fiber distributions exist, and in all three cases the fibers remain straight lines in the deformed configuration and, consequently, are also straight lines in the reference configuration.

More recently, universal displacements in fiber-reinforced anisotropic linear elastic solids were investigated in \citep{Yavari2024Fibers}. Specifically, compressible solids reinforced with a uniform distribution of inextensible straight fibers parallel to the $x_3$-axis were considered. For each of the seven anisotropy classes compatible with this reinforcement (excluding isotropy), the corresponding space of universal displacements was characterized. It was shown that the inclusion of inextensible fibers enlarges the space of universal displacements in all but the triclinic and cubic symmetry classes. 

Other examples of internal constraints include: i) the \textit{Bell constraint}, defined by $\lambda_1 + \lambda_2 + \lambda_3 = 3$, where $\lambda_i$, $i = 1, 2, 3$, are the principal stretches \citep{Bell1985}; ii) the \textit{Ericksen constraint}, given by $\lambda_1^2 + \lambda_2^2 + \lambda_3^2 = 3$ \citep{Ericksen1986}; iii) \textit{inexpansibility constraint}, in which the body is foliated by surfaces whose area elements remain unchanged under deformation \citep{Kurashige1985}; and iv) \textit{in-plane rigidity constraint}, where the body consists of parallel rigid planes bonded by an elastic matrix \citep{Tommasi1996}.\footnote{I am grateful to Giuseppe Saccomandi for bringing to my attention the references \citep{Kurashige1985, Tommasi1996}.}
See also \citep{Beatty1992-I,Beatty1992-II,Pucci1996,Martins1998,Destrade2004}.

It should be emphasized that the study of universal deformations in the presence of internal constraints has thus far been carried out systematically only for the incompressibility constraint and for elastic bodies reinforced by a family of inextensible planes. 
In the latter case, \citet{Tommasi1996} identified three families of universal deformations: (i) combined torsion about an axis normal to the reinforcing planes and uniform extension along that axis, (ii) combined shearing and bending, where the bending resembles that of classical beam theory in which planar cross-sections normal to the axis remain planar after deformation, and (iii) combined uniform extension and generalized shear, where the shear varies along the direction normal to the reinforcing planes.

The characterization of universal deformations for compressible isotropic solids reinforced by a single family of inextensible fibers has not been systematically studied in the literature. For fiber-reinforced solids, known universal families from incompressible isotropic elasticity have been examined to assess whether they remain universal under added internal constraints. However, a systematic study of universal deformations does not appear to exist in the literature.
\citet{Beskos1972} remarked: ``It is of interest to attack the problem of determining all possible universal solutions for a compressible isotropic elastic material reinforced with a system of fibers." This is precisely the aim of the present work. Our goal is to formulate this problem in detail and to identify the corresponding classes of universal deformations for a system of straight fibers.

This paper is organized as follows. In \S\ref{Sec:NonlinearElasticity} we review the kinematics and governing equations of nonlinear elasticity, and introduce the model of compressible isotropic Cauchy elastic solids reinforced by a family of inextensible fibers. In \S\ref{Sec:SingleFamily} we determine the universal deformations of ideal fiber-reinforced compressible isotropic Cauchy elastic solids. In \S\ref{Sec:Hyperelastic} we investigate the corresponding problem for compressible isotropic hyperelastic solids. Conclusions are given in \S\ref{Sec:Conclusions}.

\section{Nonlinear Elasticity} \label{Sec:NonlinearElasticity}

In this section, we briefly review nonlinear elasticity before posing and formulating the problem of determining universal deformations of a compressible, nonlinear, isotropic elastic body reinforced by a single family of distributed inextensible fibers.

\subsection{Kinematics}

In nonlinear elasticity a body $\mathcal{B}$ is identified with a flat Riemannian manifold $(\mathcal{B},\mathbf{G})$, which is a submanifold of the Euclidean $3$-space $(\mathcal{S},\mathbf{g})$ \citep{MarsdenHughes1994}. 
$\mathbf{G}$ is the material metric, which is induced from the ambient space metric $\mathbf{g}$. A deformation is a mapping $\varphi:\mathcal{B}\rightarrow \mathcal{S}$. 
The deformation gradient is the tangent map (or derivative) of $\varphi$ and is denoted by $\mathbf{F}=T\varphi$. 
The deformation gradient at each material point $\mathbf{X} \in \mathcal{B}$ is a linear map $\mathbf{F}(\mathbf{X}):T_{\mathbf{X}}\mathcal{B}\rightarrow T_{\varphi(\mathbf{X})}\mathcal{S}$. 
With respect to local (curvilinear) coordinates $\{x^a\}:\mathcal{C}\rightarrow\mathbb{R}^n$ and $\{X^A\}:\mathcal{B}\rightarrow\mathbb{R}^n$ on $\mathcal{S}$ and $\mathcal{B}$, respectively ($n=2$ or $3$), the deformation gradient has the components $F^a{}_{A}(\mathbf{X})=\frac{\partial \varphi^a}{\partial X^A}(\mathbf{X})$.
The dual of the deformation gradient, $\mathbf{F}^\star(\mathbf{X}) : T_{\varphi_t(\mathbf{X})}\mathcal{C}_t \to T_{\mathbf{X}}\mathcal{B}$, is defined by
\begin{equation}
    \mathbf{F}^\star = F^a{}_A\,dX^A \otimes \frac{\partial}{\partial x^a} \,.
\end{equation}
The transpose of deformation gradient is defined as
\begin{equation}
	\mathbf{F}^{\textsf{T}}:T_{\mathbf{x}}\mathcal{S}  
	\rightarrow T_{\mathbf{X}}\mathcal{B},\qquad 
	\llangle \mathbf{FV},\mathbf{v} \rrangle_{\mathbf{g}}    
	=\llangle\mathbf{V},\mathbf{F}^{\textsf{T}}\mathbf{v}\rrangle_{\mathbf{G}}, \qquad
	\forall \mathbf{V} \in T_{\mathbf{X}}\mathcal{B},~\mathbf{v} \in T_{\mathbf{x}} \mathcal{S}\,,
\end{equation}
which in components reads $(F^{\textsf{T}}(\mathbf{X}))^A{}_{a}=g_{ab}(\mathbf{x})\,F^b{}_{B}(\mathbf{X})\,G^{AB}(\mathbf{X})$.
Another measure of strain is the right Cauchy-Green deformation tensor (or strain), which is defined as $\mathbf{C}(X)=\mathbf{F}^{\textsf{T}}(\mathbf{X})\, \mathbf{F}(\mathbf{X}):T_{\mathbf{X}} \mathcal{B}\rightarrow T_{\mathbf{X}} \mathcal{B}$ and has components $C^A_{~B}=(F^{\textsf{T}})^A{}_{a}F^a{}_{B}$. 
Note that $C_{AB}=(g_{ab}\circ \varphi)\, F^a{}_{A}F^b{}_{B}$, which implies that the right Cauchy-Green strain is the pulled-back metric, i.e., $\mathbf{C}^\flat=\varphi^*\mathbf{g}=\mathbf{F}^{\star}\mathbf{G} \mathbf{F}$, where $\flat$ is the flat operator induced by the metric $\mathbf{g}$, and is used for lowering indices.
The left Cauchy-Green strain is defined as $\mathbf{B}^{\sharp}=\varphi^*(\mathbf{g}^{\sharp})=\mathbf{F}^{-1}\mathbf{g}^\sharp \mathbf{F}^{-\star}$, and has components $B^{AB}=F^{-A}{}_a\,F^{-B}{}_b\,g^{ab}$, where $F^{-A}{}_a$ are components of $\mathbf{F}^{-1}$. 
Note that $\mathbf{B}=\mathbf{C}^{-1}$.
The spatial analogues of $\mathbf{C}^\flat$ and $\mathbf{B}^{\sharp}$ are denoted by $\mathbf{c}^\flat$ and $\mathbf{b}^{\sharp}$, respectively, and are defined as
\begin{equation} \label{c-b-def}
\begin{aligned}
  	\mathbf{c}^\flat & =\varphi_*\mathbf{G}=\mathbf{F}^{-\star}\mathbf{G}^\sharp \mathbf{F}^{-1}\,, && \quad c_{ab}
 	=F^{-A}{}_a\,F^{-B}{}_b\,G_{AB}\,, \\
	\mathbf{b}^{\sharp} &=\varphi_*(\mathbf{G}^{\sharp})=\mathbf{F}\,\mathbf{G}^\sharp \mathbf{F}^\star\,,
	&& \quad b^{ab}=F^a{}_{A}F^b{}_{B}G^{AB}\,.
\end{aligned}
\end{equation}
$\mathbf{b}^{\sharp}$ is called the Finger deformation tensor. 
The tensors $\mathbf{C}$ and $\mathbf{b}$ have the same principal invariants $I_1$, $I_2$, and $I_3$, which are defined as \citep{Ogden1984,MarsdenHughes1994}
\begin{equation}\label{Principal-Invariants}
    I_1 =\operatorname{tr}\mathbf{b}=b^{ab}\,g_{ab}\,,\qquad
    I_2 =\frac{1}{2}\left(I_1^2-\operatorname{tr}\mathbf{b}^2\right)
    =\frac{1}{2}\left(I_1^2-b^{ab}b^{cd}\,g_{ac}\,g_{bd}\right)\,,\qquad
    I_3 =\det \mathbf{b}.
\end{equation}

\subsection{Balance of linear and angular momenta}

The balance of linear and angular momenta in the absence of inertial effects in material form read
\begin{equation}
  \operatorname{Div}\mathbf{P}+\rho_0\mathbf{B}=\mathbf{0}\,,\quad\quad
  \mathbf{P}\mathbf{F}^{\star}=\mathbf{F}\mathbf{P}^{\star}\,,
\end{equation}
where $\mathbf{B}$ is body force per unit undeformed volume, $\rho_0$ is the material mass density, and $\mathbf{P}$ is the first Piola-Kirchhoff stress. 
In components, these are written as
\begin{equation}
	P^{aA}{}_{|A}+\rho_0 B^a= \frac{\partial P^{aA}}{\partial X^A}+\Gamma^A{}_{AB} P^{aB}+\gamma^a{}_{bc} F^b{}_A P^{cA}+\rho_0 B^a=0 \,,\qquad
	P^{aA} F^b{}_{A}=F^a{}_{A} P^{aA}\,,
\end{equation}
where $\Gamma^C{}_{AB}$ and $\gamma^c{}_{ab}$ are the the Christoffel symbols of the material metric $\mathbf{G}$ and ambient space metric $\mathbf{g}$, respectively, and are defined as 
\begin{equation}
	\Gamma^A{}_{BC}=\frac{1}{2}G^{AK}\left(G_{KB,C}+G_{KC,B}-G_{BC,K}\right) \,,\qquad
	\gamma^a{}_{bc}=\frac{1}{2}g^{ak}\left(g_{kb,c}+g_{kc,b}-g_{bc,k}\right) \,.
\end{equation}

$\mathbf{P}$ is related to the Cauchy stress $\boldsymbol{\sigma}$ as $J\sigma^{ab}=P^{aA}F^b{}_A$, where $J$ is the Jacobian of deformation that relates the material ($dV$) and spatial ($dv$) Riemannian volume forms as $dv=J dV$, and is defined as 
\begin{equation}
	J=\sqrt{\frac{\det\mathbf{g}}{\det\mathbf{G}}}\det\mathbf{F} \,.
\end{equation}
In terms of the Cauchy stress $\boldsymbol{\sigma}$ the balance of linear and angular momenta read
\begin{equation}
  \operatorname{div}\boldsymbol{\sigma}+\rho\mathbf{b}=\mathbf{0}\,,\quad\quad
  \boldsymbol{\sigma}^{\star}=\boldsymbol{\sigma}\,,
\end{equation}
where $\mathbf{b}=\mathbf{B}\circ\varphi_t^{-1}$, and $\rho=J^{-1}\rho_0$ is the spatial mass density. In components, balance of linear momentum reads $\sigma^{ab}{}_{|b}+\rho b^a=0$, where
\begin{equation}
	\sigma^{ab}{}_{|b}=\sigma^{ab}{}_{,b}+\gamma^a{}_{bc}\sigma^{cb}
	+\gamma^b{}_{bc}\sigma^{ac}\,.
\end{equation}
Balance of angular momentum in components reads $\sigma^{ab}=\sigma^{ba}$.

\subsection{Constitutive equations of hyperelasticity}

In the case of an inhomogeneous isotropic hyperelastic (Green elastic) solid the energy function (per unit undeformed volume) is written as $W=\hat{W}(\mathbf{X},\mathbf{C}^\flat,\mathbf{G})$. For an isotropic solid, the energy function can be rewritten as $W=W(\mathbf{X},I_1,I_2,I_3)$, where $I_1, I_2$, and $I_3$ are the principal invariants of the right Cauchy-Green deformation tensor that are given in \eqref{Principal-Invariants}. The Cauchy stress has the following representation \citep{DoyleEricksen1956}
\begin{equation} \label{Cauchy-Compressible}
	\sigma^{ab}=\frac{2}{\sqrt{I_3}}\left[W_1b^{ab}+(I_2W_2+I_3W_3)g^{ab}-I_3W_2\,c^{ab}\right]
	\,,
\end{equation}  
where 
\begin{equation}
	W_i=W_i(\mathbf{X},I_1,I_2,I_3)=\frac{\partial W(\mathbf{X},I_1,I_2,I_3)}{\partial I_i},\quad i=1,2,3 \,,
\end{equation}  
and $c^{ab}=F^{-M}{}_m\,F^{-N}{}_n\,G_{MN}\,g^{am}\,g^{bn}$.

\subsection{Constitutive equations of Cauchy elasticity}

In Cauchy elasticity, the stress at a point and at a given moment in time is explicitly a function of the strain at that point and that particular moment in time \citep{Cauchy1828,Truesdell1952,TruesdellNoll2004}. However, an energy function does not necessarily exist.\footnote{It is important to note that Cauchy elasticity does not encompass all elastic solids. In recent years, there has been some interest in implicit constitutive equations, e.g., constitutive equations of the form $\boldsymbol{\mathcal{F}}(\boldsymbol{\sigma}, \mathbf{b}) = \mathbf{0}$ \citep{Morgan1966,Rajagopal2003,Rajagopal2007}. Cauchy elasticity is a subset of this broader class of solids.}
In terms of the first Piola--Kirchhoff stress, one has \citep{Truesdell1952,TruesdellNoll2004,Ogden1984}
\begin{equation}
	\mathbf{P}=\hat{\mathbf{P}}(X,\mathbf{F},\mathbf{G},\mathbf{g})\,.
\end{equation}
One can show that objectivity implies that the second Piola--Kirchhoff stress must have the following functional form \citep{TruesdellNoll2004}:
\begin{equation}
	\mathbf{S}=\hat{\mathbf{S}}(X,\mathbf{C}^\flat,\mathbf{G})\,.
\end{equation}
For an isotropic solid, one obtains the following classical representation \citep{RivlinEricksen1955,Wang1969,Boehler1977}:
\begin{equation}
	\mathbf{S}=\chi\, \mathbf{G}^\sharp + \xi\, \mathbf{C}^\sharp +\eta \, \mathbf{C}^{-\sharp}\,,
\end{equation}
where $\chi$, $\xi$, and $\eta$ are functions of $(X, I_1, I_2, I_3)$, and $\sharp$ denotes the sharp operator induced by the metric $\mathbf{G}$ (i.e., it raises indices).
For a recent study of Cauchy elasticity, see \citep{YavariGoriely2025Cauchy}.

\subsection{Compressible isotropic Cauchy elastic bodies reinforced by inextensible fibers} 

Let us consider a body $\mathcal{B}$ made of a compressible isotropic Cauchy elastic material that is reinforced by a single family of inextensible fibers. The unit tangent vector to the fiber at $\mathbf{X}\in\mathcal{B}$ is denoted by $\mathbf{N}=\mathbf{N}(\mathbf{X})$ and has components $N^A$ with respect to a coordinate chart $\{X^A\}$ in the reference configuration (note that $\mathbf{N}\cdot\mathbf{N} = \llangle \mathbf{N},\mathbf{N} \rrangle_{\mathbf{G}}=N^A\,N^B\,G_{AB}=1$). In the deformed configuration tangent to the same fiber at $\mathbf{x}=\varphi(\mathbf{X})$ is $\mathbf{n}=\mathbf{F}\mathbf{N}$, or $\mathbf{n}=\varphi_*\mathbf{N}$.
With respect to coordinate charts $\{X^A\}$ and $\{x^a\}$ in the reference and current configurations, respectively, $\mathbf{n}$ has components $n^a=F^a{}_A N^A$. Let us denote the length of an infinitesimal fiber at $\mathbf{X}\in\mathcal{B}$ by $d\ell_0$ and its length in the deformed configuration by $d\ell$. Parametrizing the curve that represents the fiber at $\mathbf{X}\in\mathcal{B}$ by a parameter $S$, one has
\begin{equation} 
	d\ell_0^2=N^AN^BG_{AB}\,dS^2=dS^2\,,\qquad d\ell^2=n^an^bg_{ab}\,dS^2    \,.
\end{equation}
For inextensible fibers $d\ell=d\ell_0$, and hence $\llangle \mathbf{n},\mathbf{n} \rrangle_{\mathbf{g}}=n^an^bg_{ab}=1$, i.e., $\mathbf{n}$ is a unit vector in the deformed configuration. 
In terms of the right Cauchy-Green strain the inextensibility constraints reads
\begin{equation} \label{Inextensibility}
	n^a\,n^b\,g_{ab}=F^a{}_A\,F^b{}_B\,g_{ab}\,N^A\,N^B=C_{AB}\,N^A\,N^B=1    \,.
\end{equation}
Thus, $\llangle \mathbf{N},\mathbf{N} \rrangle_{\mathbf{C}^{\flat}}=1$. Deformations that satisfy this constraint are called $\mathbf{N}$\emph{-isometric deformations}.

The Lagrange multiplier corresponding to the internal constraint \eqref{Inextensibility} is denoted by $T=T(\mathbf{x})$ and is called the \textit{tension field}. The Cauchy stress has the following representation \citep{Adkins1955,TruesdellNoll2004,Saccomandi2002}
\begin{equation} \label{Cauchy-Fiber}
	\boldsymbol{\sigma} =T \mathbf{n}\otimes\mathbf{n}+\bar{\boldsymbol{\sigma}}
	\,, 
\end{equation}
where $\bar{\boldsymbol{\sigma}}$ is the constitutive part of the Cauchy stress. For a homogeneous isotropic Cauchy elastic solid, it has the following representation 
\begin{equation} \label{Cauchy-Stress-Isotropic}
	\bar{\boldsymbol{\sigma}} =\alpha\, \mathbf{g}^\sharp+\beta \mathbf{b}^\sharp
	+\gamma\mathbf{c}^\sharp\,, 
\end{equation}
where $\alpha=\alpha(I_1,I_2,I_3)$, $\beta=\beta(I_1,I_2,I_3)$, and $\gamma=\gamma(I_1,I_2,I_3)$ are some arbitrary response functions.
In components, $\sigma^{ab}=T\,n^a n^b+\bar{\sigma}^{ab}$.

The representation \eqref{Cauchy-Fiber} is rewritten in terms of the second Piola-Kirchhoff stress as
\begin{equation} \label{S-Fiber}
	\mathbf{S} = \mathring{T} \, \mathbf{N} \otimes \mathbf{N} + \bar{\mathbf{S}} \,,
\end{equation}
where $\mathring{T} = J T$ and $\mathring{T} = \mathring{T}(\mathbf{X})$.
The constitutive part of the second Piola-Kirchhoff stress has the following classic representation \citep{RivlinEricksen1955,Wang1969,Boehler1977}
\begin{equation}
	\bar{\mathbf{S}}
	=\chi \mathbf{G}^\sharp+\xi \mathbf{C}^\sharp+\eta \mathbf{C}^{-\sharp}
	=\chi \mathbf{G}^\sharp+\xi \mathbf{C}^\sharp+\eta \mathbf{B}^{\sharp}\,,
\end{equation}
where $\chi=\chi(I_1,I_2,I_3)$, $\xi=\xi(I_1,I_2,I_3)$, $\eta=\eta(I_1,I_2,I_3)$.

The representation \eqref{Cauchy-Fiber} is rewritten in terms of the first Piola-Kirchhoff stress as
\begin{equation} \label{S-Fiber-1Piola}
	\mathbf{P} = \mathring{T} \, \mathbf{n} \otimes \mathbf{N} + \bar{\mathbf{P}} \,,
\end{equation}
where $\mathring{T} = J T$, $\mathring{T} = \mathring{T}(\mathbf{X})$, and $\bar{\mathbf{P}}= J \bar{\boldsymbol{\sigma}} \mathbf{F}^{-\star}$.

\section{Universal Deformations of Compressible Isotropic Cauchy Elastic Bodies Reinforced by a Single Family of Inextensible Fibers} \label{Sec:SingleFamily}

In this section, we formulate and partially solve the problem of determining the universal deformations of compressible elastic solids reinforced by inextensible fibers.

\subsection{Equilibrium equations in the absence of body forces} 

Let us consider a family of inextensible fibers that are straight lines in the undeformed configuration.\footnote{This is the simplest case. We are not considering other cases, as our goal is to obtain concrete, explicit results rather than pursue a more abstract or general formulation in this first systematic analysis of universal deformations in this class of solids.}
We choose a Cartesian coordinate system $\{X^1,X^2,X^3\}=\{X,Y,Z\}$ for the reference configuration such that fibers are parallel to the $Z$-axis. 
Equilibrium equations in the absence of body forces $\operatorname{Div}\mathbf{P}=\mathbf{0}$ are simplified to read
\begin{equation}
	\langle d\,\mathring{T},\mathbf{N} \rangle\,\mathbf{n}
	+\mathring{T}\, \nabla^{\mathbf{G}}_{\mathbf{N}} \mathbf{n} 
	+\mathring{T}\,(\operatorname{Div}\mathbf{N})\,\mathbf{n}
	+\operatorname{Div} \bar{\mathbf{P}}
	=\mathring{T}_{,Z}\,\mathbf{n}+\mathring{T}\,\mathbf{n}_{,Z} 
	+ J \operatorname{div}\bar{\boldsymbol{\sigma}}
	=\mathbf{0}\,,
\end{equation}  
where $\langle.,.\rangle$ is the natural pairing of $1$-forms and vectors, $\mathbf{n}$ is the (unit) fiber direction in the deformed configuration, $\mathbf{N} = \partial_Z$ is the reference fiber direction, and the fact that $\operatorname{Div}\mathbf{N} =0$ for straight fibers was used.
Equilibrium equations can be recast as a first-order partial differential equation (PDE):
\begin{equation} \label{Tension-Field-Equlibrium}
	(\mathring{T}\,\mathbf{n})_{,Z} =\mathbf{f}\,,\qquad \mathbf{f} 
	:= - J \operatorname{div}\bar{\boldsymbol{\sigma}} \,.
\end{equation}  
This is an overdetermined system of PDEs.
Eq.~\eqref{Tension-Field-Equlibrium}$_1$ represents a system of three scalar PDEs for a single unknown scalar tension field $\mathring{T}$.
Here, $\mathbf{n}$ is a unit vector field defined on the deformed configuration that encodes the orientation of inextensible fibers, and $\mathbf{f}$ is a deformation-dependent vector field.
Since $\mathbf{f}$ and $\mathbf{n}$ both depend on the deformation gradient $\mathbf{F}$, this system of three PDEs for the single unknown $\mathring{T}$ is generally overdetermined. The compatibility conditions associated with this system impose differential constraints on $\mathbf{F}$, namely, that the vector field $\mathbf{n} = \mathbf{F}\mathbf{N}$ must satisfy certain integrability conditions.

The force vector $\mathbf{f}$ is explicitly calculated as follows. From \eqref{Cauchy-Stress-Isotropic}, we have
\begin{equation} \label{Equilibrium-Compressible}
	\bar{\sigma}^{ab}{}_{|b}=
	\beta\,b^{ab}{}_{|b}+\gamma\,c^{ab}{}_{|b}+
	\alpha_{,b}\,g^{ab}+\beta_{,b}\,b^{ab}+\gamma_{,b}\,c^{ab} \,. 
\end{equation}
Notice that
\begin{equation}
\begin{dcases}
	\alpha_{,b}=\frac{\partial \alpha}{\partial I_1}I_{1,b}
	+\frac{\partial \alpha}{\partial I_2}I_{2,b}
	+\frac{\partial \alpha}{\partial I_3}I_{3,b}\,,\\
	 \beta_{,b}=\frac{\partial \beta}{\partial I_1}I_{1,b}
	+\frac{\partial \beta}{\partial I_2}I_{2,b}
	+\frac{\partial \beta}{\partial I_3}I_{3,b}\,,\\
	\gamma_{,b}=\frac{\partial \gamma}{\partial I_1}I_{1,b}
	+\frac{\partial \gamma}{\partial I_2}I_{2,b}
	+\frac{\partial \gamma}{\partial I_3}I_{3,b} \,.
\end{dcases}
\end{equation}
These can be written more compactly as
\begin{equation} \label{Coefficients-Derivatives}
\begin{dcases}
	\alpha_{,b}=\alpha_1\,I_{1,b}+\alpha_2\,I_{2,b}+\alpha_3\,I_{3,b}\,,\\
	\beta_{,b}=\beta_1\,I_{1,b}+\beta_2\,I_{2,b}+\beta_3\,I_{3,b}\,,\\
	\gamma_{,b}=\gamma_1\,I_{1,b}+\gamma_2\,I_{2,b}+\gamma_3\,I_{3,b} \,,
\end{dcases}
\end{equation}
where
\begin{equation} 
	\alpha_{i}=\frac{\partial \alpha}{\partial I_i}\,, \qquad
	\beta_{i}=\frac{\partial \beta}{\partial I_i}\,, \qquad
	\gamma_{i}=\frac{\partial \gamma}{\partial I_i}\,,\qquad i=1,2,3
	\,.
\end{equation}  
Hence
\begin{equation} 
\begin{aligned}
	\bar{\sigma}^{ab}{}_{|b} & =\beta\,b^{ab}{}_{|b}+\gamma\,c^{ab}{}_{|b} \\
	& \quad + I_{1,b}\,g^{ab}\,\alpha_1+I_{2,b}\,g^{ab}\,\alpha_2+I_{3,b}\,g^{ab}\,\alpha_3 \\
	& \quad + I_{1,b}\,b^{ab}\,\beta_1+I_{2,b}\,b^{ab}\,\beta_2+I_{3,b}\,b^{ab}\,\beta_3 \\
	& \quad + I_{1,b}\,c^{ab}\,\gamma_1+I_{2,b}\,c^{ab}\,\gamma_2+I_{3,b}\,c^{ab}\,\gamma_3 \,. 
\end{aligned}
\end{equation}
In coordinate-free form we have
\begin{equation} 
	\operatorname{div}\bar{\boldsymbol{\sigma}}= \beta \operatorname{div}\mathbf{b}^\sharp
	+\gamma \operatorname{div}\mathbf{c}^\sharp 
	+\sum_{i=1}^{3} 
	\left(\alpha_i \nabla I_i + \beta_i\, \mathbf{b}\cdot\nabla I_i
	+\gamma_i\,\mathbf{c}\cdot\nabla I_i\right)
	\,.
\end{equation}  
Therefore, 
\begin{equation} \label{ForcingTerm-f}
	\mathbf{f} = -J \beta \operatorname{div}\mathbf{b}^\sharp
	-J \gamma \operatorname{div}\mathbf{c}^\sharp 
	-J \sum_{i=1}^{3} 
	\left(\alpha_i \nabla I_i + \beta_i\, \mathbf{b}\cdot\nabla I_i
	+\gamma_i\,\mathbf{c}\cdot\nabla I_i\right)
	\,.
\end{equation}  

There are two cases that we study separately: (i) $\mathbf{n}_{,Z} = \mathbf{0}$ (fibers remain straight lines in the deformed configuration), and (ii) $\mathbf{n}_{,Z} \neq \mathbf{0}$ (deformed fibers have non-vanishing curvature). 
We will fully solve the problem of determining the universal deformations in case (i). For case (ii), we derive the corresponding universality constraints and make partial progress. A complete solution for this case remains to be obtained in future work.

\subsection{Integrability equations for the tension field when $\mathbf{n}_{,Z}=\mathbf{0}$} \label{Sec:Straight-Fibers}

When $\mathbf{n}_{,Z}=\mathbf{0}$, the equilibrium equations read: $\mathring{T}_{,Z}\,\mathbf{n}=\mathbf{f}$, and hence,  $\mathring{T}_{,Z}=\mathbf{f}\cdot\mathbf{n}$. This implies that $\mathbf{f}=(\mathbf{f}\cdot\mathbf{n})\mathbf{n}$. Therefore, $\mathbf{f}=\lambda \mathbf{n}$ and the integrability equation is
\begin{equation}\label{Integrability-Z2}
	\mathbf{f} \times \mathbf{n} = \mathbf{0}\,,
\end{equation}
which implies that $\mathbf{f}\in \operatorname{span} \{\mathbf{n}\}$.
Notice that $\mathbf{n}_{,Z}=\hat{\nabla}_{\mathbf{N}}\mathbf{n}=\nabla_{\mathbf{n}}\mathbf{n}$. In components, $(\hat{\nabla}_{\mathbf{N}}\mathbf{n})^a=n^a{}_{_A} N^A=n^a{}_{_b} F^b{}_A N^A=n^a{}_{_b} n^b$.
Note that $\nabla_{\mathbf{n}}\mathbf{n}=\mathbf{0}$ implies that fibers in the deformed configuration are geodesics. It is known that geodesics of the Euclidean space are straight lines \citep{doCarmo1992}. Therefore, in this case fibers in the deformed configuration are straight lines.

Substituting \eqref{ForcingTerm-f} into the integrability equation \eqref{Integrability-Z2} and recalling that the response functions and their derivatives are arbitrary, one obtains the following set of universality constraints:
\begin{empheq}[left={\empheqlbrace }]{align} 
	\label{Universality-Constraints-Cauchy2-1}
	& \mathbf{n} \times \operatorname{div}\mathbf{b}^\sharp=\mathbf{0}\,, \\
	\label{Universality-Constraints-Cauchy2-2}
	& \mathbf{n} \times \operatorname{div}\mathbf{c}^\sharp=\mathbf{0}\,, \\
	\label{Universality-Constraints-Cauchy2-3}
	& \mathbf{n} \times \nabla I_i=\mathbf{0}\,, \quad\qquad i=1,2,3\,, \\
	\label{Universality-Constraints-Cauchy2-4}
	& \mathbf{n} \times (\mathbf{b}\cdot\nabla I_i) = \mathbf{0}\,,\quad i=1,2,3\,, \\
	\label{Universality-Constraints-Cauchy2-5}
	& \mathbf{n} \times (\mathbf{c}\cdot\nabla I_i) = \mathbf{0}\,,\quad i=1,2,3\,,
\end{empheq}
which are equivalent to
\begin{empheq}[left={\empheqlbrace }]{align} 
	\label{Universality-Constraints-Cauchy3-1}
	& \operatorname{div}\mathbf{b}^\sharp= \beta \mathbf{n}\,, \\
	\label{Universality-Constraints-Cauchy3-2}
	& \operatorname{div}\mathbf{c}^\sharp= \gamma \mathbf{n}\,, \\
	\label{Universality-Constraints-Cauchy3-3}
	& \nabla I_i= \lambda_i \mathbf{n}\,, \qquad i=1,2,3\,, \\
	\label{Universality-Constraints-Cauchy3-4}
	& \mathbf{b}\cdot\nabla I_i= \xi_i \mathbf{n}\,,\quad i=1,2,3\,, \\
	\label{Universality-Constraints-Cauchy3-5}
	& \mathbf{c}\cdot\nabla I_i= \eta_i \mathbf{n}\,,\quad i=1,2,3\,,
\end{empheq}
for some scalar fields $\beta$, $\gamma$, $\lambda_i$, $\xi_i$, and $\eta_i$.
If $\nabla I_i\neq \mathbf{0}$, from \eqref{Universality-Constraints-Cauchy3-3} and \eqref{Universality-Constraints-Cauchy3-4} we conclude that $\mathbf{b}\cdot\mathbf{n} = \frac{\xi_i}{\lambda_i} \mathbf{n}$.
Similarly, from \eqref{Universality-Constraints-Cauchy3-3} and \eqref{Universality-Constraints-Cauchy3-5} we conclude that $\mathbf{c}\cdot \mathbf{n} = \frac{\eta_i}{\lambda_i} \mathbf{n}$.
Thus, $\mathbf{n}$ is an eigenvector of both $\mathbf{b}$ and $\mathbf{c}$, and hence $\eta_i \xi_i=\lambda_i^2$.
In summary, either all the principal invariants are constant or $\mathbf{n}$ is an eigenvector of both $\mathbf{b}$ and $\mathbf{c}$.

\subsubsection{At least one principal invariant is not constant}

When $\mathbf{n}$ is an eigenvector of $\mathbf{b}$, in components one has $b^a{}_b\,n^b = \mu^2\,n^a$ for some scalar $\mu$. 
Recall that in components, $n^a = F^a{}_A\,N^A$ and $b^{ab} = F^a{}_A\,F^b{}_B\,G^{AB}$. Let us start with $b^{ab}\,n_b = \mu^2\,n^a$.
Substituting $n_b = g_{bc}\,n^c = g_{bc}\,F^c{}_C\,N^C$, we get $b^{ab}\,g_{bc}\,F^c{}_C\,N^C = \mu^2\,F^a{}_A\,N^A$. Thus, $F^a{}_A\,F^b{}_B\,G^{AB}\,g_{bc}\,F^c{}_C\,N^C = \mu^2\,F^a{}_A\,N^A$.
This implies that $F^b{}_B\,g_{bc}\,F^c{}_C\,G^{AB}\,N^C = \mu^2\,N^A$.
Noting that $F^b{}_B\,g_{bc}\,F^c{}_C = C_{BC}$, this becomes
\begin{equation}
	G^{AB}\,C_{BC}\,N^C = \mu^2\,N^A\,,
\end{equation}
or equivalently,
\begin{equation} \label{C-Eigenvector}
	C^A{}_C\,N^C = \mu^2\,N^A\,.
\end{equation}
Thus, $\mathbf{N}$ is an eigenvector of $\mathbf{C}$ with eigenvalue $\mu^2$.
Therefore, $\mathbf{N} = \partial_Z$ is an eigenvector of $\mathbf{C}$. 
Now using \eqref{C-Eigenvector} and the inextensibility constraint we can write
\begin{equation}
	1=N_A\,C^A{}_C\,N^C = C_{BC} N^B N^C = 1 = \mu^2\,N^AN_A= \mu^2\,,
\end{equation}
and hence $\mu^2=1$.
Therefore, $\mathbf{C}^\flat$ has the following representation:
\begin{equation} \label{Fiber-Constrained-Strain}
	\mathbf{C}^\flat = \begin{bmatrix}
	C_{11}(X,Y,Z) & C_{12}(X,Y,Z) & 0 \\
	C_{12}(X,Y,Z) & C_{22}(X,Y,Z) & 0 \\
	0 & 0 & 1
	\end{bmatrix}\,,
\end{equation}
i.e., $C_{13}=C_{23}=0$.
This also implies that $\lambda_3=1$ is an eigenvalue of $\mathbf{C}$ (and of $\mathbf{b}$). The principal invariants 
$I_1=\lambda_1^2+\lambda_2^2+\lambda_3^2$, $I_2=\lambda_1^2\lambda_2^2+\lambda_1^2\lambda_3^2+\lambda_2^2\lambda_3^2$, and 
$I_3=\lambda_1^2\lambda_2^2\lambda_3^2$ are therefore functionally dependent. 
Since $\lambda_3 = 1$, each invariant depends only on $\lambda_1$ and $\lambda_2$, and at most two of $I_1$, $I_2$, and $I_3$ are functionally independent.

The universality constraint \eqref{Universality-Constraints-Cauchy3-3} in components reads $I_{i,b} \,g^{ab} = \lambda_i \,n^a$ or equivalently,  $I_{i,b}  = \lambda_i \,n^a \,g_{ab}$. Thus, $I_{i,B} \,F^{-B}{}_b = \lambda_i \,n^a \,g_{ab} =\lambda_i \,F^a{}_A \,N^A \,g_{ab}$. Therefore
\begin{equation}
	I_{i,B}= \lambda_i \,F^a{}_A \,g_{ab} \,F^b{}_B \,N^A= \lambda_i \,C_{BA} \,N^A
	= \lambda_i \,C_{BA} \,\delta^A_3= \lambda_i \,C_{B3}\,.
\end{equation}
This implies that $I_{i,X}=I_{i,Y}=0$, and hence\footnote{If either $I_1$ or $I_2$ is constant they are functionally dependent. If $\nabla I_1 \neq \mathbf{0}$ and $\nabla I_2 \neq \mathbf{0}$, from \eqref{Universality-Constraints-Cauchy3-3} we have $\nabla I_1= \lambda_1 \mathbf{n}$ and $\nabla I_2= \lambda_2 \mathbf{n}$, which implies that $\nabla I_1$ and $\nabla I_2$ are parallel, and hence $I_1$ and $I_2$ are functionally dependent. This is consistent with \eqref{Principal-Invariants-Z}.}
\begin{equation} \label{Principal-Invariants-Z}
	I_i=I_i(Z)\,,\qquad i=1,2,3\,.
\end{equation}
When the principal invariants only depend on $Z$, one concludes that the principal stretches of $\mathbf{C}^\flat$ depend only on $Z$ as well. This implies that
\begin{equation} 
	\begin{bmatrix}
	C_{11} & C_{12}  \\
	C_{12} & C_{22} 
	\end{bmatrix}
	=\begin{bmatrix}
	\cos \Theta & \sin \Theta \\
	-\sin \Theta & \cos \Theta
	\end{bmatrix}
	\begin{bmatrix}
	\lambda_1^2(Z) & 0  \\
	0 & \lambda_2^2(Z)
	\end{bmatrix}
	\begin{bmatrix}
	\cos \Theta & -\sin \Theta \\
	\sin \Theta & \cos \Theta
	\end{bmatrix}
	\,,
\end{equation}
where $\Theta=\Theta(X,Y,Z)$, and $\lambda_1(Z)$ and $\lambda_2(Z)$ are the principal stretches ($\lambda_3=1$).
Thus
\begin{equation}
\begin{dcases}
	C_{11}(X,Y,Z) = \lambda_1^2(Z)\,\cos^2\Theta(X,Y,Z) + \lambda_2^2(Z)\,\sin^2\Theta(X,Y,Z)\,, \\
	C_{12}(X,Y,Z) = \tfrac{1}{2}\left( \lambda_1^2(Z) - \lambda_2^2(Z) \right) \sin 2\Theta(X,Y,Z)\,, \\
	C_{22}(X,Y,Z) = \lambda_1^2(Z)\,\sin^2\Theta(X,Y,Z) + \lambda_2^2(Z)\,\cos^2\Theta(X,Y,Z)\,.
\end{dcases}
\end{equation}

\begin{remark}
If $\lambda_1(Z)=\lambda_2(Z)$, it is straightforward to see that
\begin{equation} 
	\begin{bmatrix}
	C_{11} & C_{12}  \\
	C_{12} & C_{22} 
	\end{bmatrix}
	=
	\lambda_1^2(Z) \begin{bmatrix}
	1 & 0  \\
	0 & 1
	\end{bmatrix}
	\,,
\end{equation}
which is only a function of $Z$. 
The compatibility equation for the right Cauchy-Green strain in a simply-connected body is the vanishing of the Riemann curvature of $\mathbf{C}^\flat$, which, in three dimensions, is equivalent to the vanishing of its Ricci curvature \citep{Berger2003,Yavari2013}.
The Ricci curvature in this case reads
\begin{equation}
	\mathbf{Ric}(\mathbf{C}^\flat)=
	\begin{bmatrix}
	\left(\lambda'\right)^2 + \lambda\, \lambda'' & 0 & 0 \\
	0 & \left(\lambda'\right)^2 + \lambda\, \lambda'' & 0 \\
	0 & 0 & \dfrac{2\, \lambda''}{\lambda}
	\end{bmatrix}\,.
\end{equation}
Compatibility equations $\mathbf{Ric}=\mathbf{0}$ imply that $\lambda'(Z)=0$, and hence, the right Cauchy-Green strain is constant. This implies that the corresponding deformations are homogeneous \citep[Theorem~1.3]{Blume1989}.
\end{remark}

\subsubsection{The universality constraints $\operatorname{div}\mathbf{b}^\sharp= \beta \mathbf{n}$ and $\operatorname{div}\mathbf{c}^\sharp= \gamma \mathbf{n}$} 

The eigenvalues of $\mathbf{b}^\sharp$ fall into the following categories: (i) all eigenvalues are distinct, (ii) $\Lambda_1 = \Lambda_2$, and (iii) either $\Lambda_1 = 1$ or $\Lambda_2 = 1$, where $\Lambda_1=\lambda_1^2$ and $\Lambda_2=\lambda_2^2$. We have already demonstrated that deformations corresponding to case (ii) are homogeneous. We now proceed to analyze cases (i) and (iii).

The symmetric $(1,1)$-tensor $\mathbf{b}$ admits the spectral decomposition
\begin{equation}
	\mathbf{b}^\sharp =  \mathbf{n} \otimes \mathbf{n}+ \Lambda_1\, \n \otimes \n 
	+ \Lambda_2\, \nn \otimes \nn 
	 \,,
\end{equation}
where $\Lambda_1,\Lambda_2>0$, and $\{\n, \nn\}$ is an orthonormal basis for the plane normal to $\mathbf{n}$.

\paragraph{Case (i) Principal stretches are distinct.}
We know that 
\begin{equation} \label{n-n1-n2-g}
	\mathbf{n}\otimes\mathbf{n} + \n \otimes \n + \nn \otimes \nn=\mathbf{g}^\sharp
	\,,
\end{equation}
and hence,
\begin{equation} \label{b-Spectral1}
	\mathbf{b}^\sharp =  (1-\Lambda_2)\, \mathbf{n} \otimes \mathbf{n}
	+ (\Lambda_1-\Lambda_2)\, \n \otimes \n 
	+ \Lambda_2\, \mathbf{g}^\sharp 
	\,.
\end{equation}
Thus
\begin{equation} \label{c-Spectral1}
	\mathbf{c}^\sharp = \frac{1}{1-\Lambda_2}\, \mathbf{n} \otimes \mathbf{n}
	+ \frac{1}{\Lambda_1-\Lambda_2}\, \n \otimes \n 
	+ \frac{1}{\Lambda_2}\, \mathbf{g}^\sharp 
	\,.
\end{equation}
Now the divergence of Finger tensor is written as
\begin{equation}
\begin{aligned}
	\operatorname{div} \mathbf{b}^\sharp 
	&= \left[ -\nabla \Lambda_2 \cdot \mathbf{n} 
	+ (1 - \Lambda_2)\, \operatorname{div} \mathbf{n} \right] \mathbf{n} 
	+ (1 - \Lambda_2)\, \nabla_{\mathbf{n}} \mathbf{n} \\
	&\quad + \left[ \nabla (\Lambda_1 - \Lambda_2) \cdot \n 
	+ (\Lambda_1 - \Lambda_2)\, \operatorname{div} \n \right] \n 
	+ (\Lambda_1 - \Lambda_2)\, \nabla_{\n} \n + \nabla \Lambda_2\,.
\end{aligned}
\end{equation}
Recall that $\nabla_{\mathbf{n}} \mathbf{n}=\mathbf{n}_{,Z}=\mathbf{0}$, and hence,
\begin{equation}
\begin{aligned}
	\operatorname{div} \mathbf{b}^\sharp 
	&= \left[ -\nabla \Lambda_2 \cdot \mathbf{n} 
	+ (1 - \Lambda_2)\, \operatorname{div} \mathbf{n} \right] \mathbf{n} 
	+ \nabla \Lambda_2  \\
	&\quad + \left[ \nabla (\Lambda_1 - \Lambda_2) \cdot \n 
	+ (\Lambda_1 - \Lambda_2)\, \operatorname{div} \n \right] \n 
	+ (\Lambda_1 - \Lambda_2)\, \nabla_{\n} \n \,.
\end{aligned}
\end{equation}
From \eqref{Universality-Constraints-Cauchy3-1}, we know that $\operatorname{div} \mathbf{b}^\sharp \cdot \n=\operatorname{div} \mathbf{b}^\sharp \cdot \nn=0$, and therefore
\begin{equation}
	\nabla \Lambda_2\cdot \n  
	+ \left[ \nabla (\Lambda_1 - \Lambda_2) \cdot \n + (\Lambda_1 - \Lambda_2)\, \operatorname{div} \n \right] =0 
	\,, \qquad
	\nabla \Lambda_2\cdot \nn  + (\Lambda_1 - \Lambda_2)\, \nabla_{\n} \n \cdot \nn =0
	\,.
\end{equation}
This is simplified to read
\begin{equation} \label{n-n2-Identity-1}
	\nabla \Lambda_1 \cdot \n + (\Lambda_1 - \Lambda_2)\, \operatorname{div} \n  =0 \,, \qquad
	\nabla \Lambda_2\cdot \nn  + (\Lambda_1 - \Lambda_2)\, \nabla_{\n} \n \cdot \nn =0
	\,.
\end{equation}
Similarly, one can write
\begin{equation}
\begin{aligned}
	\operatorname{div} \mathbf{c}^\sharp 
	&= \left[ \frac{1}{(1 - \Lambda_2)^2} \nabla \Lambda_2 \cdot \mathbf{n} 
	+ \frac{1}{1 - \Lambda_2}\, \operatorname{div} \mathbf{n} \right] \mathbf{n} 
	-\frac{1}{\Lambda_2^2} \nabla \Lambda_2  \\
	&\quad + \left[ -\frac{1}{(1 - \Lambda_2)^2} \,\nabla (\Lambda_1 - \Lambda_2) \cdot \n 
	+ \frac{1}{\Lambda_1 - \Lambda_2}\, \operatorname{div} \n \right] \n 
	+ (\Lambda_1 - \Lambda_2)\, \nabla_{\n} \n \,.
\end{aligned}
\end{equation}
From \eqref{Universality-Constraints-Cauchy3-2}, we know that $\operatorname{div} \mathbf{c}^\sharp \cdot \n=\operatorname{div} \mathbf{c}^\sharp \cdot \nn=0$, and therefore
\begin{equation}
\begin{aligned}
	& -\frac{1}{\Lambda_2^2} \nabla \Lambda_2 \cdot \n
	+ \left[ -\frac{1}{(1 - \Lambda_2)^2} \,\nabla (\Lambda_1 - \Lambda_2) \cdot \n
	+ \frac{1}{\Lambda_1 - \Lambda_2}\, \operatorname{div} \n \right]=0\,,  \\
	& -\frac{1}{\Lambda_2^2} \nabla \Lambda_2 \cdot \nn
	+ (\Lambda_1 - \Lambda_2)\, \nabla_{\n} \n \cdot \nn =0
	\,.
\end{aligned}
\end{equation}
This is simplified to read
\begin{equation} \label{n-n2-Identity-2}
\begin{aligned}
	&  -\frac{1}{(1 - \Lambda_2)^2} \,\nabla \Lambda_1  \cdot \n
	+\frac{\Lambda (2\Lambda_2 - \Lambda)}{\Lambda_2^2\, (\Lambda - \Lambda_2)^2} 
	\, \nabla \Lambda_2 \cdot \n
	+ \frac{1}{\Lambda_1 - \Lambda_2}\, \operatorname{div} \n =0\,,  \\
	& -\frac{1}{\Lambda_2^2} \nabla \Lambda_2 \cdot \nn
	+ (\Lambda_1 - \Lambda_2)\, \nabla_{\n} \n \cdot \nn =0
	\,.
\end{aligned}
\end{equation}
First let us consider \eqref{n-n2-Identity-1}$_2$ and \eqref{n-n2-Identity-2}$_2$:
\begin{equation} 
\begin{dcases}
	\nabla \Lambda_2\cdot \nn  + (\Lambda_1 - \Lambda_2)\, \nabla_{\n} \n \cdot \nn =0 \,,\\
	-\frac{1}{\Lambda_2^2} \nabla \Lambda_2 \cdot \nn
	+ (\Lambda_1 - \Lambda_2)\, \nabla_{\n} \n \cdot \nn =0
	\,.
\end{dcases}
\end{equation}
When $\Lambda_1 \neq \Lambda_2$, one concludes that $\nabla \Lambda_2 \cdot \nn = 0\,, \nabla_{\n} \n \cdot \nn = 0$.
Knowing that $\nabla_{\n} \n \cdot \n = 0$ we conclude that $\nabla_{\n} \n$ is parallel to $\mathbf{n}$, i.e.,
\begin{equation} \label{n1-n-identity}
	\nabla_{\n} \n = \ell \mathbf{n}\,,
\end{equation}
for some scalar field $\ell$.

Instead of \eqref{b-Spectral1} and \eqref{c-Spectral1}, one can equivalently use the following spectral decomposition for $\mathbf{b}^\sharp$ and $\mathbf{c}^\sharp$:
\begin{equation}
\begin{aligned}
	\mathbf{b}^\sharp & = (1 - \Lambda_1)\, \mathbf{n} \otimes \mathbf{n} 
	+ (\Lambda_2 - \Lambda_1)\, \nn \otimes \nn 
	+ \Lambda_1\, \mathbf{g}^\sharp \,,\\
	\mathbf{c}^\sharp &= \frac{1}{1 - \Lambda_1}\, \mathbf{n} \otimes \mathbf{n} 
	+ \frac{1}{\Lambda_2 - \Lambda_1}\, \nn \otimes \nn 
	+ \frac{1}{\Lambda_1}\, \mathbf{g}^\sharp \,.
\end{aligned}
\end{equation}
The universality constraints  \eqref{Universality-Constraints-Cauchy3-1} and  \eqref{Universality-Constraints-Cauchy3-2} give us
\begin{equation} \label{n-n2-Identity-12}
\begin{aligned}
	& \nabla \Lambda_1 \cdot \n + (\Lambda_2 - \Lambda_1)\, \nabla_{\nn} \nn \cdot \n = 0\,,\\
	& \nabla \Lambda_2 \cdot \nn + (\Lambda_2 - \Lambda_1)\, \operatorname{div} \nn = 0\,,\\
	&  -\frac{1}{(1 - \Lambda_1)^2} \,\nabla \Lambda_2  \cdot \nn
	+\frac{\Lambda (2\Lambda_1 - 1)}{\Lambda_1^2\, (1 - \Lambda_1)^2} 
	\, \nabla \Lambda_1 \cdot \nn
	+ \frac{1}{\Lambda_2 - \Lambda_1}\, \operatorname{div} \nn =0\,,  \\
	& -\frac{1}{\Lambda_1^2} \nabla \Lambda_1 \cdot \n
	+ (\Lambda_2 - \Lambda_1)\, \nabla_{\nn} \nn \cdot \n =0
	\,.
\end{aligned}
\end{equation}
When $\Lambda_1 \neq \Lambda_2$, from the the first and fourth constraints one concludes that $\nabla \Lambda_1 \cdot \n = 0\,, \nabla_{\nn} \nn \cdot \n = 0$.
Knowing that $\nabla_{\nn} \nn \cdot \nn = 0$ we conclude that $\nabla_{\nn} \nn$ is parallel to $\mathbf{n}$.
The remaining universality constraints are
\begin{equation} 
\begin{aligned}
	& (\Lambda_1 - \Lambda_2)\, \operatorname{div} \n  =0 \,, \\
	& (\Lambda_2 - \Lambda_1)\, \operatorname{div} \nn = 0\,,\\
	& \frac{2\Lambda_2 - 1}{\Lambda_2^2\, (1 - \Lambda_2)^2} 
	\, \nabla \Lambda_2 \cdot \n
	+ \frac{1}{\Lambda_1 - \Lambda_2}\, \operatorname{div} \n =0\,,\\
	&  \frac{2\Lambda_1 - 1}{\Lambda_1^2\, (1 - \Lambda_1)^2} \, \nabla \Lambda_1 \cdot \nn
	+ \frac{1}{\Lambda_2 - \Lambda_1}\, \operatorname{div} \nn =0
	\,.
\end{aligned}
\end{equation}
Therefore, when $\Lambda_1 \neq \Lambda_2$ we conclude that $\operatorname{div} \n=\operatorname{div} \nn=0$.
If $\Lambda_1\neq \frac{1}{2}\Lambda$ and $\Lambda_2\neq \frac{1}{2}\Lambda$, one concludes that $\nabla \Lambda_2 \cdot \n=\nabla \Lambda_1 \cdot \nn=0$.
Therefore, $\Lambda_1$ and $\Lambda_2$ can vary only along $\mathbf{n}$ in this case.

\paragraph{Case (iii) $\Lambda_2=\Lambda=1$.}
In this case, the spectral decompositions \eqref{b-Spectral1} and \eqref{c-Spectral1} are simplified to read

\begin{equation}  \label{b-case-iii}
\begin{aligned}
	\mathbf{b}^\sharp & =   \mathbf{g}^\sharp +(\Lambda_1-1)\, \n \otimes \n 
	= \Lambda_1\, \mathbf{g}^\sharp+ (1-\Lambda_1)( \mathbf{n} \otimes \mathbf{n} + \nn \otimes \nn)	\\
	\mathbf{c}^\sharp & =  \mathbf{g}^\sharp+ \frac{1}{\Lambda_1-1}\, \n \otimes \n  
	= \frac{1}{\Lambda_1}\, \mathbf{g}^\sharp+ \frac{1}{1-\Lambda_1}\,( \mathbf{n} \otimes \mathbf{n} + \nn \otimes \nn)
	\,.
\end{aligned}
\end{equation}
Thus
\begin{equation}
\begin{aligned}
	\operatorname{div} \mathbf{b}^\sharp 
	&= \left[ \nabla \Lambda_1 \cdot \n + (\Lambda_1 - 1)\, \operatorname{div} \n \right] \n \\
	&= \nabla \Lambda_1 + \left[ -\nabla  \Lambda_1\cdot \mathbf{n} 
	+ (1 - \Lambda_1)\, \operatorname{div} \mathbf{n} \right] \mathbf{n} 
	+ \left[ -\nabla  \Lambda_1\cdot \n + (1 - \Lambda_1)\, \operatorname{div} \n \right] \n  \\
	\operatorname{div} \mathbf{c}^\sharp 
	&= \left[ -\frac{1}{(\Lambda_1 - 1)^2} \nabla \Lambda_1 \cdot \n 
	+ \frac{1}{\Lambda_1 - 1}\, \operatorname{div} \n \right] \n   \\
	&= -\frac{1}{\Lambda_1^2}\,\nabla \Lambda_1 
	+ \left[ \frac{1}{\Lambda_1^2}\,\nabla  \Lambda_1\cdot \mathbf{n} + \frac{1}{1 - \Lambda_1}\, \operatorname{div} \mathbf{n} \right] \mathbf{n} 
	+ \left[ \frac{1}{\Lambda_1^2}\,\nabla  \Lambda_1\cdot \n + \frac{1}{1 - \Lambda_1}\, \operatorname{div} \n \right] \n
	\,.
\end{aligned}
\end{equation}
The universal constraint $\operatorname{div} \mathbf{b}^\sharp \cdot \n  = 0$ implies that
\begin{equation}
	\nabla \Lambda_1 \cdot \n + (\Lambda_1-1) \operatorname{div} \n = 0 \,, \qquad (1-\Lambda_1)\operatorname{div} \n = 0 \,,
\end{equation}
and hence $\nabla \Lambda_1 \cdot \n=\operatorname{div} \n = 0$.
The universality constraint $\operatorname{div} \mathbf{b}^\sharp \cdot \nn=0$ gives us $\nabla \Lambda_1 \cdot \nn=0$.
Therefore
\begin{equation}
	\operatorname{div} \mathbf{b}^\sharp =\mathbf{0}
	= \nabla \Lambda_1 + \left[ -\nabla  \Lambda_1\cdot \mathbf{n} + (1 - \Lambda_1)\, \operatorname{div} \mathbf{n} \right] \mathbf{n} 
	\,.
\end{equation}
Dot product of both sides by $\mathbf{n}$ we obtain
\begin{equation}
	0=\nabla \Lambda_1\cdot \mathbf{n} 
	+ \left[ -\nabla  \Lambda_1\cdot \mathbf{n} + (1 - \Lambda_1)\, \operatorname{div} \mathbf{n} \right]
	=(1 - \Lambda_1)\, \operatorname{div} \mathbf{n}
	\,,
\end{equation}
and thus
\begin{equation}\label{Constraint-divn}
	\operatorname{div} \mathbf{n}=0 \,.
\end{equation}

In the reference configuration, fibers induce a foliation of the undeformed body, i.e., the body is partitioned into a continuous family of non-intersecting surfaces, much like the pages of a book. It is physically reasonable to expect that this foliation structure is preserved in the deformed configuration. Therefore, the deformed fibers define a foliation by surfaces, which we represent as $\psi(\mathbf{x})=c$ for some smooth scalar function $\psi$. Assuming the body is simply-connected, the necessary and sufficient condition for the existence of such a function is that $\mathbf{n}\cdot(\operatorname{curl}\mathbf{n})=0$, where $\mathbf{n}$ is the unit tangent vector to the deformed fibers. This is the Frobenius integrability condition. A stronger condition is $\operatorname{curl}\mathbf{n}=\mathbf{0}$, which guarantees the global existence of a potential $\psi$ whose level sets are the fiber surfaces, but this is more restrictive than physically required.

For a unit vector field $\mathbf{n}$, if $\operatorname{div}\mathbf{n} = 0$ and $\operatorname{curl}\mathbf{n} = \mathbf{0}$, then $\mathbf{n}$ must be constant. As a matter of fact, any irrotational vector field is locally the gradient of a scalar potential, so $\mathbf{n} = \nabla \psi$ for some scalar function $\psi$. The divergence-free condition then implies $\Delta \psi = 0$, i.e., $\psi$ is harmonic. However, the constraint $\|\nabla \psi\| = 1$ cannot hold globally for any nontrivial harmonic function \citep{Evans2010}. Thus, $\nabla \psi$ must be constant, and therefore $\mathbf{n} = \mathbf{n}_0$ is constant. This implies that the fibers in the deformed configuration are parallel straight lines and the level sets of $\psi$ are planes. We will see shortly that this necessarily forces the deformation to be homogeneous.

\subsubsection{Deformed fibers are straight lines} 

We know that the inextensible fibers in the deformed configuration are straight lines.
We seek the most general deformation that maps each material fiber (i.e., each line of constant $(X,Y)$ and varying $Z$) to a straight line in the deformed configuration.

\begin{lem}
Let $(X,Y,Z)$ be the Cartesian coordinates in the reference configuration, and $(x,y,z)$ be the Cartesian coordinates in the deformed configuration. The most general deformation that maps vertical lines to straight lines is
\begin{equation} \label{Deformation-Z-Lines}
\begin{dcases}
	x(X,Y,Z) = a_1(X,Y) + n_1(X,Y)\,Z\,, \\
	y(X,Y,Z) = a_2(X,Y) + n_2(X,Y)\,Z\,, \\
	z(X,Y,Z) = a_3(X,Y) + n_3(X,Y)\,Z\,,
\end{dcases}
\end{equation}
where $\mathbf{a}(X,Y) = \left(a_1(X,Y), a_2(X,Y), a_3(X,Y)\right)$ and $\mathbf{n}(X,Y) = \left(n_1(X,Y), n_2(X,Y), n_3(X,Y)\right)$ are smooth functions. 
\end{lem}

\begin{proof}
Consider a fiber in the reference configuration parameterized by $Z$ with fixed $(X_0, Y_0)$. Its parametric form is $\boldsymbol{\alpha}(Z)=(X_0, Y_0, Z)$ with $Z \in \mathbb{R}$. Its image under the deformation $\varphi: (X,Y,Z) \mapsto (x,y,z)$ is $\varphi(X_0, Y_0, Z)$. For the image to be a straight line, it must be representable as
\begin{equation}
	\boldsymbol{\alpha}(Z) = \mathbf{a}(X_0, Y_0) + \mathbf{n}(X_0, Y_0) Z,
\end{equation}
where $\mathbf{a}(X_0, Y_0)$ is a point on the line and $\mathbf{n}(X_0, Y_0)$ is a fixed direction vector.
Since this holds for all $(X,Y)$ and all $Z$, the deformation map must be of the form
\begin{equation} \label{Deformation-Z-Lines-Vector}
	\boldsymbol{\varphi}(X,Y,Z) = \mathbf{a}(X,Y) + \mathbf{n}(X,Y) Z,
\end{equation}
where $\mathbf{a} : \mathbb{R}^2 \to \mathbb{R}^3$ and $\mathbf{n} : \mathbb{R}^2 \to \mathbb{R}^3$ are smooth functions.
\end{proof}

Here, $\mathbf{a}(X,Y)$ represents the base point of the fiber in the deformed configuration, and $\mathbf{n}(X,Y)$ is the direction vector of the deformed fiber. The deformation gradient is given by
\begin{equation}
	\mathbf{F} = 
	\Big[
	\mathbf{a}_{,X} + Z\,\mathbf{n}_{,X} \quad \mathbf{a}_{,Y} + Z\,\mathbf{n}_{,Y} \quad  \mathbf{n}
	\Big]
	=
	\begin{bmatrix}
		a_{1,X} + n_{1,X}\,Z & a_{1,Y} + n_{1,Y}\,Z & n_1 \\
		a_{2,X} + n_{2,X}\,Z & a_{2,Y} + n_{2,Y}\,Z & n_2 \\
		a_{3,X} + n_{3,X}\,Z & a_{3,Y} + n_{3,Y}\,Z & n_3
	\end{bmatrix}\,.
\end{equation}
The inextensibility constraint implies that
\begin{equation}
	C_{33} = \|\mathbf{n}(X,Y)\|^2 = n_1^2 + n_2^2 + n_3^2=1\,.
\end{equation}
The direction of the deformed fiber is given by the unit vector $\mathbf{n}(X,Y) = \mathbf{F}\,\mathbf{N}$, where $\mathbf{N} = (0,0,1)$ is the reference fiber direction. 
In order to automatically satisfy the inextensibility constraint, one can use the following spherical parametrization of $\mathbf{n}$:
\begin{equation}
\begin{dcases}
	n_1(X,Y) = \sin\theta(X,Y)\,\cos\phi(X,Y)\,, \\
	n_2(X,Y) = \sin\theta(X,Y)\,\sin\phi(X,Y)\,, \\
	n_3(X,Y) = \cos\theta(X,Y)\,.
\end{dcases}
\end{equation}
For this deformation to have $C_{13}=C_{23}=0$, one must have
\begin{equation}
\left\{
\begin{aligned}
	&\sin\theta(X,Y)\left[\cos\phi(X,Y)\,a_{1,X}(X,Y) + \sin\phi(X,Y)\,a_{2,X}(X,Y)\right] + \cos\theta(X,Y)\,a_{3,X}(X,Y) = 0\,, \\
	&\sin\theta(X,Y)\left[\cos\phi(X,Y)\,a_{1,Y}(X,Y) + \sin\phi(X,Y)\,a_{2,Y}(X,Y)\right] + \cos\theta(X,Y)\,a_{3,Y}(X,Y) = 0\,.
\end{aligned}
\right.
\end{equation}
These can be rewritten as the following first-order PDEs:
\begin{equation} \label{PDEs-C13-C23}
	\mathbf{n}(X,Y) \cdot \mathbf{a}_{,X}(X,Y) = 0\,, \qquad
	\mathbf{n}(X,Y) \cdot \mathbf{a}_{,Y}(X,Y) = 0\,.
\end{equation}
These equations state that the vector field $\mathbf{a}(X,Y)$ is constant along the direction of $\mathbf{n}(X,Y)$, and therefore can vary only in directions orthogonal to $\mathbf{n}(X,Y)$.
The mapping $(X,Y) \mapsto \mathbf{a}(X,Y)$ defines a surface in the deformed configuration, and the vectors $\mathbf{a}_{,X}$ and $\mathbf{a}_{,Y}$ span the tangent plane to this surface at each point. 
The constraints $C_{13}=C_{23}=0$ imply that $\mathbf{n}(X,Y)$ is normal to this surface. 
The referential coordinate $Z$ is the arc length parametrization for fibers in the deformed configuration.

\begin{remark}
For a body with inextensible fibers that are initially straight and parallel to the $Z$-axis in the reference configuration, the most general deformation that maps each material fiber (i.e., a line of constant $(X,Y)$ and varying $Z$) to a straight line in the deformed configuration is given by \eqref{Deformation-Z-Lines-Vector}. 
This family of straight lines defines a smooth one-dimensional foliation of the deformed body: for each fixed $(X,Y)$, the map traces out a straight line in space along the direction $\mathbf{n}(X,Y)$, and the collection of these disjoint lines covers the deformed body.
The surfaces defined by constant $Z=Z_0$ in the reference configuration are mapped to
\begin{equation}
	\mathbf{x}_{Z_0}(X,Y) = \mathbf{a}(X,Y) + Z_0\,\mathbf{n}(X,Y)\,,
\end{equation}
which form a smooth two-dimensional foliation of the deformed configuration, transverse to the fiber direction. Each such surface is parametrized by $(X,Y)$, and its tangent plane at any point is spanned by the vectors $\mathbf{a}_{,X}(X,Y)$ and $\mathbf{a}_{,Y}(X,Y)$. This follows from the fact that $Z$ is fixed and only $(X,Y)$ vary in the parameterization. Hence, the deformation induces a global product structure on the deformed configuration, with the one-dimensional foliation defined by fibers and the transverse two-dimensional foliation defined by the image of constant-$Z$ surfaces.
\end{remark}

Because $\lambda_3 = 1$, we need to consider only two of the principal invariants, as the third one is functionally dependent on the other two. We work with $I_1$ and $I_3$. 
We have the following possibilities: (i) $I_1 = I_1(Z)$ and $I_3 = I_3(Z)$, (ii) $I_1 = I_1(Z)$ and $I_3$ is constant, (iii) $I_3 = I_3(Z)$ and $I_1$ is constant, and (iv) both $I_1$ and $I_3$ are constant.
$I_1$ is written as
\begin{equation}
	I_1 = 1 + \|\mathbf{a}_{,X}\|^2 + \|\mathbf{a}_{,Y}\|^2  
	+ 2Z\,\left( \mathbf{a}_{,X} \cdot \mathbf{n}_{,X} + \mathbf{a}_{,Y} \cdot \mathbf{n}_{,Y} \right) 
	+ Z^2\left( \|\mathbf{n}_{,X}\|^2 + \|\mathbf{n}_{,Y}\|^2 \right)\,.
\end{equation}
For $I_1 = I_1(Z)$ to hold, each coefficient in the polynomial expansion of $I_1$ in powers of $Z$ must be constant. Therefore, 
\begin{equation}
	\|\mathbf{a}_{,X}\|^2 + \|\mathbf{a}_{,Y}\|^2=c_1\,, \qquad
	\mathbf{a}_{,X} \cdot \mathbf{n}_{,X} + \mathbf{a}_{,Y} \cdot \mathbf{n}_{,Y}=c_2\,, \qquad
	\|\mathbf{n}_{,X}\|^2 + \|\mathbf{n}_{,Y}\|^2=c_3\,,
\end{equation}
where $c_1$, $c_2$, and $c_3$ are constants.\footnote{$I_1$ is constant if and only if $c_2=c_3=0$.} 
It is straightforward to see that
\begin{equation}
	J = \sqrt{I_3} = \left( \mathbf{a}_{,X} \times \mathbf{a}_{,Y} \right) \cdot \mathbf{n}
	+ Z\,\left( \mathbf{a}_{,X} \times \mathbf{n}_{,Y} + \mathbf{n}_{,X} \times \mathbf{a}_{,Y} \right) \cdot \mathbf{n}
	+ Z^2\,\left( \mathbf{n}_{,X} \times \mathbf{n}_{,Y} \right) \cdot \mathbf{n}\,.
\end{equation}
For $J = J(Z)$ to hold, we must have 
\begin{equation}
	\left( \mathbf{a}_{,X} \times \mathbf{a}_{,Y} \right) \cdot \mathbf{n} = c_4\,,\qquad
	\left( \mathbf{a}_{,X} \times \mathbf{n}_{,Y} + \mathbf{n}_{,X} \times \mathbf{a}_{,Y} \right) \cdot \mathbf{n} = c_5\,,\qquad
	\left( \mathbf{n}_{,X} \times \mathbf{n}_{,Y} \right) \cdot \mathbf{n} = c_6\,,
\end{equation}
where $c_4$, $c_5$, and $c_6$ are constants.\footnote{$I_3=J^2$ is constant if and only if $c_5=c_6=0$.} 
The deformation \eqref{Deformation-Z-Lines-Vector} is determined by a pair of vectors $(\mathbf{a}(X,Y) ,\mathbf{n}(X,Y) )$, which must satisfy the following overdetermined system of PDEs (in addition to the constraint $\|\mathbf{n}\| =1$):
\begin{empheq}[left={\empheqlbrace }]{align} 
	\label{Universality-Surface-1}
	&  \mathbf{n} \cdot \mathbf{a}_{,X} = 0\,, \\
	\label{Universality-Surface-2}
	&  \mathbf{n} \cdot \mathbf{a}_{,Y} = 0\,, \\
	\label{Universality-Surface-3}
	&  \|\mathbf{a}_{,X}\|^2 + \|\mathbf{a}_{,Y}\|^2=c_1^2\,, \\
	\label{Universality-Surface-4}
	&  \mathbf{a}_{,X} \cdot \mathbf{n}_{,X} + \mathbf{a}_{,Y} \cdot \mathbf{n}_{,Y}=c_2\,, \\
	\label{Universality-Surface-5}
	&  \|\mathbf{n}_{,X}\|^2 + \|\mathbf{n}_{,Y}\|^2=c_3^2\,, \\		
	\label{Universality-Surface-6}
	&  \left( \mathbf{a}_{,X} \times \mathbf{a}_{,Y} \right) \cdot \mathbf{n} = c_4 \,, \\
	\label{Universality-Surface-7}
	&  \left( \mathbf{a}_{,X} \times \mathbf{n}_{,Y} + \mathbf{n}_{,X} \times 
	\mathbf{a}_{,Y} \right) \cdot \mathbf{n} = c_5\,, \\
	\label{Universality-Surface-8}
	&  \left( \mathbf{n}_{,X} \times \mathbf{n}_{,Y} \right) \cdot \mathbf{n} = c_6\,,
\end{empheq}
together with the remaining universality constraints \eqref{Universality-Constraints-Cauchy3-1} and \eqref{Universality-Constraints-Cauchy3-2}, which simplify to:
\begin{equation}
\begin{aligned}
	F^{-A}{}_a\,F^{-B}{}_b\,b^{ab}{}_{|B}=\beta N^A\,,\qquad
	F^{-A}{}_a\,F^{-B}{}_b\,c^{ab}{}_{|B}=\gamma N^A\,.
\end{aligned}
\end{equation}
Therefore\footnote{
Recall that
\begin{equation}
	b^{ab}{}_{|B} = \frac{\partial b^{ab}}{\partial X^B}+ \gamma^a{}_{cd}\, F^c{}_B\, b^{db}+ \gamma^b{}_{cd}\, F^c{}_B\, b^{ad}\,.
\end{equation}
When using Cartesian coordinates in the ambient space, we always have $b^{ab}{}_{|B} =b^{ab}{}_{,B}$.
}
\begin{equation} \label{Universality-Constraints-b-c}
	F^{-A}{}_a\,F^{-B}{}_b\,b^{ab}{}_{|B} = F^{-A}{}_a\,F^{-B}{}_b\,c^{ab}{}_{|B}=0\,,	\qquad A=1,2\,.
\end{equation}

The Gaussian curvature of the surface with tangent vectors $\mathbf{a}_{,X}$ and $\mathbf{a}_{,Y}$ and normal vector $\mathbf{n}$ is written as \citep{doCarmo1976,ONeill2006}
\begin{equation}
	K = \frac{ \left( \mathbf{n}_{,X} \times \mathbf{n}_{,Y} \right) \cdot \mathbf{n}}
	{\left(\mathbf{a}_{,X} \times \mathbf{a}_{,Y} \right) \cdot \mathbf{n}} \,.
\end{equation}
From \eqref{Universality-Surface-6} and \eqref{Universality-Surface-8} we observe that the Gaussian curvature is 
\begin{equation}
	K =\frac{c_6}{c_4}\,,
\end{equation}
which is everywhere a constant.
The complete,\footnote{A surface is called complete if all geodesics can be extended indefinitely, or equivalently, if it is complete as a metric space with respect to the induced Riemannian distance. In the context of nonlinear elasticity, it is reasonable to assume that the deformed surface normal to inextensible fibers is complete, provided the deformation is smooth and the body has no cracks, tears, or nonsmooth boundaries.} connected, embedded surfaces in $\mathbb{R}^3$ with constant Gaussian curvature $K$ are classified as follows \citep{doCarmo1976,Spivak1979,Kuhnel2006}:
\begin{itemize}[topsep=0pt,noitemsep, leftmargin=10pt]
\item $K = 0$: planes, cylinders, cones (i.e., developable surfaces),
\item $K > 0$: portions of spheres,
\item $K < 0$: no complete, smooth, embedded surfaces exist; only local models such as the pseudosphere.
\end{itemize}

\vskip 0.1in
\noindent
The mean curvature of the surface with tangent vectors $\mathbf{a}_{,X}$ and $\mathbf{a}_{,Y}$ and normal vector $\mathbf{n}$ is given by \citep{doCarmo1976,ONeill2006}
\begin{equation}
	H = -\frac{1}{2} \left( \mathbf{a}_{,X} \cdot \mathbf{n}_{,X} 
	+ \mathbf{a}_{,Y} \cdot \mathbf{n}_{,Y} \right)\,.
\end{equation}
The constraint \eqref{Universality-Surface-4} implies that
\begin{equation}
	H = -\frac{1}{2} c_2\,,
\end{equation}
i.e., the mean curvature is constant. 
Among planes, cylinders, cones, and spheres, cones do not have constant mean curvature. Therefore, the surfaces normal to fibers in the deformed configuration can only be planes, cylinders, or spheres.

\begin{remark}
In his classification of universal deformations in incompressible isotropic hyperelasticity with $I_1$, $I_2$ not both constant ($I_3 = 1$), \citet{Ericksen1954} showed that the surfaces orthogonal to one of the eigenvectors of $\mathbf{b}$ must have constant mean and Gaussian curvatures. Interestingly, in our setting, the surfaces orthogonal to the fiber direction in the deformed configuration—which is also an eigenvector of $\mathbf{b}$—exhibit the same geometric property. This parallels Ericksen’s observation, although it arises in a different context.
In a related but distinct problem, \citet{Ericksen1967} studied the universal orientation patterns of liquid crystals. In a liquid crystal, each point $\mathbf{x}$ is associated with a preferred direction defined by a unit vector field $\mathbf{h} = \mathbf{h}(\mathbf{x})$, and the stored energy density depends on both $\mathbf{h}$ and its gradient: $W = W(\mathbf{h}, \nabla\mathbf{h})$. Ericksen showed that the integral curves of $\mathbf{h}$ are straight lines and that the surfaces orthogonal to $\mathbf{h}$ have constant mean and Gaussian curvatures.
\end{remark}

\paragraph{Surfaces normal to fibers are planes.}
Let us assume that $\mathbf{a}_{,X}$ and $\mathbf{a}_{,Y}$ define a surface that is a portion of a plane in $\mathbb{R}^3$. This implies that $\mathbf{a}$ is an affine function of $(X,Y)$:
\begin{equation}
	\mathbf{a}(X,Y) = \mathbf{a}_0 + \mathbf{p}_0\,X + \mathbf{q}_0\,Y\,,
\end{equation}
for some constant vectors $\mathbf{a}_0$, $\mathbf{p}_0$, and $\mathbf{q}_0$ in $\mathbb{R}^3$. It follows that $\mathbf{a}_{,X} = \mathbf{p}_0$ and $\mathbf{a}_{,Y} = \mathbf{q}_0$ are constant vectors. Therefore, the surface normal vector
\begin{equation}
	\mathbf{n} = \frac{\mathbf{p}_0 \times \mathbf{q}_0}{\|\mathbf{p}_0 \times \mathbf{q}_0\|}=\mathbf{n}_0 \,,
\end{equation}
is also a constant unit vector. Thus, if the surface defined by $\mathbf{a}(X,Y)$ is planar, then $\mathbf{n}$ must be constant.
Substituting the affine form of $\mathbf{a}(X,Y)$ and the constant unit vector $\mathbf{n}$ into the general deformation \eqref{Deformation-Z-Lines}, we obtain a homogeneous deformation:
\begin{equation} \label{Family-0Z}
	\varphi(X,Y,Z) = \mathbf{a}_0 + \mathbf{p}_0\,X + \mathbf{q}_0\,Y + \mathbf{n}_0\,Z \,.
\end{equation}
Note that the constraints \eqref{Universality-Surface-1}-\eqref{Universality-Surface-8} are all satisfied.
For a homogeneous deformation $\mathbf{b}^\sharp$ and $\mathbf{c}^\sharp$ are (covariantly) constant tensors, and hence, \eqref{Universality-Constraints-b-c} are trivially satisfied.

\begin{defi}
A deformation is called \emph{Z-isometric} if, with respect to the Cartesian coordinates $(X, Y, Z)$, the $ZZ$-component of its right Cauchy-Green deformation tensor satisfies $C_{ZZ} = 1$. 
\end{defi}

We have shown that, for isotropic compressible Cauchy elastic solids reinforced by a family of inextensible fibers parallel to the $Z$-axis in the undeformed configuration, all $Z$-isometric homogenous deformations are universal. We call this \textit{Family $0Z$ Universal Deformations}.

\paragraph{Surfaces normal to fibers are cylinders.}
We make the following observations when the surfaces normal to $\mathbf{n}$ are circular cylinders:
\begin{itemize}[topsep=0pt,noitemsep, leftmargin=10pt]
\item Fibers are vertical in the reference configuration, i.e., are parallel to the $Z$-axis.
\item In the deformed configuration, fibers become straight radial lines in $\mathbb{R}^3$.
\item Horizontal planes $Z = \textsf{const.}$ in the reference configuration are mapped to circular cylinders centered on the $z$-axis in the deformed configuration.
\end{itemize}

\vskip 0.1in
\noindent
In the deformed configuration the cylindrical surfaces have axes parallel to the $z$-axis, and hence, $n_3(X,Y) = 0$.
We assume that $a_3(X,Y) = a_3(Y)$.\footnote{We argue that it is not restrictive to assume that the third component of the base point depends only on $Y$, i.e., $a_3 = a_3(Y)$, and that the fiber direction lies entirely in the $xy$-plane, i.e., $n_3 = 0$. Any admissible deformation mapping of the form \eqref{Deformation-Z-Lines} that satisfies the constraint $C_{ZZ} = 1$ can be transformed, without loss of generality, to such a form via an appropriate rigid motion and reparametrization of the undeformed configuration.  Indeed, if the fibers are mapped to straight lines in the deformed configuration and $C_{ZZ} = 1$, then the image of each vertical fiber must be a unit-speed curve. One can then choose a coordinate system in which the fiber directions lie in the $xy$-plane and the surfaces normal to these directions become circular cylinders. In this adapted frame, the out-of-plane component of the fiber direction vanishes, and the vertical position in the deformed configuration can be absorbed into the base surface, justifying $n_3 = 0$ and reducing $a_3$ to a function of $Y$ alone. This representation makes the geometry explicit: the deformation maps horizontal planes $Z = \textsf{const.}$ in the reference configuration to cylindrical surfaces of radius $r(Z)$ in the deformed configuration, with fibers directed radially outward and tangent to these cylinders. Such reparametrizations preserve the essential geometry of the deformation while simplifying the analytic structure.}
Knowing that $\mathbf{n}$ is a unit vector, one can use the following change of variables
\begin{equation}
	n_1(X,Y) = \cos\theta(X,Y)\,, \qquad n_2(X,Y) = \sin\theta(X,Y)\,,
\end{equation}
for some function $\theta=\theta(X,Y)$.
The squared radial distance of a generic material point from the $z$-axis is $r^2(X,Y,Z) = x^2(X,Y,Z) + y^2(X,Y,Z)$.
Substituting from \eqref{Deformation-Z-Lines} we have
\begin{equation}
\begin{aligned}
	r^2(X,Y,Z) &= \left[a_1(X,Y)+ Z\,\cos\theta(X,Y)\right]^2 + \left[a_2(X,Y) + Z\,\sin\theta(X,Y)\right]^2 \\
	& = a_1^2(X,Y) + a_2^2(X,Y) + 2Z \left[a_1(X,Y)\cos\theta(X,Y) + a_2(X,Y)\sin\theta(X,Y) \right] + Z^2\,.
\end{aligned}
\end{equation}
We know that $r=r(Z)$. Therefore, we must have
\begin{equation}
	a_1^2(X,Y) + a_2^2(X,Y) = a_0^2\,, \qquad a_1(X,Y)\cos\theta(X,Y) + a_2(X,Y)\sin\theta(X,Y) = c_0\,,
\end{equation}
where $a_0$ and $c_0$ are constants. 
From the first condition we have $a_1(X,Y)= a_0\,cos\phi(X,Y)$, $a_2(X,Y)=a_0 \,\sin\phi(X,Y)$, for some function $\phi=\phi(X,Y)$, and substituting into the second condition we obtain
\begin{equation}
	a_0\left[\cos\theta(X,Y) \cos\phi(X,Y)+\sin\theta(X,Y)\sin\phi(X,Y)\right]=a_0\cos(\theta(X,Y)-\phi(X,Y))\,.
\end{equation}
Therefore, $\phi(X,Y)=\theta(X,Y)+\theta_0$. 
Thus, $r^2(Z)=a_0^2+2Z a_0\cos\theta_0+Z^2=(Z+a_0\cos\theta_0)^2+a_0^2\sin^2\theta_0=(Z+Z_0)^2+r_0^2$, where $Z_0=a_0\cos\theta_0$ and $r_0^2=a_0^2\sin^2\theta_0$. Therefore
\begin{equation}
	r(Z) = \sqrt{(Z+Z_0)^2+r_0^2}\,.
\end{equation}
We observe that each material plane $Z = \textsf{const.}$ is mapped to a circular cylinder of radius $r(Z)$.
This describes a deformation that maps horizontal planes to concentric cylinders and vertical fibers to radial lines. The fiber direction field $\mathbf{n}(X,Y)$ lies entirely in the $xy$-plane and has unit norm. Moreover, the deformation satisfies $C_{ZZ} = 1$.

To summarize, we have the family of deformations \eqref{Deformation-Z-Lines-Vector}, where
\begin{equation}
\begin{aligned}
	\mathbf{a}(X,Y) &= a_0 \cos(\theta(X,Y) + \theta_0)\,\mathbf{e}_x 
	+ a_0 \sin(\theta(X,Y) + \theta_0)\,\mathbf{e}_y 
	+ a_3(Y)\,\mathbf{e}_z\,, \\
	\mathbf{n}(X,Y) &= \cos\theta(X,Y)\,\mathbf{e}_x 
	+ \sin\theta(X,Y)\,\mathbf{e}_y\,.
\end{aligned}
\end{equation}
Thus
\begin{equation}
\begin{aligned}
	\mathbf{a}_{,X} &= a_0\,\theta_{,X}\,\left( -\sin(\theta + \theta_0),\, \cos(\theta + \theta_0),\, 0 \right)
	\,, \\
	\mathbf{a}_{,Y} &= a_0\,\theta_{,Y}\,\left( -\sin(\theta + \theta_0),\, \cos(\theta + \theta_0),\, 0 \right) 	+ \left( 0,\, 0,\, a_3'(Y) \right)\,, \\
	\mathbf{n}_{,X} &= \theta_{,X}\,\left( -\sin\theta,\, \cos\theta,\, 0 \right)\,, \\
	\mathbf{n}_{,Y} &= \theta_{,Y}\,\left( -\sin\theta,\, \cos\theta,\, 0 \right)\,.
\end{aligned}
\end{equation}
Next, we need to verify that the constraints \eqref{Universality-Surface-1}-\eqref{Universality-Surface-8} are satisfied.
Note that $ \mathbf{n}_{,X} \times \mathbf{n}_{,Y} =0$, and hence \eqref{Universality-Surface-8} is trivially satisfied ($c_6=0$).
From \eqref{Universality-Surface-1} and \eqref{Universality-Surface-2} we must have
\begin{equation}
	\mathbf{n} \cdot \mathbf{a}_{,X} = -a_0\,\theta_{,X}\,\sin\theta_0=0\,,\qquad
	\mathbf{n} \cdot \mathbf{a}_{,Y} = -a_0\,\theta_{,Y}\,\sin\theta_0=0\,.
\end{equation}
Assuming $\theta_0=0$, these two constraints are trivially satisfied.\footnote{If $\theta_0 \ne 0$, then we must have $\theta(X,Y) = \textsf{const.}$, and the remaining universality constraints still yield \eqref{a3-Linear}. The resulting deformation is homogeneous.}
The remaining universality constraints are
\begin{empheq}[left={\empheqlbrace }]{align} 
	&  \|\mathbf{a}_{,X}\|^2 + \|\mathbf{a}_{,Y}\|^2 = a_0^2\left( \theta_{,X}^2 + \theta_{,Y}^2 \right) + \left(a_3'(Y) \right)^2 
	=c_1^2\,, \\
	&  \mathbf{a}_{,X} \cdot \mathbf{n}_{,X} + \mathbf{a}_{,Y} \cdot \mathbf{n}_{,Y} = a_0\left( \theta_{,X}^2 + \theta_{,Y}^2 
	\right)  =c_2\,, \\
	&  \|\mathbf{n}_{,X}\|^2 + \|\mathbf{n}_{,Y}\|^2  = \theta_{,X}^2 + \theta_{,Y}^2 = c_3^2\,, \\		
	&  \left( \mathbf{a}_{,X} \times \mathbf{a}_{,Y} \right) \cdot \mathbf{n} = r_0\,\theta_{,X}\,a_3'(Y)= c_4 \,, \\
	&  \left( \mathbf{a}_{,X} \times \mathbf{n}_{,Y} + \mathbf{n}_{,X} \times 
	\mathbf{a}_{,Y} \right) \cdot \mathbf{n} =  \theta_{,X}\,a_3'(Y) =c_5\,.
\end{empheq}
From the first and third constraints we conclude that $a_3'(Y)$ must be constant, and hence
\begin{equation} \label{a3-Linear}
	a_3(Y) = k_1 Y+k_0\,.
\end{equation}
The last two constraints imply that $\theta_{,X}$ must be constant, and hence $\theta(X,Y)=\alpha_0 X+\bar{\theta}(Y)$. But from the third constraint $\alpha_0^2+\bar{\theta}'(Y)^2$ must be a constant, and hence $\bar{\theta}'(Y)$ is constant.\footnote{Note that $c_5=\alpha_0\,k_1$.} Therefore
\begin{equation}
	\theta(X,Y) = \alpha_0 X+\beta_0 Y+\gamma_0\,.
\end{equation}
The remaining universality constraints \eqref{Universality-Constraints-b-c} are trivially satisfied.
Therefore, we have found the following family of universal deformations
\begin{equation}
\begin{dcases}
x(X,Y,Z) = (Z + Z_0)\,\cos\left( \alpha_0 X + \beta_0 Y + \gamma_0 \right)\,, \\[4pt]
y(X,Y,Z) = (Z + Z_0)\,\sin\left( \alpha_0 X + \beta_0 Y + \gamma_0 \right)\,, \\[4pt]
z(X,Y,Z) = k_1 Y + k_0\,.
\end{dcases}
\end{equation}
In cylindrical coordinates:
\begin{equation}
\begin{dcases}
	r(X,Y,Z) = Z + Z_0\,, \\
	\theta(X,Y,Z) = \alpha_0 X + \beta_0 Y + \gamma_0\,, \\
	z(X,Y,Z) = k_1 Y + k_0\,.
\end{dcases}
\end{equation}

\begin{figure}[t!]
\centering
\includegraphics[width=0.75\textwidth]{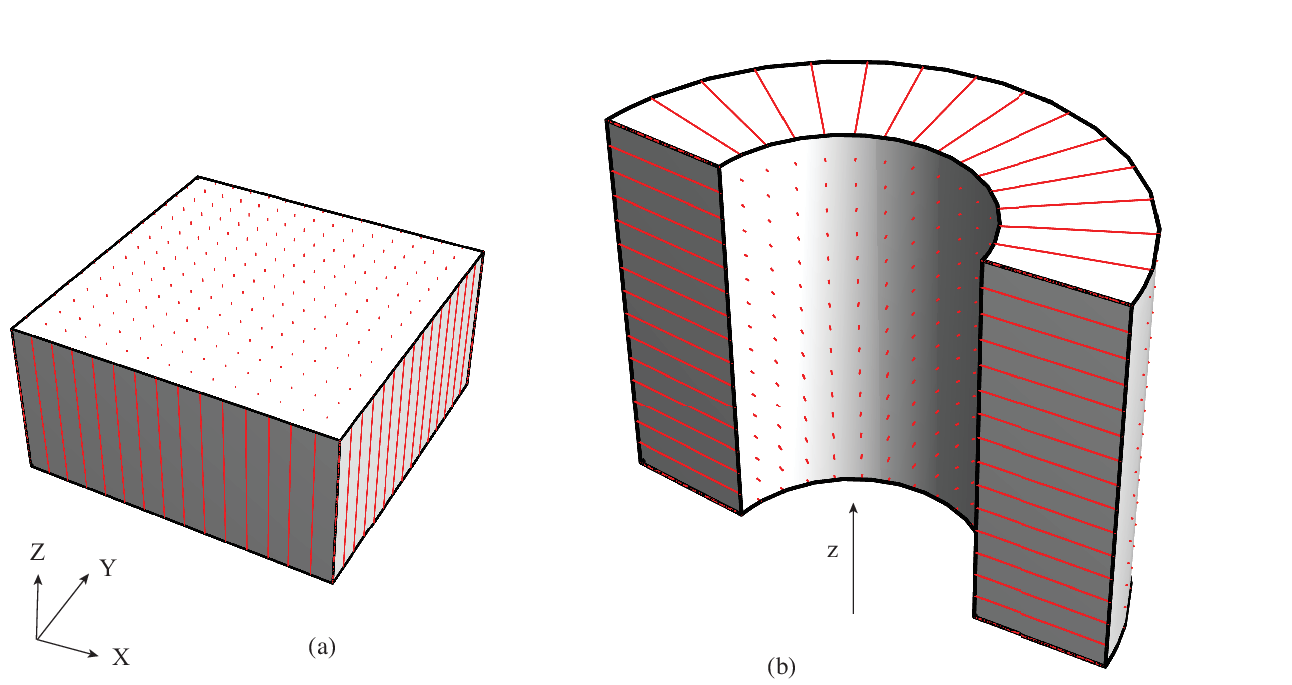}
\vspace*{0.20in}
\caption{A universal deformation from Family $Z_1$ combining bending about the $z$-axis and uniaxial stretch along the $z$-axis.  
(a) Undeformed configuration: a rectangular block with inextensible fibers parallel to the $Z$-axis.  
(b) Deformed configuration ($\alpha_0 = 1.5$, $\beta_0 = 0$, $\theta_0 = 1.5$, $z_0 = 1$, $k_1 = 1.5$, $A = B = C = 1$): horizontal planes $Z = \textsf{const.}$ are bent into concentric circular cylinders centered on the $z$-axis, and material lines along the $Y$-axis are stretched uniformly into the $z$-direction. Planes normal to the $Z$-axis are mapped to cylindrical surfaces, and the deformation preserves fiber inextensibility along $Z$.}
\label{Fig:Family-Z1}
\end{figure}
Finally, we define the \textit{Family $Z_1$ universal deformations} by
\begin{equation}
\begin{dcases}
	r(X,Y,Z) = Z + Z_0\,, \\
	\theta(X,Y,Z) = \alpha_0 X + \beta_0 Y + \theta_0\,, \\
	z(X,Y,Z) = k_1 Y + z_0\,.
\end{dcases}
\end{equation}
Note that $J = \alpha_0\,k_1\,(Z + Z_0)$, and therefore these deformations are non-isochoric.

Let us consider a block with side lengths $2A$, $2B$, and $2C$ in the undeformed configuration, i.e., $\mathcal{B} = \left\{	-A \leq X \leq A,\ -B \leq Y \leq B,\ -C \leq Z \leq C \right\}$. The undeformed configuration is the following thick shell:
\begin{equation}
	\mathcal{C} = \left\{ 
	Z_0 - C \leq r \leq Z_0 + C,\ 
	\theta_0 - \alpha_0 A - \beta_0 B \leq \theta \leq \theta_0 + \alpha_0 A + \beta_0 B,\ 
	z_0 - k_1 B \leq z \leq z_0 + k_1 B 
	\right\}\,.
\end{equation}
Fig.~\ref{Fig:Family-Z1} shows a schematic of this family of deformations.

\begin{remark}
The principal stretches are
\begin{equation}
\begin{aligned}
	\lambda_{1,2}^2 = \frac{1}{2} \left[ (Z+Z_0)^2 \left(\alpha_0^2 + \beta_0^2 \right) + k_1^2 
	 \pm \sqrt{\left[(Z+Z_0)^2 (\alpha_0^2 - \beta_0^2) - k_1^2 \right]^2  + 4 (Z+Z_0)^4 \alpha_0^2 \beta_0^2	}
	\right]\,,\qquad \lambda_3^2 = 1\,.
\end{aligned}
\end{equation}
The eigenvalues $\lambda_1^2$ and $\lambda_2^2$ are equal if and only if the discriminant inside the square root vanishes, i.e., $\left((Z+Z_0)^2 (\alpha_0^2 - \beta_0^2) - k_1^2 \right)^2 + 4 (Z+Z_0)^4 \alpha_0^2 \beta_0^2 = 0$.
Since the sum of squares is zero only if each term is zero, this implies $(Z+Z_0)^2 (\alpha_0^2 - \beta_0^2) - k_1^2 = 0~ \alpha_0 \beta_0 = 0$.
Because $Z$ is a variable, the first condition cannot hold unless $\alpha_0 = \beta_0 = 0$ and $k_1 = 0$. Thus, $\lambda_1 = \lambda_2$ only in the trivial case.
\end{remark}

\paragraph{Surfaces normal to fibers are spheres.}
In this case, fibers are mapped to straight radial lines in the deformed configuration and the surfaces orthogonal to these fibers are concentric spheres. Therefore, for each fixed $(X,Y)$, the image of the fiber $Z \mapsto \varphi(X,Y,Z)$ traces out a straight line in the direction $\mathbf{n}(X,Y)$, and the surfaces $Z = \textsf{const.}$ are mapped to spherical surfaces with radius $r(Z)$.
Using the fact that $\mathbf{n}(X,Y) = \mathbf{e}_r$ in the deformed configuration, we write the deformation mapping explicitly in spherical coordinates. Let
\begin{equation}
	\mathbf{n}(X,Y) = 
	\begin{bmatrix}
		\sin\phi(X,Y)\,\cos\theta(X,Y) \\
		\sin\phi(X,Y)\,\sin\theta(X,Y) \\
		\cos\phi(X,Y)
	\end{bmatrix}\,,
\end{equation}
where $\theta(X,Y)$ and $\phi(X,Y)$ are angular functions specifying the direction of $\mathbf{n}(X,Y)$ in spherical coordinates. 
Therefore, the deformation mapping is written as $\varphi(X,Y,Z) = r(Z)\,\mathbf{n}(X,Y)$, i.e.,
\begin{equation}
\begin{dcases}
	x(X,Y,Z) = r(Z)\, \sin\phi(X,Y)\, \cos\theta(X,Y)\,, \\
	y(X,Y,Z) = r(Z)\, \sin\phi(X,Y)\, \sin\theta(X,Y)\,, \\
	z(X,Y,Z) = r(Z)\, \cos\phi(X,Y)\,.
\end{dcases}
\end{equation}
Note that
\begin{equation}
\begin{aligned}
	C_{33} &	= \left( r'(Z)\,\mathbf{n}(X,Y) \right) \cdot \left( r'(Z)\,\mathbf{n}(X,Y) \right)
	= (r'(Z))^2\,\|\mathbf{n}(X,Y)\|^2
	= (r'(Z))^2\,, \\
	C_{13} & = r(Z)\,r'(Z)\,\mathbf{n}_{,X} \cdot \mathbf{n}(X,Y)\,, \\
	C_{23} &	= r(Z)\,r'(Z)\,\mathbf{n}_{,Y} \cdot \mathbf{n}(X,Y)\,.
\end{aligned}
\end{equation}
Knowing that $C_{33} = 1$, we obtain $r'(Z) = \pm 1$. Therefore, the general solution is a linear function: $r(Z) = \pm Z + r_0$, where $r_0$ is a constant. As $\mathbf{n}$ is a unit vector, $\mathbf{n}_{,X} \cdot \mathbf{n}(X,Y)=\mathbf{n}_{,Y} \cdot \mathbf{n}(X,Y)=0$, and hence, the conditions $C_{13}=C_{23}=0$ are trivially satisfied.

The first invariant $I_1$ reads
\begin{equation}
	I_1 = \|\boldsymbol{\varphi}_{,X}\|^2 + \|\boldsymbol{\varphi}_{,Y}\|^2 + \|\boldsymbol{\varphi}_{,Z}\|^2 
	= r(Z)^2\left( \|\mathbf{n}_{,X}\|^2 + \|\mathbf{n}_{,Y}\|^2 \right) + {r'}^2(Z) \,.
\end{equation}
The Jacobian is written as
\begin{equation}
	J = r^2(Z)\,r'(Z)\,\mathbf{n}(X,Y) \cdot \left( \mathbf{n}_{,X} \times \mathbf{n}_{,Y} \right)\,.
\end{equation}
For $I_1$ and $I_3=J^2$ to be functions of only Z, we must have
\begin{equation}
	\|\mathbf{n}_{,X}\|^2 + \|\mathbf{n}_{,Y}\|^2 =\|\bar{\nabla} \mathbf{n}\|^2 =c_1^2\,,\qquad 
	\mathbf{n} \cdot \left( \mathbf{n}_{,X} \times \mathbf{n}_{,Y} \right) =c_2\,,
\end{equation}
where $c_1$ and $c_2$ are constants. Therefore, we must have
\begin{equation} \label{Sphere-PDEs}
	\phi_{,X}^2 + \phi_{,Y}^2 + \sin^2\phi\,\left( \theta_{,X}^2 + \theta_{,Y}^2 \right) = c_1^2\,, \qquad
	\sin\phi \left(\theta_{,Y}\, \phi_{,X} -\, \phi_{,Y}\, \theta_{,X} \right) = c_2\,.
\end{equation}
This is a coupled system of nonlinear second-order PDEs. 
If $c_1=0$, then $\phi_{,X} = \phi_{,Y} = \theta_{,X} = \theta_{,Y} = 0$, implying that $\mathbf{n} = \mathbf{n}_0$ is constant. However, this leads to $J=0$, which is not physically acceptable. Therefore, we must have $c_1 \neq 0$.
Also, because $J>0$, we must have $c_2 \neq 0$.

Our first attempt at solving the above system of nonlinear PDEs was the following. 
Let us use cylindrical coordinates $(R,\Theta,Z)$ in the reference configuration. For each fixed $Z$, the reference surface is a circular disk mapped to a spherical cap described in spherical coordinates $(r,\phi,\theta)$, where $0\leq \theta<2\pi$ is the azimuthal angle and $0\leq \phi<\pi$ is the polar angle. Surfaces normal to fibers are spheres, and fibers are radial lines, so $r=r(Z)$. Concentric circles (constant $R$) are mapped to concentric circles, so $\phi=\phi(R)$. No assumption is made on $\theta$, which is taken as an arbitrary function of $R$ and $\Theta$. The deformation is given by
\begin{equation} \label{disk-to-sphericalcap}
	r=r(Z)\,,\qquad \theta=\theta(R,\Theta)\,,\qquad \phi=\phi(R)\,.
\end{equation}
There is no stretch along the fibers, so $\lambda_r=1$, and planes $Z=\textsf{const.}$ are mapped to concentric spheres.
With respect to Cartesian coordinates, the deformation mapping is written as
\begin{equation}
\begin{dcases}
	x(R,\Theta,Z) = r(Z)\, \sin\phi(R)\, \cos\theta(R,\Theta)\,, \\
	y(R,\Theta,Z) = r(Z)\, \sin\phi(R)\, \sin\theta(R,\Theta)\,, \\
	z(R,\Theta,Z) = r(Z)\, \cos\phi(R)\,.
\end{dcases}
\end{equation}
The Jacobian of deformation reads
\begin{equation}
	J = \det\mathbf{F}\sqrt{\frac{\det\mathbf{g}}{\det\mathbf{G}}}
	= \frac{r^2(Z)\,r'(Z)\,\phi'(R)\,\sin\phi(R)\,\theta_{,\Theta}(R,\Theta)}{R}\,.
\end{equation}
For $J=\sqrt{I_3}$ to be a function of $Z$ only we first must have $\theta_{,\Theta}(R,\Theta)=c_2$, and hence $\theta(R,\Theta)=c_2 \Theta+f(R)$, where $f(R)$ is some function. Now $J$ is written as
\begin{equation}
	J = c_2\, r(Z)^2\, r'(Z)\, \frac{\sin\phi(R)\, \phi'(R)}{R}\,.
\end{equation}
For $J$ to be a function of $Z$ we must have
\begin{equation}
	\frac{\sin\phi(R)\, \phi'(R)}{R}=-2c_1\,.
\end{equation}
Thus, $\phi(R) = \cos^{-1}\left( c_1R^2 + c_3 \right)$.
For this deformation $C_{33}={r'}^2(Z)=1$, and hence $r'(Z)=\pm 1$. Therefore, $r(Z)=\pm Z+r_0$. Thus, $J = -2\,c_1\,c_2\,(\pm Z+r_0 )^2$.

The first invariant is written as
\begin{equation}
\begin{aligned}
	I_1 &= {r'}^2(Z)	+ r(Z)^2 \left[ {\phi'}^2(R) 
	+ \sin^2\phi(R) \left( 
	R^2 \left( \theta_{,\Theta}(R,\Theta) \right)^2 
	+ \left( \theta_{,R}(R,\Theta) \right)^2 
	\right) \right]  \\
	&= 1 + (r_0 + Z)^2 \left[ 
	\frac{4 c_1^2 R^2}{1 - \left( c_1 R^2 + c_3 \right)^2}
	+ \left( 1 - \left( c_1 R^2 + c_3 \right)^2 \right) \left( 
	c_2^2 R^2 + {f'}^2(R) \right)
	\right]\,.
\end{aligned}
\end{equation}
For $I_1$ to be only a function of $Z$, we must have
\begin{equation}
	\frac{4 c_1^2 R^2}{1 - \left( c_1 R^2 + c_3 \right)^2}
	+ \left( 1 - \left( c_1 R^2 + c_3 \right)^2 \right) \left( 
	c_2^2 R^2 + {f'}^2(R) \right)=c_4\,.
\end{equation}
Therefore, assuming that $f'(R)>0$, we obtain
\begin{equation}
f'(R) = \sqrt{
-\,\frac{c_2^2}{R^2}
- \frac{
4 c_1^2 R^2 + c_4 \left(-1 + c_3 + c_1 R^2\right)\left(1 + c_3 + c_1 R^2\right)
}{
\left(-1 + c_3 + c_1 R^2\right)^2 \left(1 + c_3 + c_1 R^2\right)^2
}
}\,.
\end{equation}
Thus
\begin{equation}
f(R) = f_0 + \int_0^R 
\sqrt{
-\,\frac{c_2^2}{\xi^2}
- \frac{
4 c_1^2 \xi^2 + c_4 \left(-1 + c_3 + c_1 \xi^2\right)\left(1 + c_3 + c_1 \xi^2\right)
}{
\left(-1 + c_3 + c_1 \xi^2\right)^2 \left(1 + c_3 + c_1 \xi^2\right)^2
}
} \, d\xi\,.
\end{equation}
The universality constraint \eqref{Universality-Constraints-b-c}$_1$ can be satisfied only if $4\, c_1\, R^2 \left[ (c_3 + c_1 R^2)^2 - 1 \right] = 0$, which implies $c_1 = 0$. This, in turn, requires $\phi(R)$ to be constant, which does not correspond to an admissible deformation.
Therefore, deformations of the form \eqref{disk-to-sphericalcap} are not universal.
We explored several other simplifying assumptions but were unable to solve the nonlinear PDEs \eqref{Sphere-PDEs}.

In the following we show that there are no universal deformations in this case.
Knowing that $\mathbf{n} = \mathbf{e}_r$, and $\mathbf{n}\cdot\n=\nc^r=0$, we conclude that $\nc^r=0$. Note that
\begin{equation} 
	\n\cdot \n=\nc_{n}\nc^n=g^{mn}\nc_{n}\nc_m
	=\frac{1}{r^2}(\nc_{\theta})^2+\frac{1}{r^2\sin^2\phi} (\nc_{\phi})^2=1\,.
\end{equation}
Hence, $\nc_{\theta}=r\cos\psi$ and $\nc_{\phi}=r\sin\phi\,\sin\psi$, for some function $\psi=\psi(r,\theta,\phi)$. Thus
\begin{equation} 
	\nc^r=0\,,\qquad \nc^{\theta}=\frac{\cos\psi}{r}\,,\qquad \nc_{\phi}=\frac{\sin\psi}{r\sin\phi}\,.
\end{equation}

First, let us assume that the principal stretches are distinct. 
The universality constraint \eqref{n1-n-identity} gives us the following two PDEs
\begin{equation}
\begin{aligned}
	&-\,\frac{1}{r^2} \left( \frac{1}{\tan\theta} 
	+ \frac{1}{\sin\theta}\,\psi_{,\phi}
	+ \frac{1}{\tan\psi}\,\psi_{,\theta} \right) \sin^2\psi = 0\,, \\
	&\frac{1}{r^2} \left[ \left( \cos\theta + \psi_{,\phi} \right) \sin\psi\, \csc\theta 
	+ \cos\psi\, \psi_{,\theta} \right] \cos\psi\, \csc\theta = 0\,.
\end{aligned}
\end{equation}
We simplify the PDE system by noting that $r>0$ and $\sin\psi$, $\cos\psi$, and $\csc\theta$ are nonzero almost everywhere and can be canceled. Thus
\begin{equation}
\begin{cases}
\displaystyle
	\frac{1}{\sin\theta}\,\psi_{,\phi} + \frac{1}{\tan\psi}\,\psi_{,\theta} = -\frac{1}{\tan\theta}\,, \\[6pt]
	\psi_{,\phi} + \cos\psi\, \sin\theta\, \psi_{,\theta} = -\cos\theta\,.
\end{cases}
\end{equation}
Multiplying the first equation by $\sin\theta$ and subtracting from the second gives us $\left( \cos\psi - \cot\psi \right)\, \psi_{,\theta} = 0$.
Thus, $\psi_{,\theta} = 0$, and the second equation implies that $\psi_{,\phi} = -\cos\theta$.
Hence, the solution of the system is
\begin{equation}
	\psi_{,\theta} = 0\,, \qquad \psi_{,\phi} = -\cos\theta\,.
\end{equation}
Taking mixed partial derivatives we obtain $\psi_{,\theta\phi} = \sin\theta \neq \psi_{,\phi\theta} = 0$, which shows that no smooth solution $\psi$ exists. Interestingly, this parallels the conclusion of \citet{Ericksen1954} in the case of incompressible hyperelasticity (see Eqs.\,(4.1)-(4.3)). 

When the principal stretches are not distinct, $\mathbf{b}$ has the spectral decomposition \eqref{b-case-iii}$_1$. Then, from \eqref{Constraint-divn} we must have $\operatorname{div} \mathbf{n}=0$. However, for a sphere of radius $r(Z)$ we have
\begin{equation}
	\operatorname{div} \mathbf{n}= \frac{2}{r(Z)}\neq 0 \,,
\end{equation}
and hence, there is no universal solution in this case.

\begin{prop} \label{Prop:FamilyZ1}
For a compressible isotropic Cauchy elastic solid reinforced by a single family of parallel inextensible straight fibers, if the deformed fibers remain straight lines (i.e., have vanishing curvature), then the only universal deformations are those belonging to the \textit{Family~$Z_1$ universal deformations}, given by
\begin{equation}
\begin{dcases}
	r(X,Y,Z) = Z + Z_0\,, \\
	\theta(X,Y,Z) = \alpha_0 X + \beta_0 Y + \theta_0\,, \\
	z(X,Y,Z) = k_1 Y + z_0\,,
\end{dcases}
\end{equation}
where $\alpha_0\,, \beta_0\,, \theta_0\,, k_1\,, z_0\,, Z_0$ are constants.
\end{prop}

\subsubsection{All principal invariants are constant}

For $I_1$ and $J=\sqrt{I_3}$ to be constant, in each expression the coefficients of $Z$ and $Z^2$ must vanish. The universality constraints \eqref{Universality-Surface-1}-\eqref{Universality-Surface-8} are modified to read
\begin{empheq}[left={\empheqlbrace }]{align} 
	\label{Universality-Surface-const1}
	&  \mathbf{n} \cdot \mathbf{a}_{,X} = 0\,, \\
	\label{Universality-Surface-const2}
	&  \mathbf{n} \cdot \mathbf{a}_{,Y} = 0\,, \\
	\label{Universality-Surface-const3}
	&  \|\mathbf{a}_{,X}\|^2 + \|\mathbf{a}_{,Y}\|^2=c_1^2\,, \\
	\label{Universality-Surface-const4}
	&  \mathbf{a}_{,X} \cdot \mathbf{n}_{,X} + \mathbf{a}_{,Y} \cdot \mathbf{n}_{,Y}=0\,, \\
	\label{Universality-Surface-const5}
	&  \|\mathbf{n}_{,X}\|^2 + \|\mathbf{n}_{,Y}\|^2=0\,, \\		
	\label{Universality-Surface-const6}
	&  \left( \mathbf{a}_{,X} \times \mathbf{a}_{,Y} \right) \cdot \mathbf{n} = c_4 \,, \\
	\label{Universality-Surface-const7}
	&  \left( \mathbf{a}_{,X} \times \mathbf{n}_{,Y} + \mathbf{n}_{,X} \times 
	\mathbf{a}_{,Y} \right) \cdot \mathbf{n} = 0\,, \\
	\label{Universality-Surface-const8}
	&  \left( \mathbf{n}_{,X} \times \mathbf{n}_{,Y} \right) \cdot \mathbf{n} = 0\,.
\end{empheq}
We also have the universality constraints \eqref{Universality-Constraints-b-c}.
The constraint \eqref{Universality-Surface-const5} implies that $\mathbf{n}_{,X}=\mathbf{n}_{,Y}=\mathbf{0}$, and hence, $\mathbf{n}(X,Y)=\mathbf{n}_0$ is a constant vector. Thus, the fibers in the deformed configuration are all parallel.
Now the constraints \eqref{Universality-Surface-const4}, \eqref{Universality-Surface-const7} and \eqref{Universality-Surface-const8} are trivially satisfied.
We choose Cartesian coordinates $(x,y,z)$ in the deformed configuration such that the fibers are aligned with the $z$-axis. Then $\mathbf{n}_0 = \mathbf{e}_z$, and the remaining universality constraints become
\begin{empheq}[left={\empheqlbrace}]{align} 
	& a^3_{,X} = a^3_{,Y} = 0\,, \\
	\label{Constant-Invariants-Straight-Fibers1}
	& (a^1_{,X})^2 + (a^2_{,X})^2 + (a^1_{,Y})^2 + (a^2_{,Y})^2 = c_1^2\,, \\
	\label{Constant-Invariants-Straight-Fibers2}
	& a^1_{,X} a^2_{,Y} - a^1_{,Y} a^2_{,X} = c_4\,.
\end{empheq}
The first constraint implies that $a^3(X,Y) = a_3$. 
Let us define a surface deformation map $\bar{\varphi}: \mathbb{R}^2 \to \mathbb{R}^2$ by
\begin{equation}
	\bar{\varphi}(X,Y) = (\alpha(X,Y), \beta(X,Y))=(a^1(X,Y), a^2(X,Y))\,.
\end{equation}
The surface deformation gradient of this map is written as
\begin{equation}
	\bar{\mathbf{F}}(X,Y) 
	=\begin{bmatrix} \alpha_{,X} & \alpha_{,Y} \\ \beta_{,X} & \beta_{,Y} \end{bmatrix}
	\,.
\end{equation}
Then the remaining constraints become
\begin{equation} \label{J-Constraints}
	\operatorname{tr} \left(\bar{\mathbf{F}}^{\mathsf{T}} \bar{\mathbf{F}} \right) = c_1^2\,, \qquad
	\det \bar{\mathbf{F}} = c_4\,.
\end{equation}
Thus, these constraints imply that $\bar{\mathbf{F}}(X,Y)$ has constant determinant and constant Frobenius norm. 
It should also be noted that $\operatorname{tr}(\bar{\mathbf{F}}^{\mathsf{T}} \bar{\mathbf{F}})$ and $\det \bar{\mathbf{F}}$ are the first and second principal invariants of the planar deformation map $\bar{\varphi}$.

In summary, the deformation mapping with respect to Cartesian coordinates has the following form
\begin{equation} 
	x(X,Y,Z) = \alpha(X,Y)\,,\qquad 
	y(X,Y,Z) = \beta(X,Y)\,,\qquad 
	z(X,Y,Z) = Z+ a_3\,.
\end{equation}
Let us now use Cylindrical coordinates in both reference and current configurations.
\begin{equation}
	r(R,\Theta,Z) = r(R,\Theta)\,, \qquad
	\theta(R,\Theta,Z) = \theta(R,\Theta)\,, \qquad
	z(R,\Theta,Z) = Z + a_3\,.
\end{equation}

Obviously, any homogeneous map satisfies the constraints \eqref{J-Constraints}.
The principal invariants of $\bar{\varphi}$ being constant is equivalent to the two principal stretches being constant. Let us denote them by $\lambda_1$ and $\lambda_2$.
Therefore
\begin{equation} 
	\begin{bmatrix}
	C_{11} & C_{12}  \\
	C_{12} & C_{22} 
	\end{bmatrix}
	=\begin{bmatrix}
	\cos \Theta & \sin \Theta \\
	-\sin \Theta & \cos \Theta
	\end{bmatrix}
	\begin{bmatrix}
	\lambda_1^2 & 0  \\
	0 & \lambda_2^2
	\end{bmatrix}
	\begin{bmatrix}
	\cos \Theta & -\sin \Theta \\
	\sin \Theta & \cos \Theta
	\end{bmatrix}
	\,,
\end{equation}
where $\Theta=\Theta(X,Y)$.
First let us assume that the principal stretches are equal, i.e., $\lambda_1=\lambda_2=\lambda$. Then 
\begin{equation} 
	\begin{bmatrix}
	C_{11} & C_{12}  \\
	C_{12} & C_{22} 
	\end{bmatrix}
	= \lambda^2
	\begin{bmatrix}
	1 & 0 \\
	0 & 1
	\end{bmatrix}
	\,,
\end{equation}
i.e., the right Cauchy-Green strain is constant, and consequently, the corresponding deformations are homogeneous \citep[Theorem~1.3]{Blume1989}.

Next we discuss the case $\lambda_1 \neq \lambda_2$.
When principal stretches are constant, deformation gradient has the following representation\footnote{Using polar decomposition and diagonalization of the stretch tensor, one can write $\mathbf{F} = \mathbf{R}\mathbf{U} = \mathbf{R}\mathbf{Q}_2\boldsymbol{\Lambda}\mathbf{Q}_2^{\mathsf{T}} = \mathbf{Q}_1\boldsymbol{\Lambda}\mathbf{Q}_2^{\mathsf{T}}$, where $\boldsymbol{\Lambda} = \operatorname{diag}(\lambda_1, \lambda_2)$. See \citet{Gevirtz1992} for a similar representation.}
\begin{equation} 
	\mathbf{F}
	=\begin{bmatrix}
	\cos \Theta & -\sin \Theta \\
	\sin \Theta & \cos \Theta
	\end{bmatrix}
	\begin{bmatrix}
	\lambda_1 & 0  \\
	0 & \lambda_2
	\end{bmatrix}
	\begin{bmatrix}
	\cos \Phi & -\sin \Phi \\
	\sin \Phi & \cos \Phi
	\end{bmatrix}
	\,,
\end{equation}
where $\Theta=\Theta(X,Y)$ and $\Phi=\Phi(X,Y)$.
Compatibility of the deformation gradient $\operatorname{Curl}\mathbf{F}=\mathbf{0}$ gives us the following system of PDEs:
\begin{equation} \label{2D-F-Compatibility}
\begin{dcases}
\cos\Theta \left[-\sin\Phi \left(\lambda_2 \Theta_{,Y} + \lambda_1 \Phi_{,Y}\right) \right]
+ \cos\Phi \left(\lambda_2 \Theta_{,X} + \lambda_1 \Phi_{,X} \right) \\[6pt]
\quad - \sin\Theta \left[
\cos\Phi \left(\lambda_1 \Theta_{,Y} + \lambda_2 \Phi_{,Y}\right)
+ \sin\Phi \left(\lambda_1 \Theta_{,X} + \lambda_2 \Phi_{,X}\right)
\right] = 0\,, \\[12pt]
\cos\Phi \left[
\cos\Theta \left(\lambda_1 \Theta_{,Y} + \lambda_2 \Phi_{,Y}\right)
+ \sin\Theta \left(\lambda_1 \Theta_{,X} + \lambda_2 \Phi_{,X}\right)
\right] \\[6pt]
\quad + \sin\Phi \left[
-\sin\Theta \left(\lambda_2 \Theta_{,Y} + \lambda_1 \Phi_{,Y}\right)
+ \cos\Theta \left(\lambda_2 \Theta_{,X} + \lambda_1 \Phi_{,X}\right)
\right] = 0\,.
\end{dcases}
\end{equation}

The universality constraint $\operatorname{div}\mathbf{b}^\sharp=\mathbf{0}$ is simplified to read
\begin{equation}
\label{Universality-Constraint-Vector}
\frac{(\lambda_1^2 - \lambda_2^2)}{\lambda_1 \lambda_2}
\begin{bmatrix}
\cos(2\Phi)\,\Theta_{,Y} + \sin(2\Phi)\,\Theta_{,X} \\
-\sin(2\Phi)\,\Theta_{,Y} + \cos(2\Phi)\,\Theta_{,X}
\end{bmatrix}
= \mathbf{0}\,.
\end{equation}
Knowing that $\lambda_1\neq \lambda_2$, this can be written as
\begin{equation}
	\mathbf{Q}^{\mathsf{T}}(2\Phi) \cdot \nabla \Theta = \mathbf{0}\,,
	\qquad
	\mathbf{Q}(2\Phi) =
	\begin{bmatrix}
	\cos(2\Phi) & -\sin(2\Phi) \\
	\sin(2\Phi) & \cos(2\Phi)
	\end{bmatrix}\,.
	\end{equation}
Since $\mathbf{Q}(2\Phi)$ is an invertible matrix (rotation), one concludes that $\nabla \Theta = \mathbf{0}$.
Knowing that $\Theta$ is constant, \eqref{2D-F-Compatibility} are simplified to read
\begin{equation}
\begin{dcases}
\left[ \lambda_1 \cos \Theta \cos \Phi - \lambda_2 \sin \Theta \sin \Phi \right] \Phi_{,X} 
- \left[ \lambda_2 \cos \Phi \sin \Theta + \lambda_1 \cos \Theta \sin \Phi \right] \Phi_{,Y} = 0 \,, \\[6pt]
\left[ \lambda_1 \cos \Phi \sin \Theta + \lambda_2 \cos \Theta \sin \Phi \right] \Phi_{,X} 
+ \left[ \lambda_2 \cos \Theta \cos \Phi - \lambda_1 \sin \Theta \sin \Phi \right] \Phi_{,Y} = 0 \,.
\end{dcases}
\end{equation}
The determinant of the coefficient matrix is $\lambda_1\lambda_2\neq0$, and hence $\Phi_{,X} =\Phi_{,Y} =0$. Therefore, $\Phi$ is constant.
In summary, we have proved the following result.

\begin{prop}\label{Prop:Straight}
If $\mathbf{n}_{,Z} = \mathbf{0}$ and all three principal invariants are constant, then no inhomogeneous universal deformations exist.
\end{prop}

\subsection{\texorpdfstring{Integrability equations for the tension field when $\mathbf{n}_{,Z} \neq \mathbf{0}$}{Integrability equations for the tension field when n,Z not zero}} 

We know that $\mathbf{n}$ is a unit vector, and hence, $\mathbf{n}\cdot \mathbf{n}_{,Z}=0$, i.e., $\mathbf{n}_{,Z}\perp \mathbf{n}$. When $\mathbf{n}_{,Z}\neq \mathbf{0}$, the set $\{\mathbf{n},\mathbf{n}_{,Z},\mathbf{n}\times\mathbf{n}_{,Z}\}$ is an orthogonal basis for $\mathbb{R}^3$. The components of the equilibrium equations $\mathring{T}_{,Z}\,\mathbf{n}+\mathring{T}\,\mathbf{n}_{,Z} =\mathbf{f}$ with respect to this basis are:
\begin{equation} \label{Integrability-Z1}
\left\{
\begin{aligned}
\mathbf{n}& : \quad \mathring{T}_{,Z} = \mathbf{f} \cdot \mathbf{n}\,, \\
\mathbf{n}_{,Z} & : \quad \mathring{T} \|\mathbf{n}_{,Z}\|^2 = \mathbf{f} \cdot \mathbf{n}_{,Z}\,, \\
\mathbf{n} \times \mathbf{n}_{,Z} & : \quad \mathbf{f} \cdot (\mathbf{n} \times \mathbf{n}_{,Z}) = 0\,. 
\end{aligned}
\right.
\end{equation}
Eq.~\eqref{Integrability-Z1}$_3$ is the integrability condition, which implies that $\mathbf{f}\in \operatorname{span} \{\mathbf{n} ,\mathbf{n} _{,Z}\}$. This condition guarantees that the overdetermined system admits a scalar solution for $\mathring{T}$.
Note that for straight lines the binormal vector $\boldsymbol{\xi}$ is not well-defined, and that is why we treated the case of $\mathbf{n}_{,Z}=\mathbf{0}$ separately in \S\ref{Sec:Straight-Fibers}.

\begin{remark}
The following is a geometric interpretation of the vector $\mathbf{n}\times \nabla_{\mathbf{n}}\mathbf{n}$. Consider a fiber in the deformed configuration and parametrize it with arc length $s$. Its unit tangent vector is $\mathbf{n}(s)$.  Its derivative $\mathbf{n}'(s)$ is derivative of $\mathbf{n}$ along the tangent vector, i.e., along $\mathbf{n}$ itself. Thus, $\mathbf{n}'(s)=\nabla_{\mathbf{n}(s)}\mathbf{n}(s)$. This vector is not necessarily of unit length; its length is the curvature of the curve $\kappa(s)$.
The binormal vector $\mathbf{b}(s)$ is defined as the cross product of the unit and the normal vectors, i.e.,  \citep{doCarmo1976} 
\begin{equation} 
	\mathbf{b}(s)=\mathbf{n}(s)\times\frac{1}{\kappa(s)}\mathbf{n}'(s)
	=\frac{1}{\kappa(s)}\mathbf{n}(s)\times\nabla_{\mathbf{n}(s)}\mathbf{n}(s)
	\,.
\end{equation}
We see that up to a scalar factor the vector $\mathbf{n}\times \nabla_{\mathbf{n}}\mathbf{n}$ is the binormal vector of the fibers in the deformed configuration.
\end{remark}

In our problem, fibers in the deformed configuration are parameterized by $Z$ and $\nabla_{\mathbf{n}}\mathbf{n}=\mathbf{n}_{,Z}= \kappa(Z)\,\mathbf{n}$, where $\kappa(Z)$ is the curvature of the deformed fiber.
Let 
\begin{equation} 
	\boldsymbol{\xi}:=\mathbf{n} \times \frac{\mathbf{n}_{,Z}}{\kappa(Z)}
	\,.
\end{equation}
Substituting \eqref{ForcingTerm-f} into the integrability equation \eqref{Integrability-Z1}$_3$ and recalling that the response functions and their derivatives are arbitrary, one obtains the following set of universality constraints:
\begin{empheq}[left={\empheqlbrace }]{align} 
	\label{Universality-Constraints-Cauchy-1}
	& \boldsymbol{\xi}\cdot\operatorname{div}\mathbf{b}^\sharp=0\,, \\
	\label{Universality-Constraints-Cauchy-2}
	& \boldsymbol{\xi}\cdot\operatorname{div}\mathbf{c}^\sharp=0\,, \\
	\label{Universality-Constraints-Cauchy-3}
	& \boldsymbol{\xi}\cdot\nabla I_i=0\,,\qquad i=1,2,3\,, \\
	\label{Universality-Constraints-Cauchy-4}
	& \boldsymbol{\xi}\cdot \mathbf{b}\cdot\nabla I_i=0\,,\quad i=1,2,3\,, \\
	\label{Universality-Constraints-Cauchy-5}
	& \boldsymbol{\xi}\cdot \mathbf{c}\cdot\nabla I_i=0\,,\quad i=1,2,3\,.
\end{empheq}
Or, equivalently
\begin{empheq}[left={\empheqlbrace }]{align} 
	\label{Universality-Constraints-Cauchy4-1}
	& \operatorname{div}\mathbf{b}^\sharp= \beta_1 \mathbf{n} + \beta_2 \mathbf{n}_{,Z}\,, \\
	\label{Universality-Constraints-Cauchy4-2}
	& \operatorname{div}\mathbf{c}^\sharp= \gamma_1 \mathbf{n} + \gamma_2 \mathbf{n}_{,Z}
	\,, \\
	\label{Universality-Constraints-Cauchy4-3}
	& \nabla I_i= \lambda_i \mathbf{n} + \kappa_i \mathbf{n}_{,Z}\,,\qquad i=1,2,3\,, \\
	\label{Universality-Constraints-Cauchy4-4}
	&  \mathbf{b}\cdot\nabla I_i=\xi_i \mathbf{n} + \nu_i \mathbf{n}_{,Z}\,,\quad i=1,2,3\,, \\
	\label{Universality-Constraints-Cauchy4-5}
	& \mathbf{c}\cdot\nabla I_i=\eta_i \mathbf{n} + \chi_i \mathbf{n}_{,Z}\,,\quad i=1,2,3\,.
\end{empheq}
Note that the universality constraints \eqref{Universality-Constraints-Cauchy3-1}–\eqref{Universality-Constraints-Cauchy3-5} represent a special case of the above universality constraints when $\mathbf{n}_{,Z} = \mathbf{0}$.

\begin{remark}
A comment is in order here. The universality constraints derived above are more intricate than those of incompressible isotropic elasticity. This is because the incompressibility constraint is, in a sense, isotropic—it does not depend on any preferred direction. In contrast, for a solid reinforced with inextensible fibers, the inextensibility constraint is inherently anisotropic: it depends explicitly on the direction of the deformed fiber. Consequently, the resulting universality constraints encode this directional dependence in terms of the Frenet frame associated with the deformed fiber.
\end{remark}

While we are not yet able to fully characterize the universal deformations corresponding to deformed fibers with non-vanishing curvature, we present a detailed formulation of the problem and offer several partial results. These include conditions on the functional dependence of the principal invariants, the binormal vector being an eigenvector of the Finger tensor $\mathbf{b}$, and other geometric constraints. A complete solution to this case remains an open problem and is left for future work.

\subsubsection{At least one principal invariant is not constant} \label{Sec:Curved-Fibers}

The principal invariants $I_1$, $I_2$, and $I_3$ are functionally dependent if there exists a non-trivial function (a function that is not identically zero) such that $F(I_1,I_2,I_3)=0$. Taking derivatives one obtains
\begin{equation}\label{I-Jacobian}
\begin{bmatrix}
  I_{1,1} & I_{2,1} & I_{3,1}   \\
  I_{1,2} & I_{2,2} & I_{3,2}  \\
  I_{1,3} & I_{2,3} & I_{3,3}  
\end{bmatrix}
\begin{pmatrix} \frac{\partial F}{\partial I_1} \\ \frac{\partial F}{\partial I_2} \\ \frac{\partial F}{\partial I_3} \end{pmatrix}
=\begin{pmatrix} 0 \\ 0 \\ 0 \end{pmatrix}.
\end{equation}
For $F$ to be non-trivial, the Jacobian matrix in \eqref{I-Jacobian} must have rank less than $3$.
Let us define the $3 \times 3$ matrix $\mathbf{M} = \big[ \nabla I_1 \quad \nabla I_2 \quad \nabla I_3 \big]$.
From \eqref{Universality-Constraints-Cauchy4-3}, each column lies in the subspace $\operatorname{span}\{\mathbf{n},\,\mathbf{n}_{,Z}\}$.
Since $\{\mathbf{n},\,\mathbf{n}_{,Z}\}$ span at most a two-dimensional subspace of $\mathbb{R}^3$, it follows that the rank of $\mathbf{M}$ is at most $2$.
Therefore, $I_1$, $I_2$, and $I_3$ are functionally dependent. 
This means that there exist independent variables $(\zeta, \chi)$ such that $I_i = I_i(\zeta, \chi)$ for $i = 1, 2, 3$.

Let us first assume that the vectors $\nabla I_1$ and $\nabla I_2$ are linearly independent. Then the set $\{\nabla I_1, \nabla I_2\}$ spans the plane $\mathcal{P} = \operatorname{span} \{ \mathbf{n}, \mathbf{n}_{,Z} \}$. 
Since $\mathbf{b}\cdot\nabla I_i \in \mathcal{P}$ for $i=1,2$, it follows that the plane $\mathcal{P}$ is invariant under $\mathbf{b}$.
Therefore, the orthogonal complement $\mathcal{P}^\perp = \operatorname{span} \{ \mathbf{n} \times \mathbf{n}_{,Z} \}$ is also invariant under $\mathbf{b}$.\footnote{We know that $\mathbf{b}$ is a symmetric $(1,1)$-tensor on the Riemannian manifold $(\mathcal{C},\mathbf{g})$, i.e.,
\begin{equation}
	\llangle \mathbf{b} \cdot \mathbf{u},\, \mathbf{v} \rrangle_{\mathbf{g}} = \llangle \mathbf{u},\, \mathbf{b} \cdot \mathbf{v} \rrangle_{\mathbf{g}}\,, \qquad \text{for all } \mathbf{u}, \mathbf{v} \in T_x\mathcal{C}\,.
\end{equation}
We also know that the subspace $\mathcal{P} \subset T_x\mathcal{C}$ is invariant under $\mathbf{b}$, i.e., $\mathbf{b} \cdot \mathbf{w} \in \mathcal{P}$ for all $\mathbf{w} \in \mathcal{P}$. 
Let $\mathbf{v} \in \mathcal{P}^\perp$. For any $\mathbf{u} \in \mathcal{P}$, we have
\begin{equation}
	\llangle \mathbf{b} \cdot \mathbf{v},\, \mathbf{u} \rrangle_{\mathbf{g}} 
	= \llangle \mathbf{v},\, \mathbf{b} \cdot \mathbf{u} \rrangle_{\mathbf{g}}
	= 0\,,
\end{equation}
where we used symmetry of $\mathbf{b}$ and the assumption that $\mathbf{b} \cdot \mathbf{u} \in \mathcal{P}$, and $\mathbf{v} \in \mathcal{P}^\perp$. Hence, $\mathbf{b} \cdot \mathbf{v} \in \mathcal{P}^\perp$, and so $\mathcal{P}^\perp$ is also invariant under $\mathbf{b}$.} That is, $\mathbf{b}$ maps both $\mathcal{P}$ and $\mathcal{P}^\perp$ to themselves.
This implies that the tensor $\mathbf{b}$ admits a block-diagonal form in the orthonormal basis $\left\{ \mathbf{n},\, \frac{\mathbf{n}_{,Z}}{\kappa(Z)},\, \mathbf{n} \times \frac{\mathbf{n}_{,Z}}{\kappa(Z)} \right\}$. In particular, the off-diagonal terms coupling $\mathcal{P}$ and $\mathcal{P}^\perp$ must vanish. Hence,
\begin{equation}
	b_{13} = b_{23} = 0\,,
\end{equation}
and $\boldsymbol{\xi} =\mathbf{n} \times \frac{\mathbf{n}_{,Z}}{\kappa(Z)}$ is an eigenvector of $\mathbf{b}$ corresponding to the one-dimensional eigenspace associated with $\mathcal{P}^\perp$, i.e., $\mathbf{b}\boldsymbol{\xi}=\Upsilon\boldsymbol{\xi}$.
Therefore, the symmetric $(1,1)$-tensor $\mathbf{b}$ admits the spectral decomposition
\begin{equation}\label{b-Spectral}
	\mathbf{b}^\sharp = \Lambda_1\, \n \otimes \n + \Lambda_2\, \nn \otimes \nn 
	+ \Upsilon\, \boldsymbol{\xi} \otimes \boldsymbol{\xi}\,,
\end{equation}
where $\Lambda_1,\Lambda_2,\Upsilon>0$, $\{\n, \nn\}$ is an orthonormal basis for the plane $\mathcal{P} = \operatorname{span} \{\mathbf{n}, \mathbf{n}_{,Z} \}$, and $\boldsymbol{\xi} =\mathbf{n} \times \frac{\mathbf{n}_{,Z}}{\kappa(Z)}$ is a unit vector in $\mathcal{P}^\perp$. 
It should be noted that $\n \otimes \n + \nn \otimes \nn + \boldsymbol{\xi} \otimes \boldsymbol{\xi}=\mathbf{g}^\sharp$.

Substituting $\boldsymbol{\xi} \otimes \boldsymbol{\xi}=\mathbf{g}^\sharp-\n \otimes \n - \nn \otimes \nn$ into \eqref{b-Spectral}, one obtains
\begin{equation}\label{b-Spectral2}
	\mathbf{b}^\sharp =  \Upsilon\, \mathbf{g}^\sharp+
	(\Lambda_1-\Upsilon)\, \n \otimes \n + (\Lambda_2-\Upsilon)\, \nn \otimes \nn 
	\,.
\end{equation}
Thus
\begin{equation}\label{c-Spectral2}
	\mathbf{c}^\sharp = \frac{1}{\Upsilon}\, \mathbf{g}^\sharp+
	 \frac{1}{\Lambda_1-\Upsilon}\, \n \otimes \n + \frac{1}{\Lambda_2-\Upsilon}\, \nn \otimes \nn 
	\,.
\end{equation}
Therefore
\begin{equation}
\begin{aligned}
	\operatorname{div} \mathbf{b}^\sharp 
	&= \left[ \nabla (\Lambda_1 - \Upsilon) \cdot \n + (\Lambda_1 - \Upsilon)\, \operatorname{div} \n \right] \n 
	+ (\Lambda_1 - \Upsilon)\, \nabla_{\n} \n \\
	&\quad + \left[ \nabla (\Lambda_2 - \Upsilon) \cdot \nn + (\Lambda_2 - \Upsilon)\, \operatorname{div} \nn \right] \nn 
	+ (\Lambda_2 - \Upsilon)\, \nabla_{\nn} \nn + \nabla \Upsilon\,.
\end{aligned}
\end{equation}
Since $\operatorname{div} \mathbf{b}^\sharp \in \operatorname{span} \{ \n, \nn \}$, we must have $\operatorname{div} \mathbf{b}^\sharp \cdot \boldsymbol{\xi} = 0$. Therefore
\begin{equation} \label{Projecttion1}
	(\Lambda_1 - \Upsilon)\, \nabla_{\n} \n \cdot \boldsymbol{\xi} 
	+ (\Lambda_2 - \Upsilon)\, \nabla_{\nn} \nn \cdot \boldsymbol{\xi} 
	+ \nabla \Upsilon \cdot \boldsymbol{\xi} = 0\,.
\end{equation}
We also know that $\operatorname{div} \mathbf{c}^\sharp \cdot \boldsymbol{\xi} = 0$, and hence
\begin{equation} \label{Projecttion2}
	\frac{1}{\Lambda_1 - \Upsilon}\, \nabla_{\n} \n \cdot \boldsymbol{\xi} 
	+ \frac{1}{\Lambda_2 - \Upsilon}\, \nabla_{\nn} \nn \cdot \boldsymbol{\xi} 
	-\frac{1}{\Upsilon^2} \nabla \Upsilon \cdot \boldsymbol{\xi} = 0\,.
\end{equation}

Substituting $\nn \otimes \nn =\mathbf{g}^\sharp -\n \otimes \n  - \boldsymbol{\xi} \otimes \boldsymbol{\xi}$ into \eqref{b-Spectral}, one obtains
\begin{equation}\label{b-Spectral3}
	\mathbf{b}^\sharp = (\Lambda_1-\Lambda_2)\, \n \otimes \n 
	+ (\Upsilon-\Lambda_2)\, \boldsymbol{\xi}  \otimes \boldsymbol{\xi}  
	+ \Lambda_2\, \mathbf{g}^\sharp\,,
\end{equation}
and hence
\begin{equation}\label{c-Spectral3}
	\mathbf{c}^\sharp = \frac{1}{\Lambda_1-\Lambda_2}\, \n \otimes \n 
	+ \frac{1}{\Upsilon-\Lambda_2}\, \boldsymbol{\xi}  \otimes \boldsymbol{\xi}  + \frac{1}{\Lambda_2}\, \mathbf{g}^\sharp\,.
\end{equation}
Thus
\begin{equation}
\begin{aligned}
	\operatorname{div} \mathbf{b}^\sharp 
	&= \left[ \nabla (\Lambda_1 - \Lambda_2) \cdot \n + (\Lambda_1 - \Lambda_2)\, \operatorname{div} \n \right] \n 
	+ (\Lambda_1 - \Lambda_2)\, \nabla_{\n} \n \\
	&\quad + \left[ \nabla (\Upsilon - \Lambda_2) \cdot \boldsymbol{\xi} 
	+ (\Upsilon - \Lambda_2)\, \operatorname{div} \boldsymbol{\xi} \right] \boldsymbol{\xi}
	+ (\Upsilon - \Lambda_2)\, \nabla_{\boldsymbol{\xi}} \boldsymbol{\xi} 
	+ \nabla \Lambda_2\,, \\
	\operatorname{div} \mathbf{c}^\sharp 
	&= \left[ -\frac{1}{(\Lambda_1 - \Lambda_2)^2}\,\nabla (\Lambda_1 - \Lambda_2) \cdot \n 
	+ \frac{1}{\Lambda_1 - \Lambda_2}\, \operatorname{div} \n \right] \n 
	+ \frac{1}{\Lambda_1 - \Lambda_2}\, \nabla_{\n} \n \\
	&\quad + \left[ -\frac{1}{(\Upsilon - \Lambda_2)^2}\,\nabla (\Upsilon - \Lambda_2) \cdot \boldsymbol{\xi} 
	+ \frac{1}{\Upsilon - \Lambda_2}\, \operatorname{div} \boldsymbol{\xi} \right] \boldsymbol{\xi} 
	+ \frac{1}{\Upsilon - \Lambda_2}\, \nabla_{\boldsymbol{\xi}} \boldsymbol{\xi} 
	- \frac{1}{\Lambda_2^2}\, \nabla \Lambda_2\,.
\end{aligned}
\end{equation}
The universality constraints $\operatorname{div} \mathbf{b}^\sharp \cdot \boldsymbol{\xi}=\operatorname{div} \mathbf{c}^\sharp \cdot \boldsymbol{\xi}=0$ are simplified to read
\begin{equation} 
\begin{aligned}
	& (\Lambda_1 - \Lambda_2)\, \nabla_{\n} \n \cdot \boldsymbol{\xi} 
	+ \nabla \Upsilon \cdot \boldsymbol{\xi} 
	+ (\Upsilon - \Lambda_2)\, \operatorname{div} \boldsymbol{\xi} = 0\,, \\
	& \frac{1}{\Lambda_1 - \Lambda_2}\, \nabla_{\n} \n \cdot \boldsymbol{\xi} 
	- \frac{1}{(\Upsilon - \Lambda_2)^2} \, \nabla \Upsilon \cdot \boldsymbol{\xi} 
	+ \left[ \frac{1}{(\Upsilon - \Lambda_2)^2} - \frac{1}{\Lambda_2^2} \right]\, \nabla \Lambda_2 
	\cdot \boldsymbol{\xi} 
	+ \frac{1}{\Upsilon - \Lambda_2}\, \operatorname{div} \boldsymbol{\xi} = 0\,.
\end{aligned}
\end{equation}

Alternatively, substituting $\boldsymbol{\xi} \otimes \boldsymbol{\xi} =\mathbf{g}^\sharp -\n \otimes \n  - \nn \otimes \nn$ into \eqref{b-Spectral}, one obtains
\begin{equation}
\begin{aligned}
	\mathbf{b}^\sharp &= \Upsilon\,\mathbf{g}^\sharp 
	+(\Lambda_1- \Upsilon)\,\n\otimes\n+(\Lambda_2- \Upsilon)\,\nn\otimes\nn \,, \\
	\mathbf{c}^\sharp &= \frac{1}{\Upsilon}\,\mathbf{g}^\sharp
	+\frac{1}{\Lambda_1- \Upsilon}\,\n\otimes\n+\frac{1}{\Lambda_2- \Upsilon}\,\nn\otimes\nn
	\,.
\end{aligned}
\end{equation}
Thus
\begin{equation}
\begin{aligned}
	\operatorname{div} \mathbf{b}^\sharp 
	&= \left[ \nabla (\Lambda_1 - \Upsilon) \cdot \n 
	+ (\Lambda_1 - \Upsilon)\, \operatorname{div} \n \right] \n 
	+ (\Lambda_1 - \Upsilon)\, \nabla_{\n} \n \\
	&\quad + \left[ \nabla (\Lambda_2 - \Upsilon) \cdot \nn 
	+ (\Lambda_2 - \Upsilon)\, \operatorname{div} \nn \right] \nn 
	+ (\Lambda_2 - \Upsilon)\, \nabla_{\nn} \nn 
	+ \nabla \Upsilon\,, \\
	\operatorname{div} \mathbf{c}^\sharp 
	&= \left[ -\frac{1}{(\Lambda_1 - \Upsilon)^2}\,\nabla (\Lambda_1 - \Upsilon) \cdot \n 
	+ \frac{1}{\Lambda_1 - \Upsilon}\, \operatorname{div} \n \right] \n 
	+ \frac{1}{\Lambda_1 - \Upsilon}\, \nabla_{\n} \n \\
	&\quad + \left[ -\frac{1}{(\Lambda_2 - \Upsilon)^2}\,\nabla (\Lambda_2 - \Upsilon) \cdot \nn 
	+ \frac{1}{\Lambda_2 - \Upsilon}\, \operatorname{div} \nn \right] \nn 
	+ \frac{1}{\Lambda_2 - \Upsilon}\, \nabla_{\nn} \nn 
	- \frac{1}{\Upsilon^2}\, \nabla \Upsilon\,.
\end{aligned}
\end{equation}
In this case, the universality constraints $\operatorname{div} \mathbf{b}^\sharp \cdot \boldsymbol{\xi}=\operatorname{div} \mathbf{c}^\sharp \cdot \boldsymbol{\xi}=0$ are simplified to read
\begin{equation}
\begin{aligned}
	& (\Lambda_1 - \Upsilon)\, \nabla_{\n} \n  \cdot \boldsymbol{\xi}
	+ (\Lambda_2 - \Upsilon)\, \nabla_{\nn} \nn \cdot \boldsymbol{\xi}
	+ \nabla \Upsilon \cdot \boldsymbol{\xi} =0\,, \\
	& \frac{1}{\Lambda_1 - \Upsilon}\, \nabla_{\n} \n \cdot \boldsymbol{\xi}
	+ \frac{1}{\Lambda_2 - \Upsilon}\, \nabla_{\nn} \nn \cdot \boldsymbol{\xi}
	- \frac{1}{\Upsilon^2}\, \nabla \Upsilon \cdot \boldsymbol{\xi}=0\,.
\end{aligned}
\end{equation}

If $\Lambda_1=\Lambda_2=\Lambda$, then using the decomposition \eqref{b-Spectral2} we have
\begin{equation}
	\mathbf{b}^\sharp = \Upsilon\, \mathbf{g}^\sharp+ (\Lambda-\Upsilon)\, \left(\n \otimes \n + \nn \otimes \nn \right) \,.
\end{equation}
In this case, deformation restricted to the plane $\operatorname{span} \{ \n, \nn \}$ is pure dilatational. This implies that both $\mathbf{n}$ and $\mathbf{n}_{,Z}$ are eigenvectors of $\mathbf{b}^\sharp$ with eigenvalue $\Lambda$.
Divergence of $\mathbf{b}^\sharp$ is calculated as
\begin{equation}
\begin{aligned}
	\operatorname{div} \mathbf{b}^\sharp 
	&= (\Lambda - \Upsilon)\left[ (\operatorname{div} \n)\, \n + \nabla_{\n} \n 
	+ (\operatorname{div} \nn)\, \nn + \nabla_{\nn} \nn \right] + \nabla \Upsilon\,.
\end{aligned}
\end{equation}
Therefore, $\operatorname{div} \mathbf{b}^\sharp \cdot \boldsymbol{\xi} = 0$ is simplified to read
\begin{equation}
	(\Lambda - \Upsilon)\left( \nabla_{\n} \n + \nabla_{\nn} \nn \right) \cdot \boldsymbol{\xi} 
	+ \nabla \Upsilon \cdot \boldsymbol{\xi} =0\,.
\end{equation}
Similarly, $\operatorname{div} \mathbf{c}^\sharp \cdot \boldsymbol{\xi} = 0$ is simplified to read
\begin{equation}
	\frac{1}{\Lambda - \Upsilon}\left( \nabla_{\n} \n + \nabla_{\nn} \nn \right) \cdot \boldsymbol{\xi} 
	-\frac{1}{\Upsilon^2}\, \nabla \Upsilon \cdot \boldsymbol{\xi} =0\,.
\end{equation}
From the above two equations after eliminating $\left( \nabla_{\n} \n + \nabla_{\nn} \nn \right) \cdot \boldsymbol{\xi}$, we obtain
\begin{equation}
	\left[\frac{(\Lambda - \Upsilon)^2}{\Upsilon^2} + 1 \right] \nabla \Upsilon \cdot \boldsymbol{\xi} = 0\,.
\end{equation}
Since the prefactor is strictly positive, it follows that
\begin{equation}
\nabla \Upsilon \cdot \boldsymbol{\xi} = 0\,,
\end{equation}
and therefore
\begin{equation}
(\nabla_{\n} \n + \nabla_{\nn} \nn) \cdot \boldsymbol{\xi} = 0\,.
\end{equation}
In particular, we conclude that the transverse eigenvalue $\Upsilon$ is constant along $\boldsymbol{\xi}$.
From $\mathbf{b}^\sharp \mathbf{n}=\Lambda \mathbf{n}$ one concludes that $\mathbf{C}\mathbf{N}=\Lambda \mathbf{N}$. In components, $C^A{}_B N^B=\Lambda N^A$. Thus, $N_A C^A{}_B N^B =1=\Lambda N^A N_A=\Lambda$, and hence $\Lambda=1$. This implies that there is no deformation in the oscillating plane $\operatorname{span} \{ \mathbf{n}, \mathbf{n}_{,Z} \}$.

We can alternatively, use the spectral decomposition \eqref{b-Spectral3}, which in this case is simplified to read
\begin{equation}
	\mathbf{b}^\sharp =  (\Upsilon-\Lambda)\, \boldsymbol{\xi}  \otimes \boldsymbol{\xi}  
	+ \Lambda\, \mathbf{g}^\sharp = (\Upsilon-1)\, \boldsymbol{\xi}  \otimes \boldsymbol{\xi}  
	+ \mathbf{g}^\sharp\,.
\end{equation}
The universality constraint $\operatorname{div} \mathbf{b}^\sharp \cdot \boldsymbol{\xi} = 0$ implies that $\operatorname{div}\boldsymbol{\xi}=0$.
We observe that at every point, there is stretch only normal to fibers along the binormal.

Another possibility is when $\Upsilon$ is equal to either $\Lambda_1$ or $\Lambda_2$. Since the decomposition \eqref{b-Spectral2} is symmetric in $\n$ and $\nn$, we may assume without loss of generality that $\Upsilon = \Lambda_2$.
Assume $\Upsilon = \Lambda_2$. Using the decomposition \eqref{b-Spectral2}, we have
\begin{equation}
	\mathbf{b}^\sharp = (\Lambda_1-\Upsilon)\, \n \otimes \n + \Upsilon\, \mathbf{g}^\sharp\,.
\end{equation}
Thus
\begin{equation}
\begin{aligned}
	\operatorname{div} \mathbf{b}^\sharp 
	&= (\Lambda_1 - \Upsilon) \left[ (\operatorname{div} \n)\, \n + \nabla_{\n} \n \right]  + \nabla \Upsilon\,.
\end{aligned}
\end{equation}
Hence, $\operatorname{div} \mathbf{b}^\sharp \cdot \boldsymbol{\xi} = 0$ gives
\begin{equation}
	(\Lambda_1 - \Upsilon)\, \nabla_{\n} \n \cdot \boldsymbol{\xi} 
	+ \nabla \Upsilon \cdot \boldsymbol{\xi} = 0\,.
\end{equation}
Similarly, $\operatorname{div} \mathbf{c}^\sharp \cdot \boldsymbol{\xi} = 0$ gives
\begin{equation}
	\frac{1}{\Lambda_1 - \Upsilon}\, \nabla_{\n} \n \cdot \boldsymbol{\xi} 
	- \frac{1}{\Upsilon^2} \nabla \Upsilon \cdot \boldsymbol{\xi} = 0\,.
\end{equation}
Eliminating $\nabla_{\n} \n \cdot \boldsymbol{\xi}$, we obtain
\begin{equation}
	\left( \frac{1}{\Lambda_1 - \Upsilon} + \frac{1}{\Upsilon^2} \right)
	\nabla \Upsilon \cdot \boldsymbol{\xi} = 0\,.
\end{equation}
Therefore, $\nabla \Upsilon \cdot \boldsymbol{\xi} = 0$.
This implies that the transverse eigenvalue $\Upsilon$ is constant along the $\boldsymbol{\xi}$ direction.

If $\nabla I_1$ and $\nabla I_2$ are linearly dependent, then $I_1$ and $I_2$ are functionally dependent, i.e., there exists a nontrivial smooth function $F(I_1, I_2) = 0$. This then implies that there is a variable $\zeta$ such that $I_1=I_1(\zeta)$, $I_2=I_2(\zeta)$, and $I_3=I_3(\zeta)$. $\boldsymbol{\xi}\cdot\nabla I_i=0$ implies that $\nabla \zeta\cdot \boldsymbol{\xi}=0$.

\subsubsection{All principal invariants are constant}

In this case we still have $C_{33}=1$, i.e.,
\begin{equation} \label{C-Fiber}
	\mathbf{C}^\flat=\begin{bmatrix}
	C_{11} & C_{12} & C_{13} \\
	C_{12} & C_{22} & C_{23} \\
	C_{13} & C_{23} & 1 
	\end{bmatrix}\,,
\end{equation}
where $C_{AB}=C_{AB}(X,Y,Z)$.
When $I_1$, $I_2$, and $I_3$ are constant, 
\begin{equation} 
	\bar{\sigma}^{ab}{}_{|b}=
	\beta\,b^{ab}{}_{|b}+\gamma\,c^{ab}{}_{|b}+\alpha_{,b}\,g^{ab}+\beta_{,b}\,b^{ab}+\gamma_{,b}\,c^{ab}
	=
	\beta\,b^{ab}{}_{|b}+\gamma\,c^{ab}{}_{|b} \,. 
\end{equation}
Therefore, the only universality constraints are \eqref{Universality-Constraints-Cauchy4-1} and \eqref{Universality-Constraints-Cauchy4-2}.

The determination of universal deformations reduces to finding the five unknown components of the right Cauchy-Green tensor $\mathbf{C}^\flat$, subject to the requirement that the corresponding spatial strains $\mathbf{b}^\sharp$ and $\mathbf{c}^\sharp$ satisfy the PDEs \eqref{Universality-Constraints-Cauchy-1} and \eqref{Universality-Constraints-Cauchy-2} and compatibility equations, along with the constraint that all three principal invariants remain constant.
This is analogous to Ericksen's open problem: determine all isochoric deformations (i.e., $I_3 = 1$) with constant $I_1$ and $I_2$  that satisfy the partial differential equations $b_a^{n}{}_{|bn} = b_b^{n}{}_{|an}$ and $c_a^{n}{}_{|bn} = c_b^{n}{}_{|an}$. The only known inhomogeneous universal deformations are Family $5$ deformations (inflation, bending, extension, and azimuthal shearing of an annular wedge) \citep{SinghPipkin1965,KlingbeilShield1966}.
In cylindrical coordinates $(R, \Theta, Z)$ and $(r, \theta, z)$ for the reference and current configurations, respectively, this family of deformations is given by\footnote{In a recent study \citet{Motaghian2025} examined Family~5 universal deformations in incompressible isotropic solids, extending the known examples beyond the classical geometry of an annular wedge. It was demonstrated that bending, inflation, azimuthal shearing, and changes in the major radius of a toroidal sector also belong to the Family~5 class of universal deformations. This result demonstrates that the universality of a deformation family is not restricted to a particular body geometry (in the case of Family~5 deformations, an annular wedge), but rather depends on the intrinsic structure of the deformation field itself.}
\begin{equation}  \label{Family5-Deformations}
	r(R,\Theta,Z) = C_1 R,\qquad 
	\theta(R,\Theta,Z) = C_2 \log R + C_3 \Theta + C_4,\qquad 
	z(R,\Theta,Z) = \frac{1}{C_1^2 C_3} Z + C_5\,,
\end{equation}
where the $Z$-axis corresponds to the central axis of the annular wedge. 
It was shown in Proposition~\ref{Prop:Straight} that when fibers in the deformed configuration are straight lines and the three principal invariants are constant, the only universal deformations possible are homogeneous.
Therefore, for universal deformations with constant principal invariants---if any such universal deformations exist---the fibers in the deformed configuration must have non-vanishing curvature. For the deformation \eqref{Family5-Deformations}, the right Cauchy-Green strain tensor reads
\begin{equation} 
	[C_{AB}] = 
	\begin{bmatrix}
	C_1^2 (C_2^2 + 1) & C_1^2 C_2 C_3 R & 0 \\
	C_1^2 C_2 C_3 R & C_1^2 C_3^2 R^2 & 0 \\
	0 & 0 & \dfrac{1}{C_1^4 C_3^2}
	\end{bmatrix}\,.
\end{equation}
First, note that only $C_{RR}$ or $C_{ZZ}$ can be equal to $1$. 
If we have parallel straight fibers in the undeformed configuration, they must be parallel to the $Z$-axis to be compatible with Family $5$ deformations and inextensibility dictates that $C_1^4 C_3^2=1$. However, the deformed fibers are straights lines parallel to the $z$-axis. Hence, Family $5$ deformations cannot be universal for compressible isotropic solids reinforced by a single family of straight parallel fibers.

As of now, we are unable to solve the problem of whether constant principal invariant universal deformations exist for compressible isotropic solids reinforced by a single family of parallel straight fibers that have non-vanishing curvature in the deformed configuration.

\subsection{Beatty's problem}

\citet{Beatty1978} studied the problem of determining those fiber distributions for which homogeneous deformations are universal. Here, we revisit this problem and derive and discuss his integrability conditions in a geometric setting.

The equilibrium equations in terms of the Cauchy stress read
\begin{equation}
	\Big[\langle dT,\mathbf{n} \rangle\,\mathbf{n}+T\,(\operatorname{div}\mathbf{n})\Big] \mathbf{n}
	+T\, \nabla^{\mathbf{g}}_{\mathbf{n}} \mathbf{n} 
	+ \operatorname{div}\bar{\boldsymbol{\sigma}}
	=\mathbf{0}\,.
\end{equation}  
For homogeneous deformations $\operatorname{div}\bar{\boldsymbol{\sigma}}=\mathbf{0}$, and hence
\begin{equation}
	\Big[\langle dT,\mathbf{n} \rangle\,\mathbf{n}+T\,(\operatorname{div}\mathbf{n})\Big] \mathbf{n}
	+T\, \nabla^{\mathbf{g}}_{\mathbf{n}} \mathbf{n} 
	=\mathbf{0}\,.
\end{equation}  
Knowing that $\nabla^{\mathbf{g}}_{\mathbf{n}} \mathbf{n} \perp \mathbf{n}$, one concludes that
\begin{equation} \label{Equilibrium-Hom-Def}
	\langle dT,\mathbf{n} \rangle +T\,(\operatorname{div}\mathbf{n})
	= \operatorname{div} (T\,\mathbf{n})	=0\,,\qquad
	T\, \nabla^{\mathbf{g}}_{\mathbf{n}} \mathbf{n}  =\mathbf{0} \,.
\end{equation}  
In particular, 
\begin{equation} \label{n-geodesic}
	\nabla^{\mathbf{g}}_{\mathbf{n}} \mathbf{n}  =\mathbf{0} \,,
\end{equation}  
and hence, the deformed fibers must be straight lines. This is identical to \citet{Beatty1978}'s Eq.~$(1.6)_2$. 

From $\llangle \mathbf{n},\mathbf{n}\rrangle_{\mathbf{g}} = 1$ it follows that $\nabla^{\mathbf{g}}_{\mathbf{u}}\, \llangle \mathbf{n},\mathbf{n}\rrangle_{\mathbf{g}}= 2\,\llangle \nabla^{\mathbf{g}}_{\mathbf{u}}\mathbf{n},\mathbf{n}\rrangle_{\mathbf{g}} = 0, ~\forall\, \mathbf{u}$, and hence
\begin{equation} \label{n-perp}
	\llangle \nabla^{\mathbf{g}}_{\mathbf{u}}\mathbf{n},\mathbf{n}\rrangle_{\mathbf{g}} = 0\,, 
	\quad \forall\, \mathbf{u}\,.
\end{equation}
Let us choose an oriented orthonormal frame  $\{\mathbf{n},\mathbf{e}_1,\mathbf{e}_2\}$  with dual coframe  $\{\boldsymbol{n}^\flat,\vartheta^1,\vartheta^2\}$,  so that $\boldsymbol{\mu} = \boldsymbol{n}^\flat \wedge \vartheta^1 \wedge \vartheta^2$. In this coframe, we can write $d\boldsymbol{n}^\flat 
= A\, \vartheta^1 \wedge \vartheta^2+ B\, \boldsymbol{n}^\flat \wedge \vartheta^1+ C\, \boldsymbol{n}^\flat \wedge \vartheta^2$.
From \eqref{n-perp} and \eqref{n-geodesic}, we have 
$d\boldsymbol{n}^\flat(\mathbf{n},\cdot)=0$,\footnote{From the definition of the exterior derivative, 
$d\boldsymbol{n}^\flat(\mathbf{u},\mathbf{v})
= \mathbf{u} (\llangle \mathbf{n},\mathbf{v}\rrangle_{\mathbf{g}} )
- \mathbf{v} (\llangle \mathbf{n},\mathbf{u}\rrangle_{\mathbf{g}} )
- \llangle \mathbf{n},[\mathbf{u},\mathbf{v}]\rrangle_{\mathbf{g}}$. 
Metric compatibility gives us
$\mathbf{u} (\llangle \mathbf{n},\mathbf{v}\rrangle_{\mathbf{g}} )
= \llangle \nabla^{\mathbf{g}}_{\mathbf{u}}\mathbf{n},\mathbf{v}\rrangle_{\mathbf{g}}
+ \llangle \mathbf{n},\nabla^{\mathbf{g}}_{\mathbf{u}}\mathbf{v}\rrangle_{\mathbf{g}}$, 
and torsion-freeness implies that
$\nabla^{\mathbf{g}}_{\mathbf{u}}\mathbf{v}-\nabla^{\mathbf{g}}_{\mathbf{v}}\mathbf{u}=[\mathbf{u},\mathbf{v}]$. 
These cancel the terms involving $\mathbf{n}$, and hence 
$d\boldsymbol{n}^\flat(\mathbf{u},\mathbf{v})
= \llangle \nabla^{\mathbf{g}}_{\mathbf{u}}\mathbf{n},\mathbf{v}\rrangle_{\mathbf{g}}
- \llangle \nabla^{\mathbf{g}}_{\mathbf{v}}\mathbf{n},\mathbf{u}\rrangle_{\mathbf{g}}$. 
Plugging $\mathbf{u}=\mathbf{n}$, and using \eqref{n-perp} and \eqref{n-geodesic} we obtain $d\boldsymbol{n}^\flat(\mathbf{n},\mathbf{v})=0$.} 
which eliminates the $B$ and $C$ terms. Thus, $d\boldsymbol{n}^\flat = A\, \vartheta^1 \wedge \vartheta^2$. It follows that $*\, d\boldsymbol{n}^\flat = A\, *(\vartheta^1 \wedge \vartheta^2) = A\, \boldsymbol{n}^\flat$, and hence
\begin{equation}
	\boldsymbol{n}^\flat \wedge *\, d\boldsymbol{n}^\flat
	= A\, \boldsymbol{n}^\flat \wedge \boldsymbol{n}^\flat
	= 0\,.
\end{equation}
Using the identity $(\operatorname{curl}\mathbf{n})^\flat = *\, d\boldsymbol{n}^\flat$ \citep{Abraham2012}, we find that $\left((\operatorname{curl}\mathbf{n})\times \mathbf{n}\right)^\flat= *\left( \boldsymbol{n}^\flat \wedge *\, d\boldsymbol{n}^\flat \right)$. Therefore,
\begin{equation} \label{curl-cross}
	\mathbf{n} \times  \operatorname{curl}\mathbf{n}  = \mathbf{0}\,,
\end{equation}
which is identical to \citet{Beatty1978}'s Eq.~$(1.9)$. 

Now starting from the equilibrium equation \eqref{Equilibrium-Hom-Def}$_1$, we use the Hodge star to write \citep{Abraham2012}
\begin{equation}
	\operatorname{div}(T\,\mathbf{n})
	= *\, d\left( * (T\,\boldsymbol{n}^\flat) \right)\,.
\end{equation}
Recalling that the Hodge star $*$ is an isomorphism, and  the identity $d(*\boldsymbol{n}^\flat) = (\operatorname{div}\mathbf{n})\,\boldsymbol{\mu}$, this is equivalent to
\begin{equation}
	d\left( * (T\,\boldsymbol{n}^\flat) \right) 
	= d\left( T *\boldsymbol{n}^\flat \right) 
	= dT \wedge *\boldsymbol{n}^\flat + T\, d(*\boldsymbol{n}^\flat) 
	= dT \wedge *\boldsymbol{n}^\flat + T\, (\operatorname{div}\mathbf{n})\,\boldsymbol{\mu} 
	= 0\,.
\end{equation}
If $T \neq 0$, set $u = \log|T|$, so that $dT = T\,du$. Thus
\begin{equation} \label{PDE-star}
	du \wedge *\boldsymbol{n}^\flat + (\operatorname{div}\mathbf{n})\,\boldsymbol{\mu} = 0\,.
\end{equation}
Note that $du \wedge *\boldsymbol{n}^\flat$ depends only on the component of $du$ in the direction of $\mathbf{n}$, as seen from the identity
\begin{equation}
	\boldsymbol{\alpha} \wedge *\boldsymbol{n}^\flat 
	= \llangle \boldsymbol{\alpha}, \boldsymbol{n}^\flat \rrangle_{\mathbf{g}}\, \boldsymbol{\mu}\,,
	\qquad \text{for any $1$-form $\boldsymbol{\alpha}$}\,.
\end{equation}
Applying this to $\boldsymbol{\alpha}=du$, we obtain $( \llangle du,\boldsymbol{n}^\flat \rrangle_{\mathbf{g}} + \operatorname{div}\mathbf{n} )\,\boldsymbol{\mu} = 0$, or equivalently,
\begin{equation}
	\llangle du,\boldsymbol{n}^\flat \rrangle_{\mathbf{g}} + \operatorname{div}\mathbf{n} = 0\,.
\end{equation}
For a solution $u$ to exist, the $2$-planes $\ker(\boldsymbol{n}^\flat)$ orthogonal to $\mathbf{n}$ must fit together to form smooth surfaces. By the Frobenius theorem, this integrability holds if and only if \citep{Lee2013}
\begin{equation} \label{Integrability-n}
	\boldsymbol{n}^\flat \wedge d\boldsymbol{n}^\flat = 0\,.
\end{equation}
Condition \eqref{Integrability-n} is the integrability condition for $u$ and consequently for $T$.
From the decomposition of $d\boldsymbol{n}^\flat$ established earlier, we know that  $d\boldsymbol{n}^\flat = A\,\vartheta^1 \wedge \vartheta^2$ after using \eqref{n-perp} and \eqref{n-geodesic}.  Thus, $d\, \boldsymbol{n}^\flat \wedge \boldsymbol{n}^\flat = A\, \vartheta^1 \wedge \vartheta^2 \wedge \boldsymbol{n}^\flat= A\, \boldsymbol{\mu}_{\mathbf{g}}$. Using $(\operatorname{curl}\mathbf{n})^\flat = *\, d\boldsymbol{n}^\flat$ and taking the inner product with $\mathbf{n}$, one finds  $A = \mathbf{n}\cdot \operatorname{curl}\mathbf{n}$. Thus
\begin{equation}
	d\, \boldsymbol{n}^\flat \wedge \boldsymbol{n}^\flat 
	= (\mathbf{n} \cdot \operatorname{curl}\mathbf{n})\, \boldsymbol{\mu}_{\mathbf{g}}\,.
\end{equation}
Therefore, Frobenius compatibility is written as (vanishing helicity or chirality)
\begin{equation} \label{Frobenius}
	\mathbf{n} \cdot \operatorname{curl}\mathbf{n}=0	\,.
\end{equation}

\begin{remark}
By the Frobenius theorem, the $2$-planes $\ker(\boldsymbol{n}^\flat)$ form smooth surfaces if and only if
\begin{equation} 
	\boldsymbol{n}^\flat \wedge d\boldsymbol{n}^\flat = 0\,.
\end{equation}
A sufficient condition for the integrability of the equilibrium equation \eqref{Equilibrium-Hom-Def}$_1$ is the Frobenius integrability of $\boldsymbol{n}^\flat$, though this is not necessary. From \eqref{PDE-star} we know that only the $\boldsymbol{n}^\flat$-component of $du$ is constrained, while its tangential components remain undetermined. For $u$ to exist we must require $d(du)=0$. If we restrict to the case $du = -(\operatorname{div}\mathbf{n})\,\boldsymbol{n}^\flat$, this condition reduces to $d\left((\operatorname{div}\mathbf{n})\,\boldsymbol{n}^\flat\right) = 0$, or equivalently
\begin{equation} \label{Compatibility-Beatty}
	\operatorname{curl}\left((\operatorname{div}\mathbf{n})\,\mathbf{n}\right) = \mathbf{0}\,,
\end{equation}
which is Beatty’s Eq.~(1.12). If $\operatorname{div}\mathbf{n}=0$, \eqref{Compatibility-Beatty} is trivially satisfied, as Beatty noted. He identified a class of fiber distributions in which the fibers lie in a family of parallel planes. Within each plane the fibers are straight lines, but as one moves in the direction normal to the planes, they rotate. These distributions are not Frobenius integrable, since the planes orthogonal to $\mathbf{n}$ do not fit together to form smooth surfaces, yet they still correspond to universal homogeneous deformations. One may select any plane not parallel to the fibers; each fiber intersects this plane at exactly one point. Specifying the tension field on this plane uniquely determines it everywhere, since along each fiber the tension is constant.
\end{remark}

\subsection{Similarities and differences between the universality constraints of incompressible isotropic elasticity and those of compressible isotropic elasticity reinforced by a single family of straight inextensible fibers}

First, it should be noted that incompressibility is an isotropic internal constraint, while inextensibility explicitly depends on the direction of deformed fibers and is therefore an anisotropic internal constraint. In the case of fiber-reinforced compressible elastic solids, the initial distribution of inextensible fibers is part of the data of the problem. In this paper, we considered the simplest case—namely, when fibers in the reference configuration are all parallel straight lines. We showed that even in this simplest case, the problem of determining universal deformations is quite difficult.
We considered the following two cases separately.

\begin{itemize}[topsep=0pt,noitemsep, leftmargin=10pt]
\item[\textbf{i)}] At least one principal invariant is not constant.

\begin{itemize}[topsep=0pt,noitemsep, leftmargin=10pt]
\item In both incompressible elasticity and fiber-reinforced compressible elasticity, the stress tensor has reactive and constitutive parts. Determining all universal deformations reduces to finding conditions that ensure the existence of the Lagrange multiplier field associated with the internal constraint. In incompressible elasticity, the integrability conditions are related to the exactness of the $1$-form $dp$, where $p$ is the Lagrange multiplier corresponding to the constraint $I_3 = 1$.
In fiber-reinforced compressible elasticity, the determination of universal deformations is related to the existence of a tension field $T$, which plays the role of the Lagrange multiplier associated with fiber inextensibility. The integrability conditions of this tension field are explicitly related to either the unit tangent vector to the fibers in the deformed configuration (in the case of straight deformed fibers) or the binormal vector of the deformed fibers (when deformed fibers have non-vanishing curvature).

\item \citet{Ericksen1954} showed that $\nabla I_1$ and/or $\nabla I_2$ are eigenvectors of $\mathbf{b}$ (when $\nabla I_1 \neq \mathbf{0}$ and/or $\nabla I_2 \neq \mathbf{0}$). This is not necessarily the case for fiber-reinforced compressible solids. Only when the deformed fibers are straight lines can one conclude that if $\nabla I_i \neq \mathbf{0}$ ($i = 1, 2, 3$), then $\nabla I_i$ is an eigenvector of $\mathbf{b}$. When the deformed fibers are curved and $\nabla I_1$ and $\nabla I_2$ are linearly independent, we proved that the bivector associated with the fiber direction is an eigenvector of $\mathbf{b}$.

\item \citet{Ericksen1954} proved that $I_1$ and $I_2$ are functionally dependent; that is, there exists a single variable $\zeta$ such that $I_1 = I_1(\zeta)$ and $I_2 = I_2(\zeta)$ (recall that in incompressible elasticity, $I_3 = 1$). We showed that in fiber-reinforced compressible elasticity, the invariants $I_1$, $I_2$, and $I_3$ are functionally dependent; that is, there exist independent variables $(\zeta, \chi)$ such that $I_i = I_i(\zeta, \chi)$ for $i = 1, 2, 3$.

\item \citet{Ericksen1954} demonstrated that the surfaces $\zeta = \mathsf{const.}$ are either planes, cylinders, or spheres. For fiber-reinforced compressible solids, we showed that when fibers in the deformed configuration are straight lines, the surfaces with normal $\mathbf{n}$ (the unit tangent vector to the deformed fibers) are also either planes, cylinders, or spheres.

\item We showed that homogeneous deformations compatible with the inextensibility constraint---\textit{$Z$-isometric homogeneous deformations}---are universal. We refer to this class as \textit{Family~$0Z$ universal deformations}.

\item When the deformed fibers are straight lines (i.e., have vanishing curvature), we showed that there is only one family of universal deformations, referred to as \textit{Family~$Z_1$ universal deformations} (Proposition~\ref{Prop:FamilyZ1}). 

\item Determining whether universal deformations that map straight fibers to curves with non-vanishing curvature exist remains an open problem.

\end{itemize}

\item[\textbf{ii)}] All three principal invariants are constant.

\begin{itemize}[topsep=0pt,noitemsep, leftmargin=10pt]
\item In incompressible elasticity, the only known family of inhomogeneous universal deformations with constant principal invariants is the Family 5 universal deformations. Whether other such universal deformations exist remains an open problem.

\item For compressible isotropic solids reinforced by a single family of parallel straight inextensible fibers, we proved that no inhomogeneous universal deformations with constant principal invariants exist when the deformed fibers remain straight lines (Proposition~\ref{Prop:Straight}).

\item Determining whether universal deformations with constant principal invariants exist when the deformed fibers have non-vanishing curvature remains an open problem.

\end{itemize}
\end{itemize}

\section{Universal Deformations of Compressible Isotropic Hyperelastic Bodies Reinforced by a Single Family of Inextensible Fibers} \label{Sec:Hyperelastic}

For a compressible isotropic hyper-elastic solid one has the following representation for the Cauchy stress \citep{DoyleEricksen1956}
\begin{equation}
	\bar{\boldsymbol{\sigma}}
	=\frac{2}{\sqrt{I_3}}\left[ (I_2W_2+I_3W_3) \,\mathbf{g}^{\sharp}
	+W_1 \mathbf{b}^{\sharp} -I_3 W_2\,\mathbf{c}^{\sharp} \right]
	\,.
\end{equation}  

In determining universal deformations, the only thing that changes in the presence of an energy function is the form of the vector $\mathbf{f}$ defined in \eqref{Tension-Field-Equlibrium}$_2$.
It is straightforward to show that
\begin{equation} \label{f-a}
\begin{aligned}
	-f^a= & \left(-\frac{I_{3,b}}{2I_3}\,b^{ab}+b^{ab}{}_{|b}\right) W_1
	+\left[-\frac{I_{3,b}}{2I_3}\left(I_2\,g^{ab}-I_3\,c^{ab}\right)
	+I_{2,b}\,g^{ab}-I_{3,b}\,c^{ab}-I_3\,c^{ab}{}_{|b}\right] W_2
	+\frac{1}{2}I_{3,b}\,g^{ab}\, W_3 \\
	&+b^{ab}\,I_{1,b} \,W_{11}
	+ I_{2,b}\left(I_2\,g^{ab}-I_3\,c^{ab}\right) \,W_{22}+I_3\,I_{3,b}\,g^{ab} \,W_{33} \\
	&+\left[ I_{1,b}\left(I_2\,g^{ab}-I_3\,c^{ab}\right)+I_{2,b}\,b^{ab} \right] \,W_{12}
	+\left(b^{ab}\,I_{3,b}+g^{ab}\, I_{1,b}\,I_3 \right) \,W_{13} \\	
	&+\left[ I_{3,b}\left(I_2\,g^{ab}-I_3\,c^{ab}\right)+I_3\,I_{2,b}\,g^{ab} \right]\,W_{23}\,,
\end{aligned}
\end{equation}
where
\begin{equation}
	W_A=\frac{\partial W}{\partial I_A}\,,\quad A=1,2,3,\quad\quad  W_{AB}=\frac{\partial^2 W}{\partial I_A\partial I_B}\,,
	\,\quad 1\leq A\leq 3\,.
\end{equation}  
Substituting \eqref{f-a} into \eqref{Integrability-Z1}$_3$ one obtains
\begin{equation} 
\begin{aligned}
	& \xi_a\left(-\frac{I_{3,b}}{2I_3}\,b^{ab}+b^{ab}{}_{|b}\right) W_1
	+\xi_a\left[-\frac{I_{3,b}}{2I_3}\left(I_2\,g^{ab}-I_3\,c^{ab}\right)
	+I_{2,b}\,g^{ab}-I_{3,b}\,c^{ab}-I_3\,c^{ab}{}_{|b}\right] W_2
	+\frac{1}{2}\xi_a\,I_{3,b}\,g^{ab}\, W_3 \\
	&+\xi_a\,b^{ab}\,I_{1,b} \,W_{11}
	+ \xi_a\,I_{2,b}\left(I_2\,g^{ab}-I_3\,c^{ab}\right) \,W_{22}+\xi_a\,I_3\,I_{3,b}\,g^{ab} \,W_{33} \\
	&+\xi_a\left[ I_{1,b}\left(I_2\,g^{ab}-I_3\,c^{ab}\right)+I_{2,b}\,b^{ab} \right] \,W_{12}
	+\xi_a\left(b^{ab}\,I_{3,b}+g^{ab}\, I_{1,b}\,I_3 \right) \,W_{13} \\	
	&+\xi_a\left[ I_{3,b}\left(I_2\,g^{ab}-I_3\,c^{ab}\right)+I_3\,I_{2,b}\,g^{ab} \right]\,W_{23}=0\,.
\end{aligned}
\end{equation}
As the derivatives of the energy function are independent, the coefficient of each of the nine derivatives must vanish independently. Therefore, we have the following universality constrains
\begin{equation} \label{Universality-Constraints1}
\begin{aligned}
	& \xi_a\left(-\frac{I_{3,b}}{2I_3}\,b^{ab}+b^{ab}{}_{|b}\right) =0\,, \\
	& \xi_a\left[-\frac{I_{3,b}}{2I_3}\left(I_2\,g^{ab}-I_3\,c^{ab}\right)
	+I_{2,b}\,g^{ab}-I_{3,b}\,c^{ab}-I_3\,c^{ab}{}_{|b}\right] =0\,, \\
	& \xi_a\,I_{3,b}\,g^{ab}=0\,, \\
	&\xi_a\,b^{ab}\,I_{1,b} =0\,, \\
	& \xi_a\,I_{2,b}\left(I_2\,g^{ab}-I_3\,c^{ab}\right) =0\,, \\
	&\xi_a\,I_3\,I_{3,b}\,g^{ab} =0\,, \\
	& \xi_a\left[ I_{1,b}\left(I_2\,g^{ab}-I_3\,c^{ab}\right)+I_{2,b}\,b^{ab} \right] =0\,, \\
	& \xi_a\left(b^{ab}\,I_{3,b}+g^{ab}\, I_{1,b}\,I_3 \right) =0\,, \\	
	& \xi_a\left[ I_{3,b}\left(I_2\,g^{ab}-I_3\,c^{ab}\right)+I_3\,I_{2,b}\,g^{ab} \right]=0\,.
\end{aligned}
\end{equation}
Note that \eqref{Universality-Constraints1}$_3$ and \eqref{Universality-Constraints1}$_6$ are equivalent, and this simplifies \eqref{Universality-Constraints1} to read
\begin{equation} \label{Universality-Constraints2}
\begin{dcases}
	\xi_a\left(-\frac{I_{3,b}}{2I_3}\,b^{ab}+b^{ab}{}_{|b}\right) =0\,, \\
	\xi_a\left[I_{2,b}\,g^{ab}-\frac{1}{2}I_{3,b}\,c^{ab}-I_3\,c^{ab}{}_{|b}\right] =0\,, \\
	\xi_a\,g^{ab}\,I_{3,b}=0\,, \\
	\xi_a\,b^{ab}\,I_{1,b} =0\,, \\
	\xi_a\,I_{2,b}\left(I_2\,g^{ab}-I_3\,c^{ab}\right) =0\,, \\
	\xi_a\left[ I_{1,b}\left(I_2\,g^{ab}-I_3\,c^{ab}\right)+I_{2,b}\,b^{ab} \right] =0\,, \\
	\xi_a\left(b^{ab}\,I_{3,b}+g^{ab}\, I_{1,b}\,I_3 \right) =0\,, \\	
	\xi_a\left[ I_{2,b}\,g^{ab}-I_{3,b}\,c^{ab} \right]=0\,.
\end{dcases}
\end{equation}
In coordinate-free form these universality constraints read
\begin{empheq}[left={\empheqlbrace }]{align} 
	\label{Univ-1}
	& \boldsymbol{\xi}\cdot \left(-\frac{1}{2I_3}\,\mathbf{b}\cdot\nabla I_3
	+\operatorname{div}\mathbf{b}^\sharp\right) =0\,, \\
	\label{Univ-2}
	& \boldsymbol{\xi}\cdot \left( \nabla I_2 -\frac{1}{2} \mathbf{c}\cdot\nabla I_3
	-I_3\,\operatorname{div}\mathbf{c}^\sharp\right) =0\,, \\
	\label{Univ-3}
	& \boldsymbol{\xi}\cdot \nabla I_3=0\,, \\
	\label{Univ-4}
	& \boldsymbol{\xi}\cdot \nabla I_1 =0\,, \\
	\label{Univ-5}
	& \boldsymbol{\xi}\cdot \left( I_2 \nabla I_2 -I_3 \,\mathbf{c}\cdot\nabla I_2\right) =0\,, \\
	\label{Univ-6}
	& \boldsymbol{\xi}\cdot \left( I_2 \nabla I_1 -I_3\, \mathbf{c}\cdot\nabla I_1
	+\mathbf{b}\cdot\nabla I_2 \right)=0\,, \\
	\label{Univ-7}
	& \boldsymbol{\xi}\cdot \left( I_3 \nabla I_1 + \mathbf{b}\cdot\nabla I_3 \right) =0\,, \\	
	\label{Univ-8}
	& \boldsymbol{\xi}\cdot \left( \nabla I_2 - \mathbf{c}\cdot\nabla I_3 \right)=0\,.
\end{empheq}

First, let us consider the case when $\mathbf{n}_{,Z}=\mathbf{0}$, and at least one principal invariant is not constant.
We showed that in this case $\lambda_3 = 1$. Therefore, the principal invariants of the right Cauchy-Green tensor are given by:
\begin{equation}
	I_1 = \lambda_1^2 + \lambda_2^2 + 1\,, \qquad
	I_2 = \lambda_1^2\lambda_2^2 + \lambda_1^2 + \lambda_2^2\,, \qquad
	I_3 = \lambda_1^2\lambda_2^2\,.
\end{equation}
It follows that, $I_2 = I_1 + I_3 - 1$, and hence 
\begin{equation} \label{Nabla-Constraints}
	\nabla I_2 = \nabla I_1 + \nabla I_3\,.
\end{equation}
Therefore, \eqref{Univ-3} and \eqref{Univ-4} are equivalent to \eqref{Universality-Constraints-Cauchy-3}.
Now the above universality constraints are simplified to read
\begin{empheq}[left={\empheqlbrace }]{align} 
	\label{Univ2-1}
	& \boldsymbol{\xi}\cdot \left(-\frac{1}{2I_3}\,\mathbf{b}\cdot\nabla I_3
	+\operatorname{div}\mathbf{b}^\sharp\right) =0\,, \\
	\label{Univ2-2}
	& \boldsymbol{\xi}\cdot \left(\frac{1}{2} \mathbf{c}\cdot\nabla I_3+I_3\,\operatorname{div}\mathbf{c}^\sharp\right) =0\,, \\
	\label{Univ2-5}
	& \boldsymbol{\xi}\cdot \left(\mathbf{c}\cdot\nabla I_2\right) =0\,, \\
	\label{Univ2-6}
	& \boldsymbol{\xi}\cdot \left(-I_3\, \mathbf{c}\cdot\nabla I_1
	+\mathbf{b}\cdot\nabla I_2 \right)=0\,, \\
	\label{Univ2-7}
	& \boldsymbol{\xi}\cdot \left(\mathbf{b}\cdot\nabla I_3 \right) =0\,, \\	
	\label{Univ2-8}
	& \boldsymbol{\xi}\cdot \left(\mathbf{c}\cdot\nabla I_3 \right)=0\,.
\end{empheq}
The constraints \eqref{Univ2-5} and \eqref{Univ2-8} together with \eqref{Nabla-Constraints} are equivalent to \eqref{Universality-Constraints-Cauchy-5}.
Thus, the remaining universality constraints are
\begin{empheq}[left={\empheqlbrace }]{align} 
	\label{Univ3-1}
	& \boldsymbol{\xi}\cdot \left(-\frac{1}{2I_3}\,\mathbf{b}\cdot\nabla I_3
	+\operatorname{div}\mathbf{b}^\sharp\right) =0\,, \\
	\label{Univ3-2}
	& \boldsymbol{\xi}\cdot \left(\operatorname{div}\mathbf{c}^\sharp\right) =0\,, \\
	\label{Univ3-6}
	& \boldsymbol{\xi}\cdot \left(\mathbf{b}\cdot\nabla I_2 \right)=0\,, \\
	\label{Univ3-7}
	& \boldsymbol{\xi}\cdot \left(\mathbf{b}\cdot\nabla I_3 \right) =0\,.
\end{empheq}
The constraints \eqref{Univ3-6} and \eqref{Univ3-7} together with \eqref{Nabla-Constraints} are equivalent to \eqref{Universality-Constraints-Cauchy-4}.
Using these in \eqref{Univ3-1}, we recover the universality constraint \eqref{Universality-Constraints-Cauchy-1}, and \eqref{Univ3-2} is identical to \eqref{Universality-Constraints-Cauchy-2}.

Next, let us consider the case when $\mathbf{n}_{,Z} \neq \mathbf{0}$, and at least one principal invariant is not constant.
We showed in \S\ref{Sec:Curved-Fibers} that the principal invariants are functionally dependent, i.e., the set $\{\nabla I_1 , \nabla I_2 , \nabla I_3 \}$ is linearly dependent. Without loss of generality, let us assume that $\nabla I_3=a \nabla I_1+b \nabla I_2$.
If $a=0$, then the universality constraints involving $\nabla I_3$ and $\nabla I_2$ are equivalent. Similarly, if $b=0$, the universality constraints involving $\nabla I_3$ and $\nabla I_1$ are equivalent. 
Without loss of generality, let us assume that $a=0$. Then, the universality constraints are reduced to $ \boldsymbol{\xi}\cdot \nabla I_1 = \boldsymbol{\xi}\cdot \nabla I_3 =0$ (identical to \eqref{Universality-Constraints-Cauchy-3}), and 
\begin{empheq}[left={\empheqlbrace }]{align} 
	\label{Const-Univ-1}
	& \boldsymbol{\xi}\cdot \left(-\frac{1}{2I_3}\,\mathbf{b}\cdot\nabla I_3
	+\operatorname{div}\mathbf{b}^\sharp\right) =0\,, \\
	\label{Const-Univ-2}
	& \boldsymbol{\xi}\cdot \left(-\frac{1}{2} \mathbf{c}\cdot\nabla I_3
	-I_3\,\operatorname{div}\mathbf{c}^\sharp\right) =0\,, \\
	\label{Const-Univ-3}
	& \boldsymbol{\xi}\cdot \mathbf{c}\cdot\nabla I_2 =0\,, \\
	\label{Const-Univ-4}
	& \boldsymbol{\xi}\cdot \left( -I_3\, \mathbf{c}\cdot\nabla I_1
	+\mathbf{b}\cdot\nabla I_2 \right)=0\,, \\
	\label{Const-Univ-5}
	& \boldsymbol{\xi}\cdot \mathbf{b}\cdot\nabla I_3  =0\,, \\	
	\label{Const-Univ-6}
	& \boldsymbol{\xi}\cdot  \mathbf{c}\cdot\nabla I_3 =0\,.
\end{empheq}
Eqs.~\eqref{Const-Univ-3} and \eqref{Const-Univ-6} are equivalent to \eqref{Universality-Constraints-Cauchy-5}. Now the universality constraints are reduced to
\begin{empheq}[left={\empheqlbrace }]{align} 
	\label{Const-Univ1-1}
	& \boldsymbol{\xi}\cdot \left(-\frac{1}{2I_3}\,\mathbf{b}\cdot\nabla I_3
	+\operatorname{div}\mathbf{b}^\sharp\right) =0\,, \\
	\label{Const-Univ1-2}
	& \boldsymbol{\xi}\cdot \operatorname{div}\mathbf{c}^\sharp =0\,, \\
	\label{Const-Univ1-3}
	& \boldsymbol{\xi}\cdot \mathbf{b}\cdot\nabla I_2 =0\,, \\
	\label{Const-Univ1-4}
	& \boldsymbol{\xi}\cdot \mathbf{b}\cdot\nabla I_3  =0\,.
\end{empheq}
Eqs.~\eqref{Const-Univ1-3} and \eqref{Const-Univ1-4} are equivalent to \eqref{Universality-Constraints-Cauchy-4},  and \eqref{Const-Univ1-1} and \eqref{Const-Univ1-2} are reduced to \eqref{Universality-Constraints-Cauchy-1} and \eqref{Universality-Constraints-Cauchy-2}, respectively. 
Therefore, the two sets of universality constraints are equivalent.

Now suppose $a\neq 0$ and $b\neq 0$. In this case, \eqref{Univ-3} is written as $\boldsymbol{\xi}\cdot \nabla I_3= a\,\boldsymbol{\xi}\cdot \nabla I_1+b\,\boldsymbol{\xi}\cdot \nabla I_2=0$, and using \eqref{Univ-4}, one obtains $\boldsymbol{\xi}\cdot \nabla I_2=0$. Thus, \eqref{Univ-3} and \eqref{Univ-4} are equivalent to \eqref{Universality-Constraints-Cauchy-3}. The universality constraints are now simplified to read
\begin{empheq}[left={\empheqlbrace }]{align} 
	\label{UnivN-1}
	& \boldsymbol{\xi}\cdot \left(-\frac{1}{2I_3}\,\mathbf{b}\cdot\nabla I_3
	+\operatorname{div}\mathbf{b}^\sharp\right) =0\,, \\
	\label{UnivN-2}
	& \boldsymbol{\xi}\cdot \left(-\frac{1}{2} \mathbf{c}\cdot\nabla I_3
	-I_3\,\operatorname{div}\mathbf{c}^\sharp\right) =0\,, \\
	\label{UnivN-3}
	& \boldsymbol{\xi}\cdot \mathbf{c}\cdot\nabla I_2 =0\,, \\
	\label{UnivN-4}
	& \boldsymbol{\xi}\cdot \left( -I_3\, \mathbf{c}\cdot\nabla I_1
	+\mathbf{b}\cdot\nabla I_2 \right)=0\,, \\
	\label{UnivN-5}
	& \boldsymbol{\xi}\cdot  \mathbf{b}\cdot\nabla I_3  =0\,, \\	
	\label{UnivN-6}
	& \boldsymbol{\xi}\cdot  \mathbf{c}\cdot\nabla I_3 =0\,.
\end{empheq}
Eqs.~\eqref{UnivN-3} and \eqref{UnivN-6} imply that $\boldsymbol{\xi}\cdot \mathbf{c}\cdot\nabla I_1 =0$, i.e., we recover \eqref{Universality-Constraints-Cauchy-5}. The universality constraints are now reduced to
\begin{empheq}[left={\empheqlbrace }]{align} 
	\label{UnivNN-1}
	& \boldsymbol{\xi}\cdot \left(-\frac{1}{2I_3}\,\mathbf{b}\cdot\nabla I_3
	+\operatorname{div}\mathbf{b}^\sharp\right) =0\,, \\
	\label{UnivNN-2}
	& \boldsymbol{\xi}\cdot \operatorname{div}\mathbf{c}^\sharp =0\,, \\
	\label{UnivNN-3}
	& \boldsymbol{\xi}\cdot \mathbf{b}\cdot\nabla I_2  =0\,, \\
	\label{UnivNN-4}
	& \boldsymbol{\xi}\cdot  \mathbf{b}\cdot\nabla I_3  =0\,.
\end{empheq}
Eqs.~\eqref{UnivNN-3} and \eqref{UnivNN-4} imply that $\boldsymbol{\xi}\cdot \mathbf{b}\cdot\nabla I_1 =0$, i.e., we recover \eqref{Universality-Constraints-Cauchy-4}.
 Finally, the remaining two universality constraints are reduced to $\boldsymbol{\xi}\cdot \operatorname{div}\mathbf{b}^\sharp = \boldsymbol{\xi}\cdot \operatorname{div}\mathbf{c}^\sharp$, i.e, \eqref{Universality-Constraints-Cauchy-1} and \eqref{Universality-Constraints-Cauchy-2}. 
 Therefore, again the two sets of universality constraints are equivalent.

Finally, when principal invariants are constant, the universality constraints \eqref{Univ-1}-\eqref{Univ-8} are reduced to $\boldsymbol{\xi}\cdot \operatorname{div}\mathbf{b}^\sharp =0$ and $\boldsymbol{\xi}\cdot \operatorname{div}\mathbf{c}^\sharp =0$, which are identical to those of Cauchy elasticity.

We observe that the universality constraints for isotropic hyperelastic solids reinforced by a single family of straight inextensible fibers are identical to those for isotropic Cauchy elastic solids with the same reinforcement. This parallels similar observations made for both compressible and incompressible hyperelastic solids and their corresponding Cauchy elastic counterparts \citep{Yavari2024Cauchy}.
Therefore, we arrive at the following result.

\begin{prop}
The universal deformations of compressible isotropic hyperelastic solids reinforced by a single family of inextensible fibers parallel to a fixed direction are identical to those of isotropic Cauchy elastic solids with the same reinforcement.
\end{prop}

Table~\ref{table:UniversalFiberReinforced} summarizes the universal deformations identified in this work for compressible isotropic Cauchy (and also hyperelastic) solids reinforced by a single family of inextensible fibers parallel to the $Z$-axis.

\vskip 0.2in
\begin{table}[hbt!]
\small
\renewcommand{\arraystretch}{1.6}
\begin{center}
\begin{tabular}{|c|c|c|}
\hline 
Family & Universal Deformations & $\mathbf{C}^{\flat}$ \\
\hline 
$0Z$ & $\begin{dcases}
x = a_{11} X + a_{12} Y + a_{13} Z \\
y = a_{21} X + a_{22} Y + a_{23} Z \\
z = a_{31} X + a_{32} Y + a_{33} Z
\end{dcases}$
& \rule{0pt}{4ex}$\begin{bmatrix}
C_{11} & C_{12} & C_{13} \\
C_{12} & C_{22} & C_{23} \\
C_{13} & C_{23} & 1
\end{bmatrix}$\rule[-2ex]{0pt}{0pt} \\  
\hline
$Z_1$ & $\begin{dcases}
	r(X,Y,Z) = Z + Z_0\,, \\
	\theta(X,Y,Z) = \alpha_0 X + \beta_0 Y + \theta_0\,, \\
	z(X,Y,Z) = k_1 Y + z_0\,,
\end{dcases}$
& \rule{0pt}{4ex}$\begin{bmatrix}
(\alpha_0^2 + \beta_0^2)(Z + Z_0)^2 & \alpha_0 \beta_0 (Z + Z_0)^2 & 0 \\
\alpha_0 \beta_0 (Z + Z_0)^2 & \beta_0^2 (Z + Z_0)^2 + k_1^2 & 0 \\
0 & 0 & 1
\end{bmatrix}$\rule[-2ex]{0pt}{0pt} \\
\hline
\end{tabular}
\end{center}
\vspace{-0.1in}
\caption[]{Universal deformations for compressible isotropic Cauchy elastic solids reinforced by a single family of inextensible fibers aligned with the $Z$-axis. For Family $0Z$, we have the constraint $a_{13}^2 + a_{23}^2 + a_{33}^2 = 1$.}
\label{table:UniversalFiberReinforced}
\end{table}

\section{Conclusions} \label{Sec:Conclusions}

In this paper, we studied universal deformations in compressible isotropic Cauchy elastic solids reinforced by a single family of inextensible fibers.
First, we noted that the distribution of fibers in the reference configuration is part of the given data of the problem. To remain concrete and obtain explicit results, we focused on a distribution of straight fibers parallel to the Cartesian $Z$-axis.
The problem of identifying universal deformations reduces to finding the conditions under which a Cauchy elastic solid with reinforcement can sustain a given deformation for arbitrary stored energy functions.
Unlike incompressibility, inextensibility is an anisotropic internal constraint that explicitly depends on the orientation of fibers in the deformed configuration. We showed that the universality constraints explicitly depend on the geometry of the deformed fibers.

We considered two cases: (i) deformed fibers are straight lines, and (ii) deformed fibers have non-vanishing curvature. The classification of universal deformations for case (i) was completely solved.
Assuming at least one principal invariant is not constant, we showed that the deformed fiber tangent vector is an eigenvector of the Finger tensor. Moreover, the principal invariants depend only on the fiber arclength parameter ($Z$). We demonstrated that the universality constraints force the surfaces orthogonal to the deformed fibers to have constant mean and Gaussian curvatures and must therefore be planes, circular cylinders, or spheres.
We showed that planar surfaces correspond to homogeneous universal deformations (compatible with inextensibility)---\textit{Family $0Z$ universal deformations}. The only other non-trivial universal deformation family is the combined uniform extension and bending deformations, which are generally non-isochoric---\textit{Family Z$_1$ universal deformations}. Finally, we proved that when all principal invariants are constant and fibers remain straight lines in the deformed configuration, the only universal deformations are homogeneous deformations.

When deformed fibers have non-vanishing curvature, the universality constraints become significantly more complicated. These constraints can be expressed as the requirement that certain vectors have zero component along the binormal of the deformed fibers. We showed that the three principal invariants are always functionally dependent. If the gradients of two principal invariants are linearly independent, then the binormal vector of the deformed fibers must be an eigenvector of the Finger tensor, with the corresponding eigenvalue remaining constant along the binormal direction.
The question of whether universal deformations exist in the case of curved deformed fibers remains open. The case where all three principal invariants are constant is analogous to Ericksen's open problem. We showed that Family $5$ universal deformations of incompressible elasticity, when made consistent with the inextensibility constraint, are no longer universal in ideal fiber-reinforced elasticity. 
The existence of constant–principal invariant universal deformations in this setting remains an open problem.

Finally, we demonstrated that the universality constraints for compressible isotropic Cauchy elastic solids reinforced by inextensible fibers are identical to those for compressible isotropic hyperelastic solids with the same reinforcement. This result parallels analogous findings in the absence of reinforcement for both compressible and incompressible elasticity.
To our knowledge, this is the first systematic classification of universal deformations for compressible isotropic fiber-reinforced solids. 

Universal deformations are exact solutions of the equilibrium equations and have been used as benchmark problems in computational mechanics \citep{Dragoni1996,Saccomandi2001,Chi2015,Shojaei2018}.  
It is well known that in the numerical solution of continuum problems with internal constraints, one may encounter locking phenomena. In the context of the finite element method, solving problems subject to the incompressibility constraint has proven challenging, and considerable effort has been devoted to addressing locking and numerical instability issues. One common approach is the use of mixed finite element methods. Similarly, in modeling anisotropic solids reinforced with inextensible or nearly inextensible fibers, mixed finite element formulations have been developed to accurately capture large deformations of such solids \citep{Wriggers2016,Bohm2023}.  
The universal solutions presented in this paper can serve as benchmark problems for current and future mixed finite element formulations of composites with inextensible fibers.

In determining the universal deformations of fiber-reinforced solids, the distribution of inextensible fibers is part of the problem data. In this paper, we considered the simplest case, namely, a distribution of straight parallel fibers. For any other fiber distribution, the determination of universal deformations follows the same general approach as presented here. The only modification lies in the form of the integrability condition for the tension field. Studying other fiber distributions---such as circular, circumferential, or radial patterns---may be an interesting problem for future work. The study of universal deformations in solids reinforced by multiple families of inextensible fibers will be the subject of a future communication.
Determining the universal deformations of compressible Cauchy elastic solids reinforced by a family of incompressible elastic surfaces will be another extension of this work.

%

\bibliographystyle{abbrvnat}
\bibliography{ref}

\end{document}